\documentclass[12pt]{article}

\usepackage{amsfonts}
\usepackage{amsmath}
\usepackage{amsthm}
\usepackage{amssymb}
\usepackage[normalem]{ulem}
\usepackage{hyperref}

\usepackage{longtable}
\usepackage{subfigure}
\usepackage{pdflscape}

\usepackage{multirow}
\usepackage{color, colortbl}
\definecolor{Gray}{gray}{0.9}
\newenvironment{proofsketch}{\noindent{{Proof sketch.}}}{\hfill}
\usepackage{wrapfig}

\newcommand{\cancel}[1]{}

\usepackage{graphicx}
\usepackage[table]{xcolor}

\newcommand{\CosnHidden}[1]{}
\newcommand{\CAhide}[1]{}

\newtheorem{theorem}{Theorem}[section]

\newtheorem{lemma}[theorem]{Lemma}

\newtheorem{fact}[theorem]{Fact}

\newtheorem{observation}[theorem]{Observation}

\def\inline#1:{\par\vskip 7pt\noindent{\bf #1:}\hskip 10pt}

\def\qed{\mbox{}\hfill $\Box$\\}

\long\def\hideabs #1\hideabsend{}

\def\abs#1{\lvert #1 \rvert}
\newcommand{\card}[1]{\lvert #1\rvert}

\def\deg{d}
\def\e{E}
\def\EE{\varepsilon}
\def\XX{\chi}

\def\cE{\mathcal{E}}

\def\cI{\mathcal{I}}
\def\cP{\mathcal{P}}
\def\cG{\mathcal{G}}
\def\cX{\mathcal{X}}

\def\bE{\mathbb{E}}

\def\NN{98}
\def\MNN{100}
\def\NND{8}

\usepackage{multibib}
\newcites{sup}{References Datasets}%  \citelatex, \nocitelatex, ...

\begin{document}

\begin{titlepage}
\title{Core-Periphery in Networks:  An Axiomatic Approach
\footnote{Supported in part by the Israel Science Foundation (grant 1549/13).}}

\author{
Chen Avin \footnotemark[4]
\and
Zvi Lotker \footnotemark[4]
\and
David Peleg \footnotemark[2]
\and
Yvonne Anne Pignolet \footnotemark[3]
\and
Itzik Turkel \footnotemark[4]
}

\def\thefootnote{\fnsymbol{footnote}}

\footnotetext[4]{
\noindent
Department of Communication Systems Engineering, Ben Gurion University of the Negev,
Beer-Sheva, Israel. 
E-mails:~{\tt \{avin,turkel,zvilo\}@cse.bgu.ac.il}.
}

\footnotetext[2]{
\noindent
Department of Computer Science, The Weizmann Institute of Science, Rehovot, Israel.  
\hbox{E-mail}:~{\tt david.peleg@weizmann.ac.il}.
Supported in part by the Israel Science Foundation (grant 1549/13),
the United States-Israel Binational Science Foundation (grant 2008348),
the I-CORE program of the Israel PBC and ISF (grant 4/11),
%the Israel Ministry of Science and Technology (infrastructures grant),
and the Citi Foundation.}

\footnotetext[3]{
\noindent
ABB Corporate Research, Switzerland. 
\hbox{E-mail}:~{\tt yvonne-anne.pignolet@ch.abb.com}.
}
%\date{}

\maketitle 
\thispagestyle{empty}

\begin{abstract}
Recent evidence shows that in many societies worldwide the relative sizes of the economic and social
\emph{elites} are continuously shrinking.
%and less and less individuals control more and more power.  
%  is awaking concerns and debates.
Is this a \emph{natural} social phenomenon?
 What are the forces that shape this process? 
%and when is it going to end?
% and how small the elite can get?
%and will a smaller and smaller group continue to poses more and more power over society? 
%and what are the forces that shape this
We try to address these questions by studying a 
%high-level
% the most high-level structure that a society can exhibit: a 
%two-tier 
Core-Periphery social structure composed of a social \emph{elite}, 
namely, a relatively small but well-connected and highly influential group 
of powerful individuals, and the rest of society, the {\em periphery}.
Herein, we present a novel axiom-based model for the 
forces governing the mutual influences between the elite and the periphery. 
Assuming a simple set of axioms, capturing the elite's 
\emph{dominance}, \emph{robustness}, \emph{compactness} and \emph{density}, 
we are able to draw strong conclusions about the elite-periphery structure. 
In particular, we show that a \emph{balance of powers} between elite and 
periphery and an elite size that is \emph{sub-linear} in the network size 
are universal properties of elites in social networks that satisfy our axioms.
We note that the latter is in controversy to the common belief that 
the elite size converges to a linear fraction of society 
(most recently claimed to be $1\%$). % \cite{piketty2014capital}). 
We accompany these findings with a large scale empirical study 
on about $\MNN$
%more than a hundred 
real-world networks, which supports our results.
\end{abstract}
\end{titlepage}

%%%%%%%%%%%%%%%%%%%%%%%
\section{Introduction}
%%%%%%%%%%%%%%%%%%%%%%%
In his book \emph{Mind and Society} \cite{pareto1935mind}, Vilfredo Pareto wrote
%DP: Previous formulation: I'm not sure this is considered relevant/appropriate
%The Italian sociologist Vilfredo Pareto wrote in his book 
%\emph{Mind and Society} \cite{pareto1935mind} 
what is by now widely accepted 
by sociologists: ``Every people is governed by an \emph{elite}, 
by a chosen element of the population''.
Indeed, with the exception of some rare examples of utopian or totally 
egalitarian societies, almost all societies exhibit an (often radically) 
uneven distribution of power, influence and wealth among their members, 
and, in particular, between the \emph{elite} and its complement, often called 
the \emph{periphery} or the \emph{masses}.
%
%
%However, defining the elite group precisely turns out to be more delicate.
%The Cambridge Dictionary defines the {\em elite} as:
%\vspace{-10pt}
%\begin{quote}
%``\emph{The richest, most powerful, best educated or best trained group
%in a society.}''
%\end{quote}
%\vspace{-10pt}
%\par\noindent
%This definition politely sweeps away two potentially disturbing properties
%often associated with elites, namely,
%%%%%%Other definitions emphasize in addition
%that they are relatively {\em small} and {\em overly} privileged;
%%%%%``well-connected.''
%consider, for contrast, the unabashed definition offered in Wikipedia
%for the elite:
%\vspace{-10pt}
%\begin{quote}
% ``\emph{An elite ...
%%%%%% in political and sociological theory,
%is a \emph{small} group of people who control a \emph{disproportionate} amount 
%of wealth or political power.}''
%\end{quote}
%\vspace{-10pt}
%\par\noindent
%
%These qualitative definitions capture the division of society 
%(represented as a social network) into two groups, one 
%
%
Typically, the elite is small, powerful and influential and the periphery is
larger, less organized and less dominant. 
This division is usually referred to as a \emph{core-periphery} 
partition \cite{borgatti2000models}\footnote{Hereafter we use the terms 
elite (mostly used for social networks) and 
core (used for any network) interchangeably.}
and the problem of identifying this partition has recently received increased 
interest  \cite{rombach2014core,Zhang2014Identification,holme2005core}.
However, while the core-periphery structure is perhaps the most high-level 
structure of society, it has so far not received a satisfactory
\emph{quantitative} definition.
%nor \emph{qualitative}  definition.
In fact, it seems unlikely that a single formal definition exists that suits 
all elite types in all social networks. 
Moreover, it is not clear that the elite must be unique.

To illustrate this last point, let us consider the following mental experiment.
Sort the members of society by decreasing order 
of influence\footnote{The exact definition of ``influence'' 
is immaterial here, and will be discussed later on.}, 
and add them to a set $\cE$ (representing the intended elite) 
one by one in this order. After the first step, $\cE$ contains only one
member of society, albeit the most influential one, hence clearly it cannot
yet be thought of as ``the elite'' - it simply has insufficient power.
This holds true also for the next few sets obtained in this way. 
On the other extreme, if the process is continued to its conclusion,
we end up with $\cE$ containing the entire society, which is clearly too large
to be considered ``the elite''. The question is therefore: 
at which point along this process does $\cE$ qualify as an elite?

Intuitively, the ``break-point'' where the process should be halted
is the point where adding new members into $\cE$ no longer serves 
to significantly strengthen the group but rather ``dilutes'' its power
(relative to its size). This point is not well-defined,
and depends to some extent on the specific circumstances of the society. 
%Moreover, the ordering of society members by influence is also not unique, 
%and depends on the influence measure used.
In this paper we propose a concrete choice for this break-point, 
referred to as the \emph{power symmetry point} of the elite.
The core and periphery are at their power symmetry point if the 
overall influence of the elite (nearly) equals that of the periphery. 

\begin{figure*}[ht]
\centering
\begin{tabular}{ccc}
\multirow{3}{*}[.5in]{\includegraphics[width=65mm]{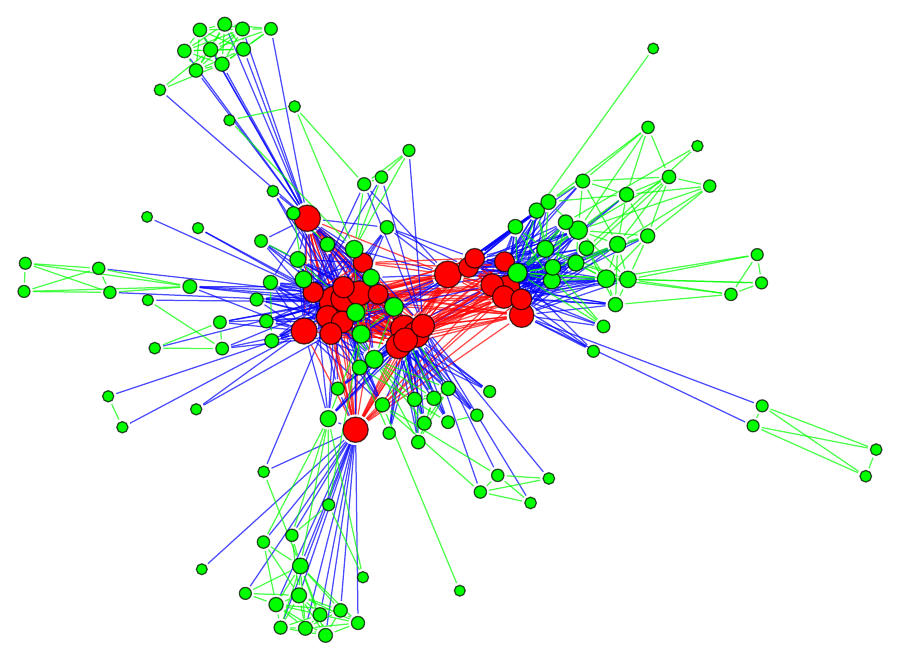}}&
\includegraphics[height=30mm]{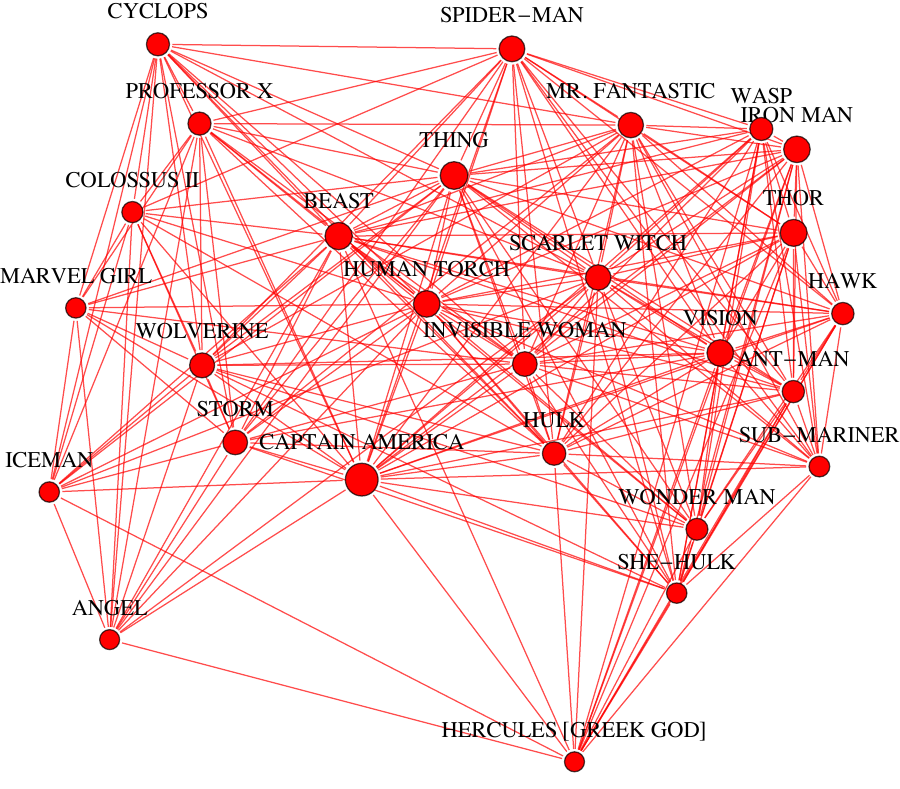}&
\includegraphics[height=30mm]{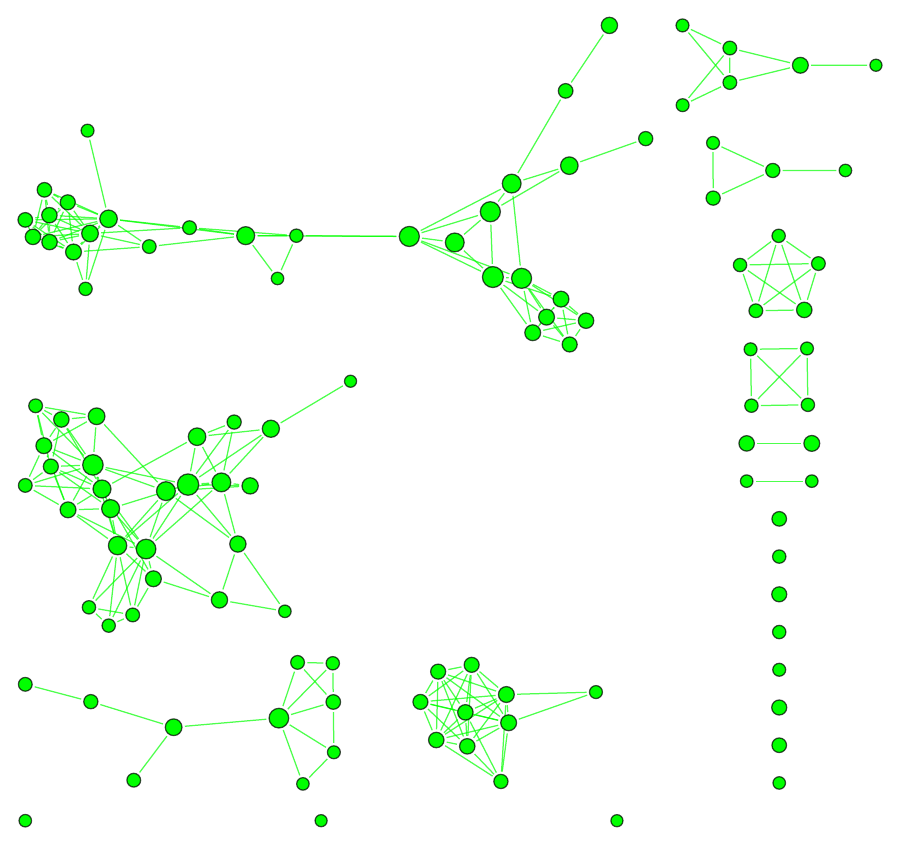}\\
	   & (b) & (c)	\\

	   &
	\includegraphics[height=25mm]{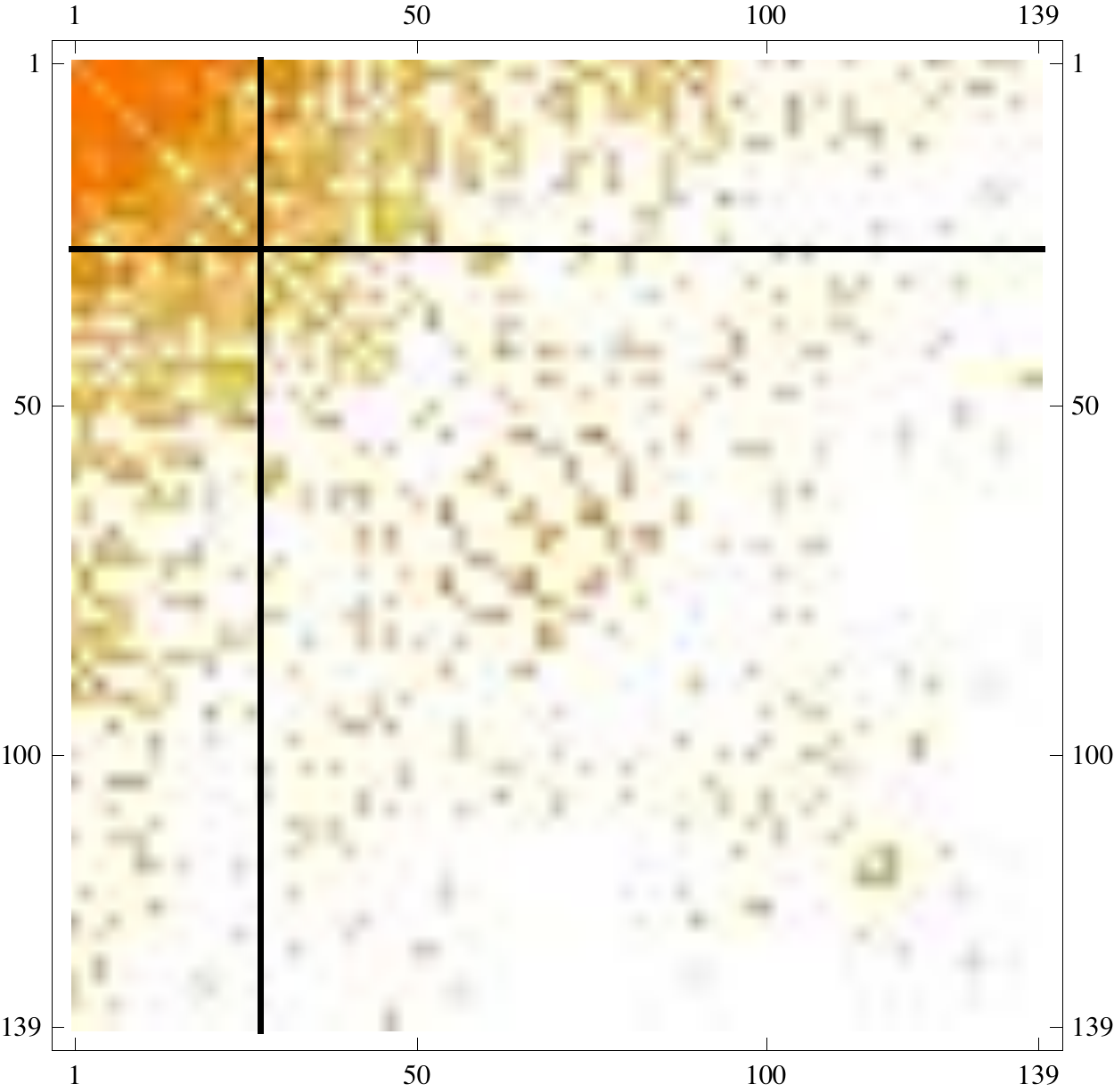} &
	     \includegraphics[height=25mm]{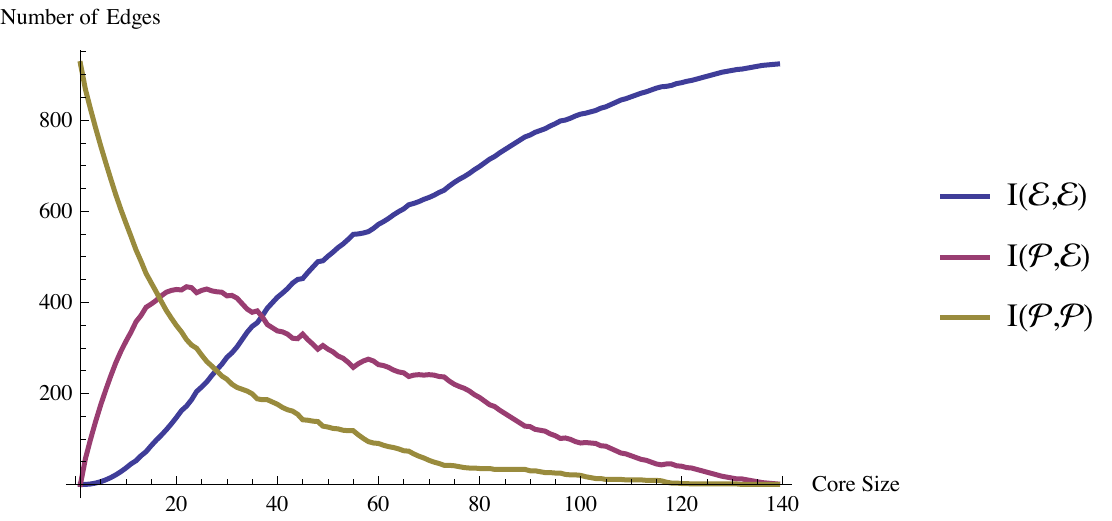} \\
	(a)& (d) & (e)
\end{tabular}
\caption{Illustration of concepts via a fictional example using the 
social network of Marvel's superheroes. There is a link between two heroes 
if they appeared together in many comic books \cite{alberich2002marvel}.
% for the meaning of our terms. 
(a) The network of the most highly connected 139 Marvel's superheroes 
and the 924 links between them, partitioned into 
a core (red vertices with red internal edges) and 
a periphery (green vertices with green internal edges). 
% a core and a periphery as shown by the colors:
% Red vertices (with red internal edges) belong to the core,
% and green vertices (with green internal edges) represent the periphery. 
Blue edges are ``crossing'' edges, connecting core and periphery vertices.
%(423 edges). 
(b) The (dense) core subgraph (27 vertices, 252 edges). 
(c) The (sparser) periphery subgraph (112 vertices, 249 edges).
(d) A representation of the elite-periphery
%Displaying this 
partition in the form of the adjacency matrix of the network
(cf. \cite{Borgatti2013Analyzing}).
(e) The {\em elite influence shift} diagram, based on adding 
the 139 vertices to the elite in order of their degree.
}
\label{fig:marvelnet}
\end{figure*}

Our main contribution is a characterization of elites,
i.e., a set of properties (formulated as ``axioms'') that any definition 
for an elite must adhere to. 
While this characterization does not lead to pinpointing a single definition
for the elite, it narrows down the range of subsets of society that are
suitable candidates to be the elite, and
moreover, it is powerful enough to allow us to derive 
several conclusions concerning basic properties of 
the elite-periphery structure of society.

Let us next describe two of our main conclusions.
The first result, stated in Theorem \ref{thm:symmetry}, is that elites 
satisfying the axioms of our model are at their power symmetry point. This 
%simple conclusion 
implies that the power of the periphery 
%(no matter whether it is measured in wealth, innovation, wisdom, 
%or some other way) 
is similar to that of the elite, or in other words, 
that as society evolves, the ``natural'' core-periphery partition 
maintains a balance between these two groups.   
 
%Our work attempts to understand the basic properties of this structure
%%%%%%%macro-level forces and dynamics that shape these groups  
%in social networks.
%%%%%%%%Can these forces be formalized and evaluated, quantitatively 
%%%%%%%%%as well as qualitatively?
%In particular, we attempt to identify
%%%%%%%%%%%%ascertain that the elite size, 
%%%%%%%%%%%%and its relation with the periphery size, are 
%{\em universal} properties of soical networks, namely, 
%properties that hold in almost every soical network. 

Our second conclusion applies to the \emph{size} of the elite.
%Wealth inequalities between the elite and the masses raise concerns due to 
Recent reports show that the gap between the richest people and the masses 
keeps increasing, and that decreasingly fewer people amass more and more wealth
\cite{Facundo2013The-World,Oxfam-International2014}.
Claims like ``The top 10 percent no longer takes in one-third of our income -- 
it now takes half,'' made by President Obama \cite{whitehouse2013} 
when recently addressing the issue, 
are interpreted as implying that the economic and political elites 
become increasingly more greedy and overbearing. 
Such claims are often used in order to criticize governments and regulatory 
financial institutions for neglecting to cope with this disturbing development. 
The question raised by us is: can society help it, or is this phenomenon 
an unavoidable by-product of some inherent natural properties of society? 
We claim that in fact, one can predict the shrinkage of elite size with time
(as a fraction of the entire society size) based on the very nature of social 
elites. In particular, in our model, such shrinkage is the natural result 
of a combination of three facts: 
First, society grows. Second, elites are \emph{denser} than peripheries 
(informally meaning that they are much better connected).
Third, 
%as proved in (???),  Chen: I don't think we need to talk about proofs here 
the size of a dense elite at the power symmetry 
point is sub-linear in the size of the society.
Combining these facts implies that the fraction of the total population size 
comprised by dense elites will decrease as the population grows with time. 
We prove this formally in Theorem \ref{thm:density}. 
%Combining these facts implies that when the elite is dense, if we express the elite size as a fraction of the total population size, then this fraction will decrease as the population grows with time.
%Specifically, we show that the elite size in a social network of $m$ links
%(or connections) must be $\Theta(\sqrt{m})$
%(implying sublinear size, except in social networks of extremal density).
%repeatedly see it size decreases for example from $10\%$ to $5\%$, and then to $1\%$ and to $0.5\%$. In other words, we claim that the elite size as a fraction of the population will become smaller and smaller as a network grows over time.
The empirical evidence we present in this work lends additional support 
to the above claim.
%and shows that a \emph{small elite size} is a universal property 
%of social networks.

A dual question we are interested in concerns the stable size of the elite
in a growing society: 
%Our hypothesis is that the elite size is a sub-linear function 
%of the population size. 
How small can the elite be while still maintaining its inherent properties? 
We prove that under our model, an elite cannot be smaller than 
$\Omega(\sqrt{m})$, where $m$ is the number of network edges.

Consequently, we assert that the elite's \emph{symmetry of powers} and 
\emph{sublinear size} properties should indeed join the growing list
of \emph{universal properties}\footnote{A property of a class of networks is universal if it keeps appearing in different types of networks and contexts.} of social networks established in recent years,
such as short average path lengths (a.k.a. the ``small world" phenomenon),
high clustering coefficients,
heavy-tailed degree distributions (e.g., scale-free networks), navigability,
and more recently dynamic properties such as
densification and a shrinking diameter
\cite{watts1998collective,albert2002statistical,leskovec2007graph,
newman2010networks}.

As a small illustrative example for the meaning of our terms we consider,
in Figure~\ref{fig:marvelnet} (a)-(c),
the network of top 139 Marvel's superheroes and the 924 links
interconnecting them (where two heroes are connected by a link if 
they appeared together in a story)~\cite{alberich2002marvel}.
We partitioned this network into a core and a periphery as shown 
by the colors in Figure~\ref{fig:marvelnet}(a)\footnote{The partitioning 
methods used, briefly described later on, are tangential to 
the current discussion.}.
%{\bf DP: How was it selected?! Can we add a sentence such as: CA: Yes
Several striking features can be clearly observed in this figure. 
First, the core 
(containing, e.g., Captain America ,Spiderman and Thor) is dense 
and organized while the periphery is much sparser and less structured. 
Second, despite their considerable size difference, 
both the elite and the periphery have almost the same number of internal edges 
($\approx 250$), thus exhibiting what we refer to 
as a \emph{symmetry} of powers. Third, the number of ``crossing'' edges
connecting the core to the periphery is almost twice as large (425), 
reaching most of the vertices in the periphery. Last, 
%but not least, 
the size of the core is ``only'' 27 vertices 
(with 112 vertices in the periphery).
%
%Are those properties an artifact of our selected example and 
%partitioning method? Or are they \emph{universal}, and should be expected 
%to recur in many (or even most) networks? 
%In particular, what are the rules that govern the core size?
%%%%%%%%\\ {\bf DP: Here's where my above concern pops up: if the partitioning was selected by us, then clearly the core size of 27 is an artifact of the partitioning method we used... \\
%%%%%%%% CA: The idea was to intrigue the reader, but feel free to rephrase } \\
%A natural human tendency is to think in fractions and percentages; so 
%our last question above can be rephrased as asking whether a core size 
%of about $19\%$, as in the above example, is a universal property. 
%We claim that it is not, and in fact, asymptotically, as the number of 
%vertices and edges in the network grows, the core of many social networks 
%will {\em not} converge to any constant fraction of the network; 
%it will often be much smaller than $10\%$ or even $1\%$. 
%
We argue, and support by evidence, that it makes sense
%a better view is 
to consider 27 as about $\sqrt{924}$, and more generally, view the core size 
as roughly the square root of the number of edges in the network. 
For almost all networks, a core of  size about $\sqrt{m}$ (where $m$ is 
the number of edges) will not grow as a linear fraction of the number of vertices\footnote{In fact, such growth can only occur in near-complete networks, 
which are rarely seen in real life.}. 
%Interestingly, as mentioned, we prove the complementary claim that 
%for undirected networks the core \emph{cannot} be much smaller than 
%the square root of the number of edges.

The rest of the paper is organized as follows. 
Section \ref{sec:axioms} presents our model of influence and axioms. 
Section \ref{sec:symmetry} discusses the property of power symmetry
(Appendix \ref{sec:symm-rg} expands on power symmetry in random graphs.)
In Section \ref{sec:size} we analyze the size properties of the elite. 
Section \ref{sec:empirical} presents our empirical results.
Related work is provided in Section \ref{sec:relwork}, and
finally, we conclude with a discussion  in Section \ref{sec:discussion}.
%%%%%%%%%%%%%%%%%%%%%%%%%%%%%%%
\section{An Axiomatic Approach}
\label{sec:axioms}
%%%%%%%%%%%%%%%%%%%%%%%%%%%%%%%
The conceptual approach we adopt towards studying core-periphery properties diverges 
from the established traditions in the field of social networks.
The common approach to explaining empirical results on social networks
is based on providing a new concrete (usually random) {\em evolutionary model}
and comparing its predictions to the observed data.
In contrast, we propose an \emph{axiomatic approach} to the questions at hand.
This approach is based on postulating a small set of axioms,
capturing certain expectations about the network structure
and the basic properties that an elite must exhibit in order
to maintain its power in the society. 
%We then use these axioms to infer some additional properties, 
%such as bounds on the elite size.
%
%Employing an axiomatic approach instead of providing an evolutionary model has
Two main advantages of the axiomatic approach are that
%for studying social and complex networks.
%While a basic random model provides us with a mechanism that generates
%networks with properties similar to the ones observed empirically,
%a mechanism alone does not necessarily advance our understanding
%of the {\em meaning} of the phenomenon. In contrast, 
%First, 
a suitable set of axioms attaches an ``interpretation'' or ``semantics'' 
to observed phenomena, and moreover,
%When describing a phenomenon in networks by a set of %axioms one can assess, 
%to some degree, whether the axioms are ``believable'' 
%by how plausible their meaning is.
%The second advantage is that 
once agreeing on the axioms, it becomes possible
to draw conclusions using logical arguments.
For example, it may become possible to infer some information on the 
\emph{asymptotic} behavior of a growing network,
which is not always clear from empirical findings.
%As a consequence, the axiomatic approach is, in some sense,
%stronger than providing a particular model, since once agreeing on the axioms
%and their implications, \emph{every} model should be consistent with them.

The framework presented here for describing the elite-periphery structure 
in a social network revolves around
%and its formation involves two main ingredients.
%which play a central role in the mechanisms by which the elite maintains 
%its power as well as the drives motivating its creation and preservation.
%The first ingredient is 
the fundamental notion of {\em influence} among groups of vertices. 
The underlying assumption is that the elite has more influence
than the rest of the population, allowing it 
on the one hand to control the rest of the population, and on the other 
to protect its members from being controlled by others outside the elite.
We refer to these two capabilities, respectively, as
{\em dominance} and {\em robustness}.

%%%%%%%%%%%%%%%%%%%%%%%%%%%%%%%%%%%
\subsection{Influence and Core-Periphery Partition} 
%\paragraph{Influence and Core-Periphery Partition} 
%%%%%%%%%%%%%%%%%%%%%%%%%%%%%%%%%%%
We consider {\em influence} to be a measurable quantity between any two 
(not necessarily distinct) {\em groups} of vertices $X$ and $Y$,
abstractly denoted by $\cI(X,Y)$.
The groups $X$ and $Y$ do not necessarily have to be distinct, 
and in particular, we are also interested in the {\em internal influence} 
exerted by the vertices of a group $X$ on themselves, denoted $\cI(X,X)$.
% or simply $\cI(X)$.

Two central strengths of an elite in a given society are its {\em dominance} 
on the rest of society on the one hand, and its {\em resistance} to influence 
by the rest of society on the other. 
We will quantify these two aspects shortly.
Denote the elite by a set $\cE$ and the rest of society (the ``periphery'') by $\cP$.
We call such a pair $(\cE, \cP)$ a {\em core-periphery} partition, and
%Given a  $(\cE, \cP)$ partition we 
will be interested in the four basic influence quantities
$\cI(\cE,\cE)$, $\cI(\cP,\cP)$, $\cI(\cE,\cP)$ and $\cI(\cP,\cE)$.

%%%%%%%%%%%%%%%%%%%%%%%
%\subsubsection{Interpretation of influence via neighborhoods}
%%%%%%%%%%%%%%%%%%%%%%%

Let us next lend the abstract notion of influence in social networks
a concrete interpretation.
A social network is modeled as a graph $G=(V,E)$, with a set $V$ 
of $n$ vertices representing the members of society, 
connected by a set $E$ of $m$ edges. 
In a social network, a network edge represents some social relation 
between the two connected vertices, such as friendship, citations, 
following on Twitter, etc. 
For our purposes, we may abstract and unify the interpretation of edges 
by simply stating that an 
%(directed) 
%DP: We have no results for the directed case, so why mention it here
edge connecting two vertices represents some kind of a channel of 
%(directed) 
{\em influence} between the two vertices.
To reflect the self-influence of every individual's opinion on itself, 
we assume that each vertex has a {\em self-loop}, namely, an edge 
connecting it to itself.

%We now use this interpretation to assign a concrete meaning to the 
%influence function $\cI$. %and the axioms.
For every vertex $v$ and set of vertices $X$, 
denote the set of edges connecting $v$ to vertices in $X$ by $\e(v,X)$.
Similarly, for vertex sets $X,Y\subseteq V$, let $\e(X,Y)$ 
denote the set of edges connecting vertices in $X$ to vertices in $Y$.
Define the {\em degree} of $v$ with respect to $X$ as $\deg_X(v)=|\e(v,X)|$.

Given a partition of $V$ to an elite $\cE$ and a periphery $\cP$ sets, 
we can now partition the edge set $E$ to four edge sets 
$\e(\cE,\cE), \e(\cE,\cP), \e(\cP,\cE)$ and $\e(\cP,\cP)$.

\[
%%% A ~=~
\left[ {\begin{array}{c|c}
              &        \\
\e(\cE,\cE)   &  \e(\cE,\cP) \\  
              &        \\
\hline
              &        \\
\e(\cP,\cE)   &  \e(\cP,\cP)  \\
              &        \\
\end{array} } \right]
\]

These sets correspond to the four basic parts of the 
block-model matrix representation \cite{FW-92} of the adjacency matrix
% $A$
of a core-periphery network \cite{Borgatti2013Analyzing}.
 The adjacency matrix of the network $G$ 
 can be now written as in the figure to the right.
See also Figure~\ref{fig:marvelnet}(d) for an example of such representation 
of the Marvel superhero network.

Now, for $X,Y\subseteq V$, define the influence of $X$ on $Y$ as 
%(and symmetrically of $Y$ on $X$) as
\begin{align}
\label{eq:influence=edges}
\cI(X,Y) ~=~ |\e(X,Y)|~.
\end{align}

%$$\cI(X,Y) ~=~ \sum_{v\in Y} \deg_X(v) ~=~ \sum_{v\in X} \deg_Y(v)~.$$
\noindent
Herein we consider only undirected graphs 
where $\cI(X,Y) = \cI(Y,X)$\footnote{A more elaborate model may allow also 
for the possibility of directed edges, representing uni-directional influence; 
this extension of our framework is left for future study.}.
Also note that due to the existence of self-loops, if an vertex $v$ belongs 
to both $X$ and $Y$, then $(v,v)\in \e(X,Y)$. Hence in particular,
$\cI(X,X)\ge\abs{X}$ for every $X\subseteq V$.

%\DPhide{
%$$\cI(X,Y) ~=~ \sum_{v\in Y} \deg_X^{in}(v)~.$$
%} %end DPhide 

%\DPrmrk{Note that I changed the definition from degrees to number of edges, 
%for compatibility with the block model representation, and to fix 
%some inconsistencies with the self loops. This entails some small changes 
%in the counting lemmas.}

%For a vertex $v\in \cE$, let $d_i(v)$ denote the internal degree of $v$ 
%within $\cE$, i.e., how many neighbors of $v$ belong to the set $\cE$; 
%analogously $d_o(v)$ denotes the number of neighbors of $v$ that are 
%outside of the set $\cE$, i.e., in $V\setminus\cE$. 
%
%For an illustration of the resulting notions of internal and external 
%influence see Fig.~\ref{fig:combined_density}.

%%%%%%%%%%%%%%%%%%%%%%%%%%%%%%%%%%
%\begin{figure}%[htb]
%\centering
%\includegraphics[width=0.6\columnwidth]{figures/eliteexample.pdf}
%\caption{Graphical demonstration of the elite and parameters in the Axioms. 
%The total number of edges in the graph is $m$, $\cE$ is the elite set 
%(colored vertices), $\cI(\cE,\cP)$ is the number of edges connecting 
%the elite to the periphery,
%and $\cI(\cE,\cE)$ is the number of edges within the elite.}
%\label{fig:combined_density}
%\end{figure}
%%%%%%%%%%%%%%%%%%%%%%%%%%%%%%%%%%

\begin{figure*}%[t!]
\begin{center}
	\begin{tabular}{cccc}
		\includegraphics[width=.2\columnwidth]{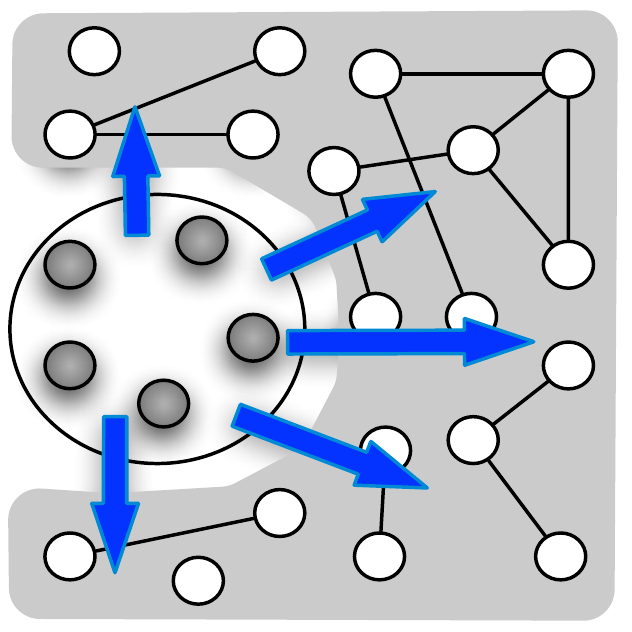} &
		\includegraphics[width=.2\columnwidth]{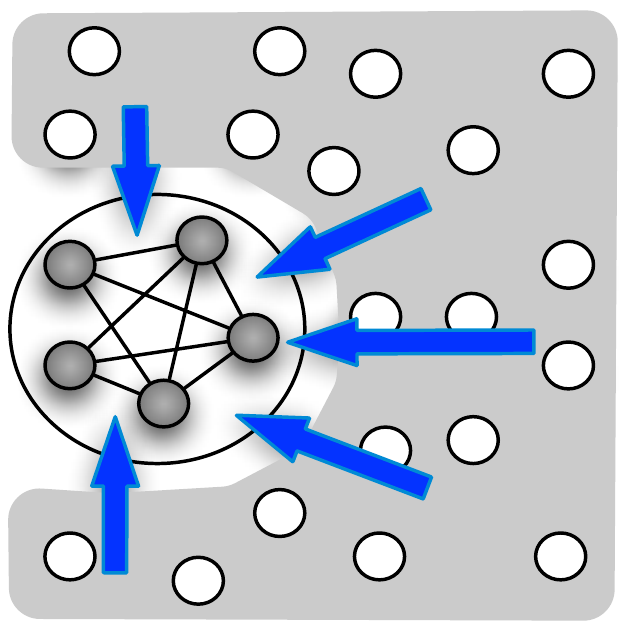} &
		\includegraphics[height=.44\columnwidth]{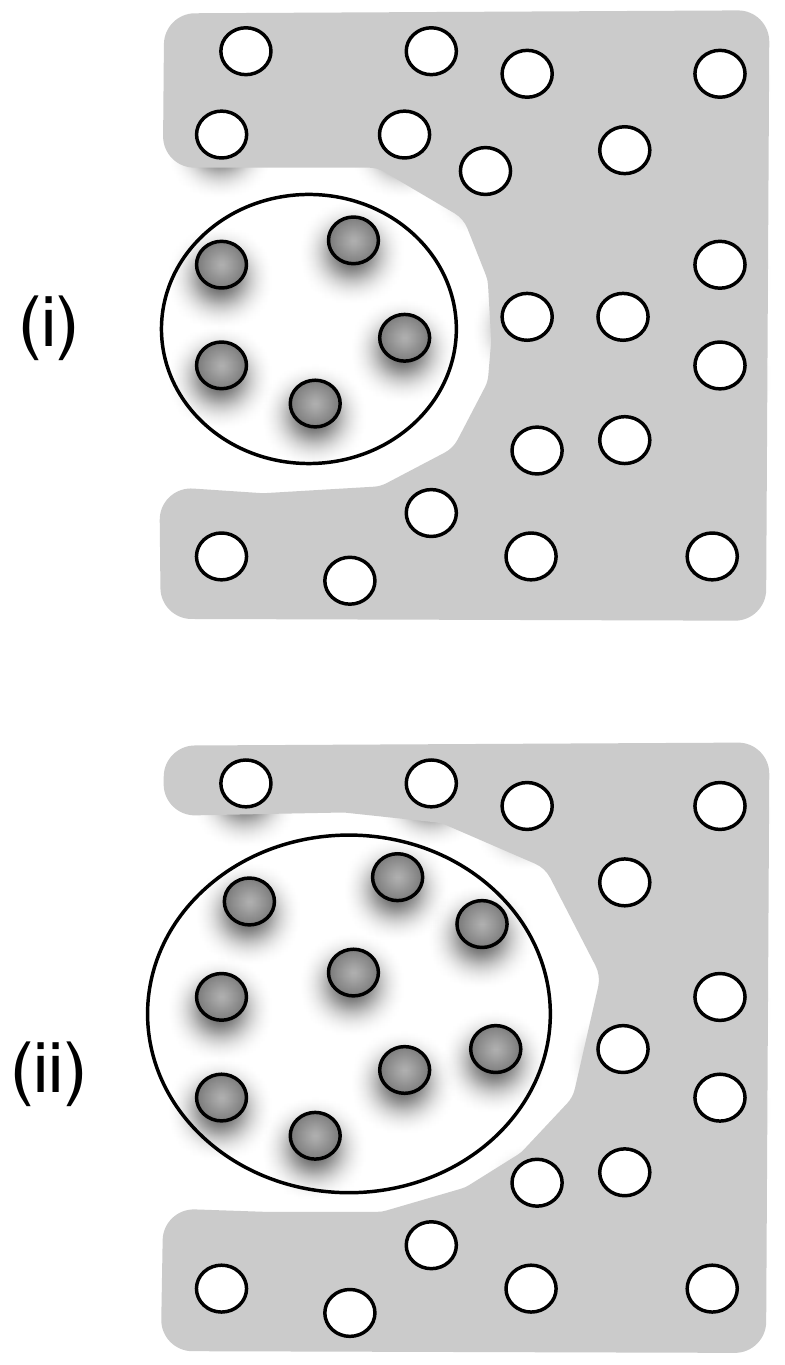} &
		\includegraphics[width=.2\columnwidth]{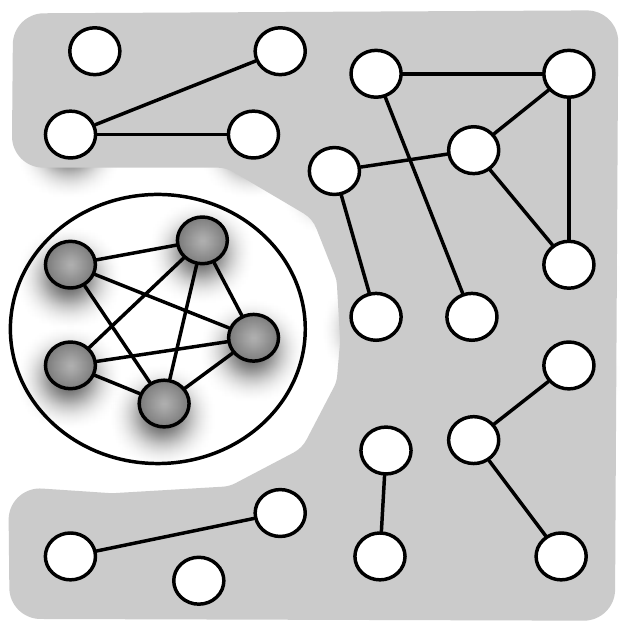} \\
		(A1) Dominance & (A2) Robustness & (A3) Compactness & (A4) Density \\
\end{tabular}
\caption{Graphical illustration of the four axioms. 
% presented in this paper. 
Elite vertices are gray.
%highlighted in gray.
(A1) The elite's external influence (blue edges) dominates the periphery's 
internal influence (black edges). 
(A2) The internal influence of the elite is robust to the periphery's 
external influence.
(A3) The elite is compact: a smaller elite (i) is preferred over a larger one (ii). 
(A4) The elite is denser than the periphery. 
%See text for formal definitions.
}
\label{fig:axioms}
\end{center}
\end{figure*}

A major question that we study is what a natural $(\cE, \cP)$ partition is 
and what are the properties of the basic influence quantities. 
We use the following additional definitions.

Define the {\em total influence} of a group $X$ in society as the sum of 
its internal and external influence on the society: 
\begin{equation}
\label{eq:I(X)}
\cI(X) ~=~ \cI(X,X)+\cI(X,V\setminus X).
\end{equation}
Alternatively, $\cI(X) = \sum_{v \in X} d_V(v)$.
\\
Note that for undirected graphs $\cI(\cE,\cP) = \cI(\cP,\cE)$, 
so by Eq. (\ref{eq:I(X)})
$$\cI(\cE) = \cI(\cP) \;\; \Longrightarrow \;\; \cI(\cE,\cE) = \cI(\cP,\cP).$$

%%%%%%%%%%%%%%%%%%%%%%%
\subsection{Core-Periphery Axioms}
%\paragraph{Core-Periphery Axioms}
%%%%%%%%%%%%%%%%%%%%%%%
We now propose and state 
%(formally and verbally) 
a set of four simple 
axioms to capture elite and periphery properties of a $(\cE,\cP)$ partition,
illustrated in Figure \ref{fig:axioms} (A1)-(A4).
Intuitively, to dominate the rest of society, 
the elite $\cE$ aspires to maintain a large amount of 
{\em external} influence on the periphery $\cP$, higher than or at least 
comparable to the {\em internal} influence that the periphery has on itself. 
Similarly, to maintain its robustness, hold its position and stick 
to its opinions, 
the elite must be able to resist ``outside'' pressure in the form of external 
influence. To achieve that, the elite $\cE$ must maintain the 
{\em internal} influence that it has on itself higher than (or at least not 
significantly lower than) the {\em external} influence exerted on it 
by the periphery. 
Both high {\em dominance} and high {\em robustness} are essential 
for the elite to be able to maintain its superior status in society.
Moreover, all other things being equal, one may expect the elite size 
to tend to be as small as possible. In social terms this may be motivated 
by the idea that if membership in the elite entails benefits, 
then maintaining the elite size as small as possible will increase the share 
coming to each of its members.
We express these requirements in the form of the following three axioms.

%\paragraph{\bf Elite-Periphery Axioms}
Let $c_d$ and $c_r$ be two positive constants. 
\begin{description}
\item[(A1) Dominance:] The elite's influence \emph{dominates} the periphery, 
or formally:
$$\cI(\cE,\cP) \ge c_d \cdot \cI(\cP,\cP).$$
\item[(A2) Robustness:]  The elite can \emph{withstand} outside influence
from the periphery, or formally:
$$\cI(\cE,\cE) \ge c_r \cdot \cI(\cP,\cE).$$
\item[(A3) Compactness:] The elite is a \emph{minimal} set satisfying 
the dominance and robustness axioms (A1) and (A2).
\end{description}

The forth axiom states that the elite members are better connected 
among themselves than the periphery members.
This assertion is justified by some of the classical elite definitions,
which state the elite is a ``clique"' where ``everyone knows everyone''. 
(In fact, having high density is a weaker requirement than being a clique.)

Formally, define the \emph{density} of a set $X \subseteq V$ as 
$\delta_X = \log \card{\e(X,X)} / \log \card{X}$. 
(Written differently, this says that 
the number of edges internal to $X$ is $\card{\e(X,X)} = \card{X}^{\delta_X}$.)

\begin{description}
\item[(A4) Density:] The elite is \emph{denser} than the whole network, namely,
$\delta_V / \delta_\cE < 1$.
\end{description}

Next, we discuss the implications of our axioms on power symmetry 
and elite size in social networks satisfying our axioms.

%%%%%%%%%%%%%%%%%%%%%%%
\section{Core-Periphery Power Symmetry}
\label{sec:symmetry}
%%%%%%%%%%%%%%%%%%%%%%%
A key notion in this paper is the {\em power symmetry point}.
%To motivate it, let us perform the following mental experiment.
Consider a social network $G(V,E)$ and assume some ordering $\pi$ 
(e.g., one reflecting influence via degrees or other centrality measures)
%$v_1,\ldots, v_n$
on the vertices of $V$.
Start with the elite defined as the empty set, and the periphery
containing all the vertices of the network, namely,
$\cE=\emptyset$ and $\cP=V$.
One by one, move the vertices from the periphery to the elite, 
according to the given ordering $\pi$. 
%\vspace {3mm}
%\hspace {-6mm}
%\begin{minipage}{0.52\textwidth}    % decrease size of minipage
As this transition evolves, the influences of the elite 
and the periphery undergo a gradual shift, where
the internal influence $\cI(\cE,\cE)$ increases, 
the internal influence $\cI(\cP,\cP)$ decreases, 
and the cross influence $\cI(\cE,\cP)$ first increases and then decreases.
This can be illustrated by an {\em elite influence shift} diagram,
such as the one in Fig. \ref{fig:marvelnet}(e).
%\end{minipage}
%\ \hspace{0.3cm} \       % generate space between the two boxes
%\parbox{3in}{                       % create the second box 
%\includegraphics[scale=0.25]{figures/elite-influence-shift.jpg}
%}
%\vspace {1mm}
%%%%%%%%%%%%%%%
%\begin{figure}%[htb]
%\centering
%\includegraphics[width=0.3\columnwidth]{figures/elite-influence-shifts.jpg}
%\caption{A schematic elite influence shift diagram.}
%\label{fig:elite influence shift}
%\end{figure}
%%%%%%%%%%%%%%%
The elite size for which the plots of $\cI(\cE,\cE)$ and $\cI(\cP,\cP)$
intersect is referred to as the {\em power symmetry point} of the network 
and the ordering.
% and denoted $x_{sym}^{\pi}$.
More formally, a given partition $(\cE, \cP)$ is said to be 
at a \emph{power symmetry point} if $$\cI(\cE) = \cI(\cP)~.$$
(Similarly, $(\cE, \cP)$ is said to be near its symmetry point if 
$\cI(\cE) \approx \cI(\cP)$, i.e., the two are equal up to a constant factor.)

Note that for a given network there are many partitions at a symmetry point.
In fact, for each ordering $\pi'$ of the vertices there is an elite 
at a symmetry point, obtained by the iterative process described above.

Assuming the first three axioms regarding an elite-periphery partition
$(\cE,\cP)$ of an undirected social network allows us to draw 
our first major result about symmetry. 

\begin{theorem}\label{thm:symmetry}
Let $(\cE,\cP)$) be a core-periphery partition that satisfy the dominance, 
robustness and compactness axioms (A1), (A2) and (A3). Then
the partition is near its symmetry point, i.e., $\cI(\cE) \approx \cI(\cP)$.
Moreover, $\cI(\cE) ~=~ \Theta(m)$ and $\cI(\cP) ~=~ \Theta(m)$.
% where $m$ is the number of edges in the network.
\end{theorem}

%\begin{corollary}
%For every $(\cE, \cP)$ partition such that the elite satisfies the dominance, 
%robustness and compactness axioms (A1), (A2) and (A3),
%\begin{enumerate}
%\item
%$\cI(\cE) ~=~ \Theta(m)~.$
%\item
%$\cI(\cP) ~=~ \Theta(m)~.$
%\item
%$(\cE, \cP)$ is near its symmetry point, i.e., $\cI(\cE) \approx \cI(\cP)$.
%\end{enumerate}
%\end{corollary}

%This justifies the use of the elite at the symmetry point, 
%where $\cI(\cE)=\cI(\cP)$ and $\cI(\cE,\cE)=\cI(\cP,\cP)$, 
%as a ``candidate'' for selecting the elite.
%It is important to recall that there are many partitions at a symmetry point
%(and actually, for each ordering $\pi$ of the vertices there is an elite 
%at a symmetry point).

This means that for any elite that satisfies Axioms (A1)-(A3), 
the overall influence of the elite, $\cI(\cE)$, is (nearly) equal 
to the overall influence of the periphery, $\cI(\cP)$, which makes
the elite-periphery \emph{power symmetry} a universal property 
and justifies the symmetry point
%where $\cI(\cE)=\cI(\cP)$ and $\cI(\cE,\cE)=\cI(\cP,\cP)$, 
as a natural ``candidate'' breakpoint for selecting the elite.
We expect this ``symmetry of powers" between the elite and the periphery 
to be recognized as a significant element in understanding the internal 
balances in social networks.

To prove Theorem \ref{thm:symmetry} we need a simple fact and two lemmas.
Recall that the number of edges in the graph is $m$.
As $(\cE,\cP)$ forms a {\em partition} of the network vertices,
we have
%elementary counting reveals the following.

\begin{fact}
\label{fct:m}
$\cI(\cE,\cE) + \cI(\cE,\cP) 
%+ \cI(\cP,\cE) 
+ \cI(\cP,\cP) = m$.
\end{fact}

%Based on the dominance and robustness axioms, we can infer the following:

\begin{lemma}
%\label{lem:lower}
\label{lem:I_lb}
If $(\cE,\cP)$
%the elite $\cE$ 
satisfies the dominance and robustness axioms (A1)-(A2),
then for some constants $c_1,c_2>0$,
\begin{enumerate}
\item
$\cI(\cE,\cE) ~\ge~ c_1 \cdot \cI(\cP,\cP)$,
\item
$\cI(\cE,\cE) ~\ge~ c_2 \cdot m$.
\end{enumerate}
%$$\cI(\cE,\cE) ~=~ \Omega(m).$$
%$$\cI(\cE,\cE) ~=~ \Omega(m/\RR).$$
\end{lemma}
\proof
By  the two axioms, we have that 
$$\cI(\cE,\cE) ~\ge~ c_r \cdot \cI(\cP,\cE) ~\ge~ c_r c_d \cdot \cI(\cP,\cP),$$
implying the first claim with $c_1=c_r c_d$.

Also, by Fact \ref{fct:m}, combined with the two axioms,
% and the definition of $\RR$,
%we have that 
$$m = \cI(\cE,\cE) + \cI(\cE,\cP) + \cI(\cP,\cP) \le 
\left(1+\frac{1}{c_r}+\frac{1}{c_rc_D}\right) \cI(\cE,\cE).$$
Hence $\cI(\cE,\cE) \ge c_2 m$ for $c_2=(1+1/c_r+1/(c_rc_D))^{-1}$.
The second claim follows.
\qed

Let us next consider the implications of the compactness axiom (A3).
%Interestingly, the most convenient way to interpret these implications
%is by considering its effect on the ``balance of powers'' it imposes
%between the elite and the periphery, as formalized next.

\begin{lemma}
\label{lem: m E-P edges}
If the elite $\cE$ satisfies also the compactness axiom (A3), 
then 
$\cI(\cE,\cP) ~=~ \Omega(m)~.$
\end{lemma}
\proof
%{\bf need to fix! DP: fixed, needs checking}
We assume $|\cP| \ge |\cE|$, or in other words $|\cP| \ge n/2$.
%{\bf DP: Check what happens otherwise.}
Since each vertex of $\cP$ has a self loop, $\cI(\cP,\cP)$ contains at least
$n/2$ self loops. 
%In addition, the requirement that the network is connected implies that 
%there are at least $n/2$ additional edges touching the vertices of $\cP$ 
%(belonging either to $\cI(\cE,\cP)$ or to $\cI(\cP,\cP)$),
%hence $\cI(\cE,\cP) + \cI(\cE,\cP) \ge n/2$. 
Combining this with Axiom (A1), 
we get that $\cI(\cE,\cP) \ge c_d \cdot n/2$. 
Hence if the network has only a linear number of edges altogether,
say, at most $4n/c_2$ edges for the constant $c_2$ of Lemma \ref{lem:I_lb},
then the lemma holds trivially. Hence herafter we consider networks where 
\begin{equation}
\label{eq:nonlinear m}
m>(4/c_2)\cdot n.
\end{equation}

Consider an elite $\cE$ that satisfies Axiom (A3), i.e., it is a minimal set 
of vertices
% in the network 
satisfying Axioms (A1) and (A2).
This implies that for every vertex $v\in\cE$, moving $v$ from $\cE$ to $\cP$
violates either (A1) or (A2).

Let us first consider the case where there exists a vertex $v\in\cE$ whose
movement from $\cE$ to $\cP$ violates the robustness Axiom (A2).
In other words, 
$$\cI(\cE,\cE) -\deg_\cE(v) ~<~ 
c_r \cdot (\cI(\cP,\cE) + (\deg_\cE(v)-1)-\deg_\cP(v)).$$
Rearranging, we get that
$$\cI(\cE,\cP) > \frac{\cI(\cE,\cE) - (1+c_r)\deg_\cE(v)}{c_r}
> \frac{\cI(\cE,\cE) - 2n}{c_r}~.$$
Applying Lemma \ref{lem:I_lb} (2) and Eq. (\ref{eq:nonlinear m}) we get that
$$\cI(\cE,\cP) ~>~ \frac{1}{c_r} \cdot \left(c_2 m - 2n\right)
~>~ \frac{c_2}{2c_r} \cdot m.$$

Next, let us consider the complementary case, where for every vertex $v\in\cE$,
moving $v$ from $\cE$ to $\cP$ does not violate the robustness Axiom (A2).
In this case, for every vertex $v\in\cE$, moving $v$ from $\cE$ to $\cP$ 
necessarily violates the dominance Axiom (A1).
This means that for every vertex $v\in\cE$, 
$$\cI(\cE,\cP) + (\deg_\cE(v)-1) - \deg_\cP(v) ~<~ 
c_d \cdot (\cI(\cP,\cP) + (\deg_\cP(v)+1)).$$
On the other hand we have by Axiom (A1) that
$$\cI(\cE,\cP) ~\ge~ c_d \cdot \cI(\cP,\cP).$$
Adding up these two inequalities and simplifying,
% we get that
$$\deg_\cE(v) < (1+c_d)\cdot \deg_\cP(v) + 2 < 2 \deg_\cP(v) + 2.$$
Summing over all $v\in\cE$,
% we get that
$2\cI(\cE,\cE) < 2\cI(\cE,\cP) + 2|\cE|$,
so
$$\cI(\cE,\cP) ~>~ \cI(\cE,\cE) - |\cE| \ge c_2 m - n ~\ge~ 
\frac{c_2}{2} \cdot m. \qquad\qquad\qquad \Box$$
%\qed

Theorem \ref{thm:symmetry} now follows by the above lemmas.

%\proof[of Theorem \ref{thm:symmetry}]
%{\bf Chen: add proof?}
%\qed

In fact, a slightly stronger observation can be made.
For a compact elite $\cE$, we say that $\cE$ is {\em over-dominant}
if for every $v\in\cE$, moving $v$ from the elite to the periphery will 
{\em not} violate the elites dominance 
(but will, necessarily, violate robustness).

\begin{lemma}
If a compact elite $\cE$ is {\em not} over-dominant, then also
$\cI(\cP,\cP) = \Omega(m)$.
%$$\cI(\cP,\cP) ~=~ \Omega(m)~.$$
\end{lemma}
\proof
Suppose $\cE$ is compact but not over-dominant. 
Let $c_3$ be the constant implied by Lemma \ref{lem: m E-P edges},
namely, such that $\cI(\cE,\cP) \ge c_3\cdot m$.
As in the proof of Lemma \ref{lem: m E-P edges}, we assume that
$|\cP| \ge n/2$, hence $\cI(\cP,\cP) \ge n/2$.
Hence if the network has only a linear number of edges altogether,
say, at most $4(1+c_d)n/c_3$ edges,
% for the constant $c_3$ of Lemma \ref{lem: m E-P edges},
then the lemma holds trivially. Hence hereafter we consider networks where 
$m>(4(1+c_d)/c_3)\cdot n$, or equivalently
\begin{equation}
\label{eq:nonlinear m 2}
n<\frac{c_3}{4(1+c_d)}\cdot m.
\end{equation}
The fact that $\cE$ is not over-dominant means that there is 
some $v\in\cE$ whose movement from the elite to the periphery will 
violate the dominance axiom (A1). Hence
$\cI(\cE,\cP) + (\deg_\cE(v)-1) - \deg_\cP(v) < 
c_d \cdot (\cI(\cP,\cP) + (\deg_\cP(v)+1))$.
Rearranging, we get that
\begin{eqnarray*}
\cI(\cP,\cP) &>& 
\frac{1}{c_d} \left(\cI(\cE,\cP) + \deg_\cE(v) - (1+c_d)(\deg_\cP(v)+1) \right)
\\ &>& \frac{1}{c_d} \left(\cI(\cE,\cP) - (1+c_d)2n\right).
\end{eqnarray*}
By Lemma \ref{lem: m E-P edges} and Eq. (\ref{eq:nonlinear m 2}) we get
\begin{eqnarray*}
\cI(\cP,\cP) &>& \frac{1}{c_d} \left(c_3m - (1+c_d)2n\right) 
\\ &>& \frac{c_3m}{c_d} - \frac{(1+c_d)2}{c_d}\cdot\frac{c_3m}{4(1+c_d)} 
~=~ \frac{c_3}{2c_d}\cdot m.
\end{eqnarray*}
% = \Omega(m).$$
The claim follows.
\qed

Moreover, one can draw an additional interesting 
%(and as of yet unmentioned) 
conclusion concerning a universal property related to the power symmetry point.

\begin{observation} 
\label{obs:max-cross-influence}
For any ordering of the vertices and corresponding elite influence shift 
diagram, the crossing influence $\cI(\cE,\cP)$
%measuring the number of crossing edges between the elite and periphery 
is maximized at the symmetry point. 
\end{observation}

%The result states that for any (e.g., random) order on the vertices 
%in social network, the number of edges that cross from the elite 
%to the periphery is maximized at the symmetry point.
This property is clearly observed in our empirical results (see for example 
Figure \ref{fig:marvelnet}(e) and Section \ref{sec:empirical}),
and we prove it formally for a very general random graph model 
known as the \emph{configuration model} (see Appendix \ref{sec:symm-rg}).

%%%%%%%%%%%%%%%%%%%%%%%
\section{The Size of the Elite}
\label{sec:size}
%%%%%%%%%%%%%%%%%%%%%%%
The above discussion about the symmetry point leaves open the question 
of the elite \emph{size} in networks. 
%Figure \ref{fig:sizeexample}, 
In fact, there are networks for which the elite can be at a symmetry point 
but with significantly different sizes varying from linear size to $\sqrt{n}$. 
(see Figure \ref{fig:sizeexample}). 
Our first contribution in this direction concerns the question of 
how small the elite can be while still preserving its properties.
We show that once satisfying the axioms, the elite cannot be smaller than 
$\Omega(\sqrt{m})$.
%where $m$ denotes the number of edges in the graph.

\begin{theorem}
\label{lem:low_sqrtm}
If $(\cE,\cP)$ 
%Let $(\cE,\cP)$ be a core-periphery partition that 
satisfies the dominance and robustness axioms (A1) and (A2), 
then $\abs{\cE} \ge \Omega(\sqrt{m})$.
%for some constant $c_2>0$,
%$$\abs{\cE} ~\ge~ \sqrt{c_2 \cdot m}~.$$
\end{theorem}

\proof
Graph-theoretical considerations dictate that 
\\
$\cI(\cE,\cE) \le {\abs{\cE} \choose 2} \le \abs{\cE}^2$, 
%\begin{equation}
%\label{eq:vertex-edge-relation}
%\cI(\cE,\cE) ~\le~ {\abs{\cE} \choose 2} 
%%%%% ~=~ \abs{\cE} (\abs{\cE}-1) / 2
%~\le~ \abs{\cE}^2 ~,
%\end{equation}
implying that
$\abs{\cE} \ge \sqrt{\cI(\cE,\cE)}$.
%$$\abs{\cE} ~\ge~ \sqrt{\cI(\cE,\cE)}~.$$
Combined with Lemma \ref{lem:I_lb}(2), 
the theorem follows.
%we get the lower bound on the elite size.
\qed

We complement this result 
by providing an example of what we call a {\em purely elitistic society}, 
where the elite reaches its minimum possible size of 
$\Theta(\sqrt{m})$ in Appendix~\ref{sec:extreme}.

In reality, however, most social networks are not purely elitistic, which 
leaves the question of an upper bound for the ``typical'' elite unanswered: 
does the ``universal'' size of elites (if exists) converge to a linear, 
or a sublinear, function of the network size? 
For illustration, consider the US population of about 314 million people. 
An elite of $0.1\%$ will consist of 314,000 people, while an elite of size
$\sqrt{314M}$ will consist of only about 18,000 people. 
These numbers differ by an order of magnitude; which of them is more plausible?

Considering also Axiom (A4), we can clarify this important point and prove 
that the elite size is sublinear.

\begin{theorem}\label{thm:density}
\label{thm:sublinear}
If $(\cE,\cP)$ 
%Let $(\cE,\cP)$ be a core-periphery partition that 
satisfies the dominance, robustness and density axioms axioms (A1), (A2) 
and (A4), then the elite size is sub-linear in the size of society, namely, 
$|\cE| \le n^{\delta_V / \delta_\cE} = n^c$ for $c<1$. 
\end{theorem}

We find it remarkable that three simple and intuitive assumptions 
lead to such a strong implication on the elite size. 
Note that Theorem \ref{thm:density} is in controversy to the common belief that 
the elite size converges to a linear fraction of the society 
(most recently claimed to be $1\%$ \cite{piketty2014capital}). 
This discrepancy may perhaps be attributed to the fact that our axiom-based
approach characterizes the elite differently than in previous approaches.
In the next section we present evidence that many social networks and complex networks tend to have sublinear elites.

Theorem \ref{thm:density} is derived from
%is a direct consequence of 
the following lemma. 
Recall that the density of a set $X \subseteq V$ is 
$\delta_X = \log \card{\e(X,X)} / \log \card{X}$,
so in particular
\begin{equation}
\label{eq:I(E)-vs-delta}
\cI(\cE,\cE) ~=~ \card{\e(X,X)} ~=~ \card{\cE}^{\delta_{\cE}}~.
\end{equation}
%for a set $X \subseteq V$, if $\cI(X,X)=\Theta(\card{X}^{\delta_X})$ for 
%$0 \le \delta_X \le 2$ then we refer to $\delta_X$ as the {\em density} of $X$.
We show that the elite density tightly determines its size.
%Assume the elite $\cE$ has density $\delta_{\cE}$, i.e., 
%$\cI(\cE,\cE) = \card{\cE}^{\delta_{\cE}}$ then,

\begin{lemma}\label{lem:elsize} 
If $(\cE,\cP)$ 
%In a network where the elite 
satisfies axioms (A1) and (A2) and $\cE$ has density $\delta_{\cE}$, then 
%the elite size satisfies
$\card{\cE} = \Theta\left(m^{\frac{1}{\delta_{\cE}}}\right)$.
%$$\card{\cE} = \Theta\left(m^{\frac{1}{\delta_{\cE}}}\right).$$
\end{lemma}
\proof
By Fact \ref{fct:m}, $\cI(\cE,\cE) \le m$.
By Lemma \ref{lem:I_lb}, $\cI(\cE,\cE) \ge m / c'$.
%(c'\cdot\RR)$.
Hence relying on Eq. \ref{eq:I(E)-vs-delta}, we get
$ m / c' ~\le~ |\cE|^{\delta_{\cE}} ~\le~ m$.
%$$ m / c' ~\le~ |\cE|^{\delta_{\cE}} ~\le~ m~.$$
%%%%%$$ m / (c'\cdot\RR) ~\le~ |\cE|^{1+\gamma} ~\le~ m~.$$
The lemma follows. 
\qed

If $\delta_X=2$ we say that $X$ is \emph{dense}, i.e., in graph terms it is 
very close to a clique. In this case, $\card{\cE} = \Theta(\sqrt{m})$.

%In other words,
%%%%%% in undirected social networks, 
%there is a tradeoff between the density parameter $\delta_{\cE}$ that defines 
%the elite density or cohesiveness and the elite size: 
%the larger the density, the smaller can the elite be.

\begin{figure*}%[!htb]
	\centering
\includegraphics[width=0.96\textwidth]{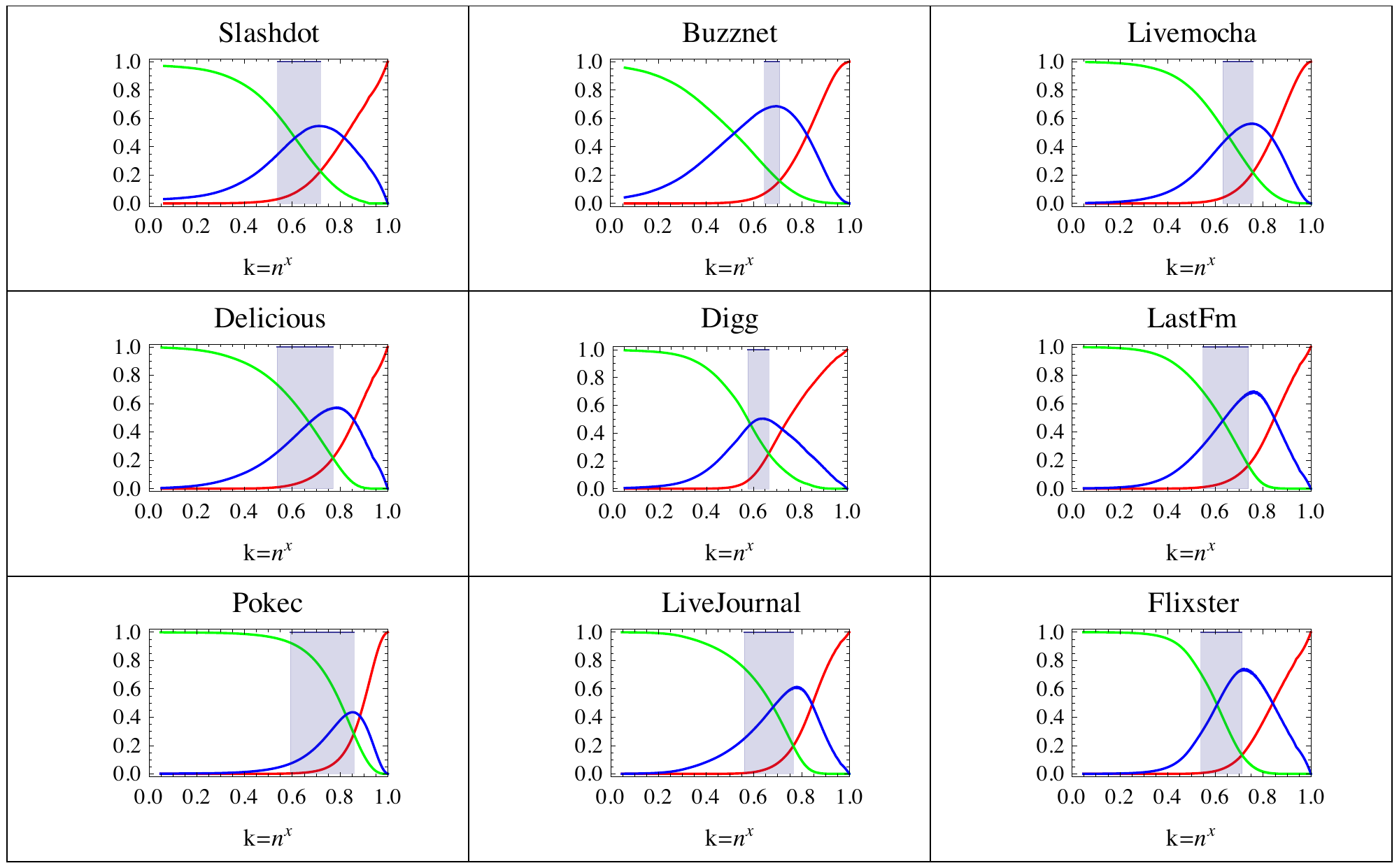}\\
(a)\\
\begin{tabular}{cc}
{\small 
\begin{tabular}{|l|r|r|}
\hline 
\textbf{Network} & \multicolumn{1}{c|}{\textbf{Nodes}} & \multicolumn{1}{c|}{\textbf{Edges}} \\ \hline 
Slashdot & 51083 & 116573  \\ \hline
Buzznet & 101163 & 2763066 \\ \hline
Livemocha & 104103 &2193083  \\ \hline
Delicious & 536408 & 1366136  \\ \hline
Digg & 771229 & 5907413  \\ \hline
LastFm & 1191812 & 4519340 \\ \hline
Pokec & 1632803 &  22301964 \\ \hline
LiveJournal & 2238731 & 12816184 \\ \hline
Flixster & 2523386 & 7918801 \\ \hline 
\multicolumn{3}{c}{ }
\end{tabular}}
 & 
\raisebox{-.5\height}{\includegraphics[width=0.5\textwidth]{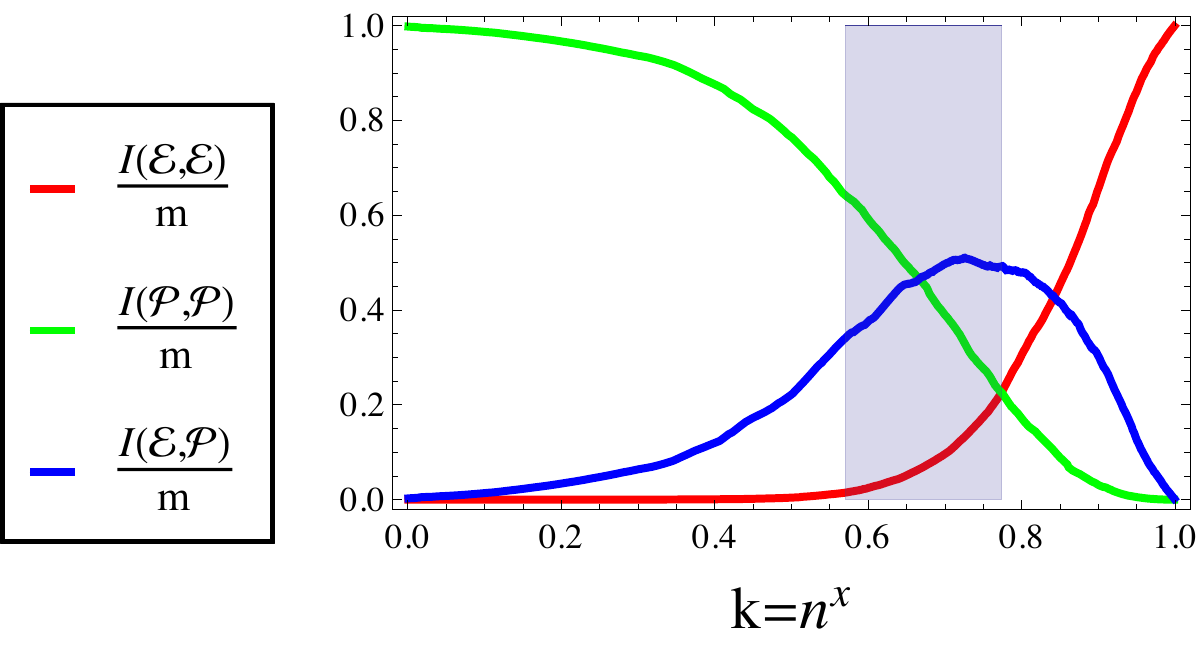} 
}\\ %figures/symmetry/symmetry_legend2.pdf}
(b) & (c)
\end{tabular}	
\caption{(a) The elite influence shift diagrams (plotting the influence ratios 
$\cI(\cE,\cE)/m$, $\cI(\cP,\cP)/m$ and $\cI(\cE,\cP)/m$) for the $k$-rich-clubs
of nine networks. The X axis is on a logarithmic scale, where a point $x$ 
(in [0,1]) represents a $k$-rich club of size $k=n^x$. 
A point $y$ on the $Y$ axis indicates the fraction $\cI(\cdot,\cdot)/m$, 
where $m$ is the total number of edges. 
The range of $x$ values between $x_0$ such that $k=\sqrt{m}=n^{x_0}$ 
and $x_1$, the symmetry point, such that $k=k_{sp}=n^{x_1}$ is highlighted. 
(b) Networks nodes and edges numbers. 
%Next to each network name we indicate its size and number of edges
%(network sizes range from $50,000$ to more than $2,000,000$).
(c) Median results of the elite influence shift diagram for the $k$-rich-club elites for the $\NN$ networks examined, for increasing $k$.
}
\label{fig:symmetry_krichclub_nine}
\end{figure*}

%\newpage
%%%%%%%%%%%%%%%%%%%%%%%%%%%%%%
\section{Empirical Results} 
\label{sec:empirical}
%%%%%%%%%%%%%%%%%%%%%%%%%%%%%%
\subsection{Symmetry Point and the Axioms}
%\paragraph{Symmetry Point and the Axioms}
%%%%%%%%%%%%%%%%%%%%%%%%%%%%%%
We investigated both the static properties of the network elites 
and their dynamics over time. In total we analyzed about $\MNN$ social 
and complex networks. 
Their observed behavior 
(w.r.t. elite and periphery relationships) is surprisingly consistent. 
While it is clear that the issue of identifying the elite members 
of a given network is of paramount importance, 
with some recent developments \cite{rombach2014core,Zhang2014Identification},
this issue is not discussed in the present work. 
Instead, in order to conduct our experiments on given networks, 
we construct an approximation of the elite, $\cE_k$, of size $k$. 
Once the $k$ members of the elite are selected, the rest of the vertices are 
considered as forming the periphery $\cP_{k}$ (of size $n-k$), and the values 
of $I(\cE_k, \cE_k)$, $I(\cE_k, \cP_k)$ and $I(\cP_k, \cP_k)$ can be calculated directly. 
We use two known methods for approximating an elite $\cE_k$ of size $k$. 
The first method
%%, described earlier informally, 
is based on the notion of the 
{\em $k$-rich-club}, and we denote the size-$k$ elite it generates 
for a given network $G$ 
%%(consisting of the $k$ vertices of highest degree)
by $\cE_k^{\mathrm{rich}}(G)$ 
(and omit $G$ and $\mathrm{rich}$ when they are clear from context). 
The second method relies on looking for a {\em $c$-core} in the network 
\cite{dorogovtsev2006k}, and it is a bit more complex.
The two methods produce different elites (albeit with some overlap). 
We present in the main text only the results for the $k$-rich-club elites, 
but similar results were obtained when we used the $c$-core method 
%for selecting the elite. 
(see Appendix~\ref{sec:kcore_results} for details on the results with the $c$-core method).

%and we denote the elite it generates for a given network $G$ using parameter $c$ by $\cE^{c}_{\mathrm{core}}(G)$. 

The $k$-rich-club \cite{zhou2004rich} is perhaps the most intuitive and natural approximation for an elite of size $k$. To build it for a given network $G$, we sort the network vertices according to their degree, and choose the $k$ highest degree vertices as the members of $\cE_{k}^{\mathrm{rich}}$ (breaking ties arbitrarily). Note that this method allows us to generate an elite for any desirable size $1 \le k \le n$.

%We conducted our measurements on the networks examined using elites of various sizes that were created in both methods. Note that although there is some overlap between the elites that were constructed by the two methods, the elites members are \emph{not} all the same, thus each method may exhibit different results. 
%Our results are described in the next sections.

%\CArmrk{Up to here: will show first the symmetry then the dom and roubst and then the size}

%\begin{table}[htdp]
%%\caption{default}
%\begin{center}
%\begin{tabular}{|c|c|c|c||c|c|c|}
%\hline
%& \multicolumn{3}{c||}{at $\sqrt{m}$} & \multicolumn{3}{c|}{at $x_{sym}^{deg}$} \\
%\cline{2-7}
% & Min & Max & Median & Min & Max & Median\\
%\hline
%$\rho_d$  & 0.01 & 1+ &  0.59 &  0.42 & 1+  &  2.28\\
%\hline
%$\rho_r$ & 0.01 & 1+ & 0.07 &  0.02 & 1+ &  0.44 \\
%\hline
%\end{tabular}
%\caption{The observed dominance and robustness ratios for $\NN$ networks with $k$-rich club method for elite selection.} 
%\end{center}
%\label{default}
%\end{table}%

%%%%%%%%%%%%%%%%%%%%%%%%%%%%%%%%%%%
%\subsection{The Empirical Symmetry Point and Elite Axioms}
%\label{ss:sym_point}
%%%%%%%%%%%%%%%%%%%%%%%%%%%%%%%%%%%

We present results from the $\NN$ networks we evaluated. (Detailed results are shown only for nine example networks, but the median values were calculated based on all the $\NN$ networks studied.)
For each network we considered an elite selection method (i.e., $k$-rich-club or $c$-core) and then examined the influence and density of the elite and periphery %(i.e., $\cI(\cE,\cE), \cI(\cP,\cP)$ and $\cI(\cE,\cP)$) 
and validated the axioms for different elite sizes and networks. 
%For each network we considered an elite selection method (i.e., $k$-rich-club or $c$-core) and then constructed the elite influence shift diagram which presents the changes in $\cI(\cE,\cE), \cI(\cP,\cP)$ and $\cI(\cE,\cP)$ as the elite (i.e., $\cE^{k}_{\mathrm{rich}}(G)$) grows from a minimum size (which is 1 for the $k$-rich-club method) to maximum size (which is $n$ for the $k$-rich-club method).
 
We first consider the elite influence shift diagrams which presents the changes in $\cI(\cE_k,\cE_k)$, $\cI(\cP_k,\cP_k)$ and $\cI(\cE_k,\cP_k)$ as the elite (i.e., $\cE^{k}_{\mathrm{rich}}(G)$) grows from its minimum size of $1$ %(which is 1 for the $k$-rich-club method)
to its maximum size %(which is $n$ for the $k$-rich-club method)
of $n$.  Figure \ref{fig:symmetry_krichclub_nine} shows the elite influence shift diagrams 
of the $k$-rich-club for $9$ networks and the median results of the $k$-rich-club for all of the $\NN$ networks in our experiments.
Note that the $X$-axis is logarithmic and presented as $k=n^x$ where $x$ goes from $0$ to $1$. The point $x=0.5$ therefore indicates a $\sqrt{n}$-rich-club.
To normalize the $Y$-axis for different networks,  each graph plots ${\cI(\cE_k,\cE_k)}/{m}$, ${\cI(\cP_k,\cP_k)}/{m}$, and ${\cI(\cE_k,\cP_k)}/{m}$.
% (note that $x=0.5$ is a $\sqrt{n}$-rich-club). 
Observe that the networks exhibit a similar pattern: 
%as the elite size grows, the number $\cI(\cE,\cE)$ of internal edges grows
%as well, and the number $\cI(\cP,\cP)$ of edges in the periphery decreases. 
%Both trends are obvious. 
the number of crossing edges between the elite and the periphery, 
$\cI(\cE_k,\cP_k)$, grows with the elite size $k$ up to some maximum, 
and then starts decreasing. 
An interesting and less obvious pattern is that $\cI(\cE_k,\cP_k)$ is larger 
than $\cI(\cE_k,\cE_k)$ right from the beginning and remains larger 
until the maximum point. The relation between these numbers changes only after 
$\cI(\cE_k,\cP_k)$ begins to decrease, while $\cI(\cE_k,\cE_k)$ continues to grow.

%\begin{figure}%[htb]
%\centering
%\includegraphics[width=0.8\columnwidth]{figures/symmetry/symmetry_Median_n.pdf}
%\caption{Median results of the elite influence shift diagram for the $k$-rich-club elites: The median of $\cI(\cE,\cE)/m$, $\cI(\cP,\cP)/m$ and $\cI(\cE,\cP)/m$ for the $\NN$ networks examined, for increasing $k$. 
%The axes are as in Figure \ref{fig:symmetry_krichclub_nine}.}
%%A point $x$ (in the range [0,1]) on the $X$ axis represents a $k$-rich club 
%%of size $k=n^x$. A point $y$ on the $Y$ axis indicates the percentage 
%%$\cI(\cdot,\cdot)/m$. 
%%%where $m$ is the total number of edges. 
%%The range of $x$ values between $x_0$ such that $k=n^{x_0}=\sqrt{m}$ 
%%and the symmetry point $x_1$ is highlighted.}
%\label{fig:symmetry_krichclub_median}
%\end{figure}

% Commented by Chen.
%(This observation is not directly explained by our theoretical model.)
%\\ {\bf DP: Is last sentence true?}

As mentioned before, a particularly interesting point in these graphs 
is the \emph{symmetry point}, namely, the elite size, denoted as $k_{sp}$, 
where the internal influence of the core and 
the internal influence of the periphery are equal, i.e.,  $\cI(\cE_{k_{sp}},\cE_{k_{sp}})=\cI(\cP_{k_{sp}},\cP_{k_{sp}})$.
Recall that  any elite that satisfies Axioms (A1), (A2) and (A3) must be 
near its symmetry point. 
An important observation (stated earlier in 
Observation \ref{obs:max-cross-influence}) is that $\cI(\cE_k,\cP_k)$ 
achieves it maximum at (or very close to) the symmetry point where $k=k_{sp}$.
%\\ {\bf Do we want to include the following in arxiv version? For now, it is in supplementary material.}\\
% This can actually be proved formally for a large range of random graphs. CA: added.
%See Theorem \ref{thm:maxflowsym}.

\begin{figure*}%[!ht]
\centering
\includegraphics[width=0.9\textwidth]{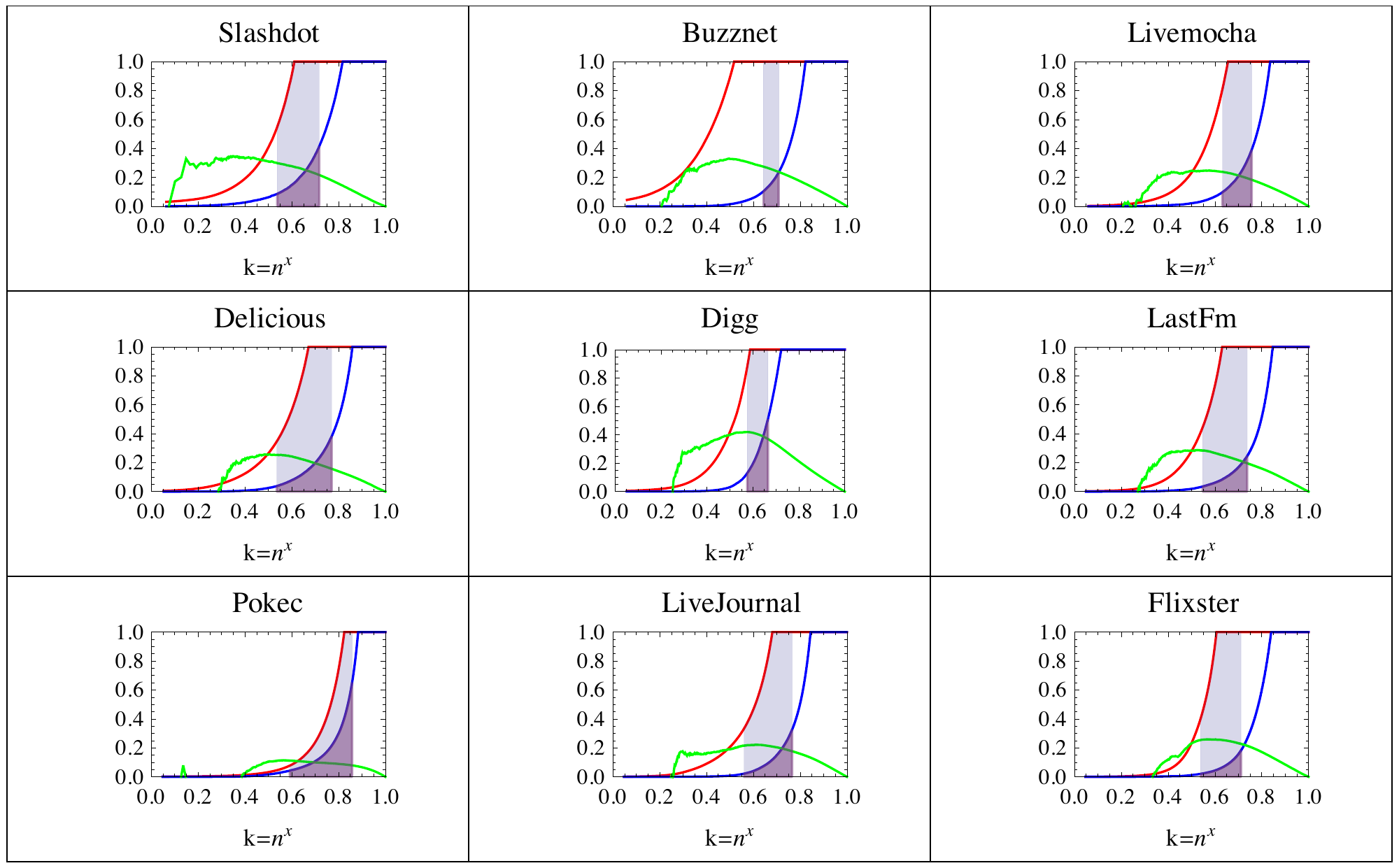} \\
(a)\\
%\vspace{-1.6cm}
\begin{tabular}{cc}
{\small 
\begin{tabular}{|c|c|c|c|}
\hline
& \multicolumn{3}{c|}{at $\sqrt{m}$} \\
\cline{2-4}
 & Min & Max & Median \\
\hline
$\mathrm{dom}(\cE_k)$  & 0.01 & 1+ &  0.59 \\
\hline
$\mathrm{rob}(\cE_k)$ & 0.01 & 1+ & 0.07 \\
\hline
$\mathrm{dns}(\cE_k)$  &  0 -  & 0.51   & 0.24   \\ 
\hline
\hline
& \multicolumn{3}{c|}{at symmetry point} \\
\cline{2-4}
 & Min & Max & Median \\
 \hline
$\mathrm{dom}(\cE_k)$ &  0.42 & 1+  &  2.28 \\ \hline
$\mathrm{rob}(\cE_k)$  &  0.02 & 1+ &  0.44 \\ \hline
$\mathrm{dns}(\cE_k)$  & 0 -   & 0.42  &  0.16  \\ \hline
\multicolumn{4}{c}{ }
\end{tabular}}
%\vspace{+1.6cm}
& 
\raisebox{-.5\height}{
\includegraphics[width=0.5\textwidth]{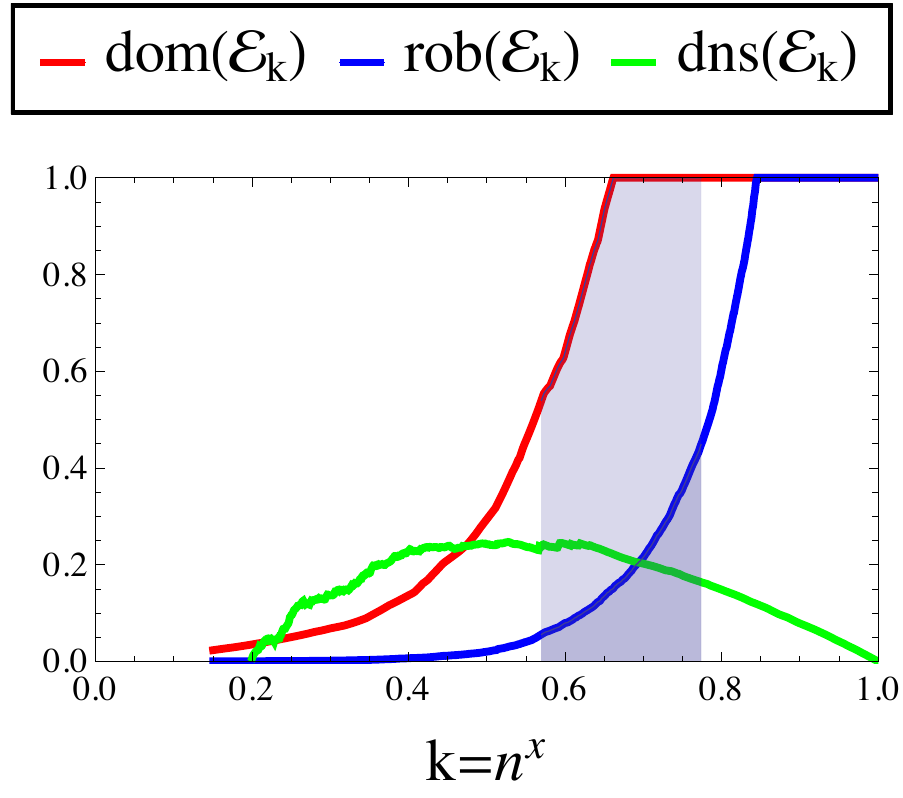}  %figures/symmetry/symmetry_legend2.pdf}
}	\\
(b) & (c)
\end{tabular}
\caption{(a) Observed dominance ratio, $\mathrm{dom}(\cE_k)$, observed robustness ratios, $\mathrm{rob}(\cE_k)$ and the observed density increase, $\mathrm{dns}(\cE_k) = \frac{\delta(\cE_k)-\delta}{\delta}$ for nine example networks. The x axis is logarithmic and marks the sizes of the $k$-rich-club elites with $k=n^x$. The highlighted area is $\cE_k$'s of sizes between $k=\sqrt{m}$ and the symmetry point $k=k_{sp}$
(b) Median, Min and Max results for $\NN$ networks at $k=\sqrt{m}$ and $k=k_{sp}$.
(c) Median results for $\NN$ networks.}
\label{fig:domrob_krichclub_nine}
\end{figure*}

One of the main contributions of this paper is the claim that elites in social networks are ``small'', that is, of sub-linear size. We address this issue in detail shortly, but for now we note that in Figure \ref{fig:symmetry_krichclub_nine}, for both the example networks and the median of all networks, the symmetry point occurs mostly at elite sizes between $n^{0.6}$ and $n^{0.85}$, and its median value is at about $n^{0.77}$.
%We present here only the results for the $k$-rich-club elite, but similar results were obtained when we used the $c$-core method for selecting the elite
%(See ~\ref{ss:kcore_results}.)
%This is a relatively small size of the core.

In all figures we highlight the range $[x_0,x_1]$ of sizes for which 
it was shown that an elite can satisfy Axioms (A1) and (A2). 
More explicitly, this range starts at the lower bound on the elite size, 
i.e., the point $x_0$ such that $k=\sqrt{m}=n^{x_0}$, and ends at the 
symmetry point, i.e., the point $x_1$ such that $k=k_{sp}=n^{x_1}$. 
%\\{\bf Maybe show $x_0$ and $x_1$ on the graphs}\\
Note that since each network has different structure and different average degree, the values $x_0$ and $x_1$ are different for each network
%the $\sqrt{m}$ (as well the symmetry point)
(hence in Figure \ref{fig:symmetry_krichclub_nine} (c), the leftmost point $x_0$ of the highlighted range represents a median value).

%{\bf Chen: Do we want to talk about the other 2 ``crossing'' points, namely: unit dominance and unit robustness.}  

%We now turn to the claim that \emph{small} elites, and in particular elites near the symmetry point, satisfy Axioms (A1) and (A2), that is, they are dominant and robust. Our empirical results supports this claim by defining the {\em observed dominance ratio} and  {\em observed robustness ratio}
%for each elite and networks and showing that elites near the symmetry point are indeed dominant and robust. (Detailed results are presented in \ref{sup:empirical})

We now turn to the claim that \emph{small} elites, and in particular that 
elites near the symmetry point $k_{sp}$ satisfy Axioms (A1), (A2) and (A3), 
that is, they are dominant, robust and denser than the whole network. 
Recall that by Thm. \ref{thm:sublinear}, if an elite satisfies these 
three axioms, then its size must be sublinear.
For a given network $G$ with elite $\cE_k$ (of size $k$) and periphery $\cP_k$, define the 
{\em observed dominance ratio} of $\cE_k$ on $G$ as 
$\mathrm{dom}(\cE_k) = \cI(\cE_k,\cP_k) / \cI(\cP_k,\cP_k)$
and the {\em observed robustness ratio} of $\cE$ on $G$ as 
$\mathrm{rob}(\cE_k) = \cI(\cE_k,\cE_k) / \cI(\cE_k,\cP_k)$.
The {\em observed density} of the $\cE_k$ is denoted as $\delta(\cE_k)$.
Note that the density of the network is $\delta(\cE_n)$ and is denoted simply as $\delta$.
We define the observed density increase as $\mathrm{dns}(\cE_k)=(\delta(\cE_k) - \delta)/\delta$, which indicates for each elite $\cE_k$ the ratio by which 
it is denser than the whole graph.
%Towards evaluating the dominance and robustness of the networks at hand, we first calculate, for every network $G$ under study and elites $\cE^{k-rich}(G)$ of different sizes $k$, their observed dominance ratio $\rho_d(G,k)$ and $\rho_r(G,k)$. 

Figure~\ref{fig:domrob_krichclub_nine} (a) shows the values of the observed dominance and robustness ratios and the observed density increase, $\mathrm{dom}(\cE_k)$, $\mathrm{rob}(\cE_k)$ and  $\mathrm{dns}(\cE_k)$ respectively, for the nine example networks $G$ with $k$-rich club elites of different sizes $k$. Figure~\ref{fig:domrob_krichclub_nine} (c) shows the median of these values for all of the $\NN$ networks.

%\begin{figure}%[htb]
%\centering
%\includegraphics[width=0.8\columnwidth]{figures/symmetry/DomRob_n_legended.pdf}
%\caption{Dominance an Robustness constants for nine example networks. The x axis is logarithmic and marks the sizes of $k$-rich-club with $k=n^x$. The shaded area is k of sizes between $\sqrt{m}$ and the 'symmetry point'.}
%\label{fig:domrob_krichclub_nine}
%\end{figure}

The X-axis for each graph is again on a logarithmic scale, where an $x$ value represents a $k$-rich-club of size $n^x$. 
%There are two plots in each graph, for the observed dominance and robustness ratios $\rho_d(G,k)$ and $\rho_r(G,k)$ respectively. 
%\\ {\bf The constants $c_d$ and $c_r$ are part of the formal definition of the axioms, and do not change from one network to another. What we calculate for a given network with a given elite is the observed ratios.} \\
We focus on values of $\mathrm{dom}(\cE_k)$ and $\mathrm{rob}(\cE_k)$ up to 1, and ignore higher values.  As before, we highlight in each figure the range of $k$ for $k$-rich-club of sizes between $k=\sqrt{m}$ and the symmetry point $k=k_{sp}$. 

Although it is not necessitated by the model or the axioms, 
%\\{\bf is that so?}\\ CA: I think it is OK
it is clear from the figures that as the elite $\cE_k$ grows in size, it attains more dominance over the rest of the network (the periphery, $\cP_k$), namely, its observed dominance ratio increases, and its observed robustness ratio grows as well. We therefore focus on the highlighted area of interest. One can see that the social networks under study exhibit high values for both  $\mathrm{dom}(\cE_k)$ and $\mathrm{rob}(\cE_k)$ at the highlighted area. Recall that by our previous analysis, elites that satisfy Axioms (A1) and (A2) are of size $k \ge \sqrt{m}$. %, but not necessarily equal to $\sqrt{m}$. 
Hence the smallest possible size for an elite is $k=\sqrt{m}$ and at that size, one might expect the elites to exhibit relatively small values of $\mathrm{dom}(\cE_k)$ and $\mathrm{rob}(\cE_k)$. Somewhat surprisingly, the calculated values are relatively high and are bounded away from zero. 
For example, At $k=\sqrt{m}$, the elite of 'Buzznet' has observed dominance 
ratio $\mathrm{dom}(\cE_k)$ way beyond 1, and the elite of 'Digg' has observed 
ratio of about 1, exhibiting high dominance over the rest of the network. 
'Pocek', which exhibits one of the lowest observed dominance ratios 
at $k=\sqrt{m}$, still has $\mathrm{dom}(\cE_k)$ of about $0.1$, 
which is also reasonably high. 
Figure~\ref{fig:symmetry_krichclub_nine}(c) shows the median value of $\mathrm{dom}(\cE_k)$ 
of all $\NN$ networks. At $k=\sqrt{m}$, this median value is very high, about $0.6$, exhibiting very high dominance of elites of this size. 
The dominance of elites of size $k_{sp}$ at the symmetry point is even higher. 
All the example networks have observed dominance ratios 
$\mathrm{dom}(\cE_{sp}) > 1$ so the median $\mathrm{dom}(\cE_{sp})$ value is also greater than 1 at the symmetry point $k_{sp}$. These empirical results show that in real social networks, elites of size greater than $\sqrt{m}$ and smaller than the symmetry point exhibit high dominance and satisfy Axiom (A1). Similar results, albeit somewhat weaker, were obtained for $c$-core elites.
%Appendix~\ref{ss:kcore_results}).   

%\begin{figure}%[htb]
%\centering
%\includegraphics[width=0.7\columnwidth]{figures/symmetry/DomRob_median_n.pdf}
%\caption{Median observed dominance and robustness ratios for $\NN$ networks. 
%The X axis is logarithmic and marks the sizes of $k$-rich-clubs with $k=n^x$. The highlighted area is of elite sizes $k$ between $\sqrt{m}$ and the symmetry point.}
%\label{fig:domrob_krichclub_median}
%\end{figure}

\begin{figure*}[h!]
\begin{center}
\includegraphics[width=0.96\textwidth]{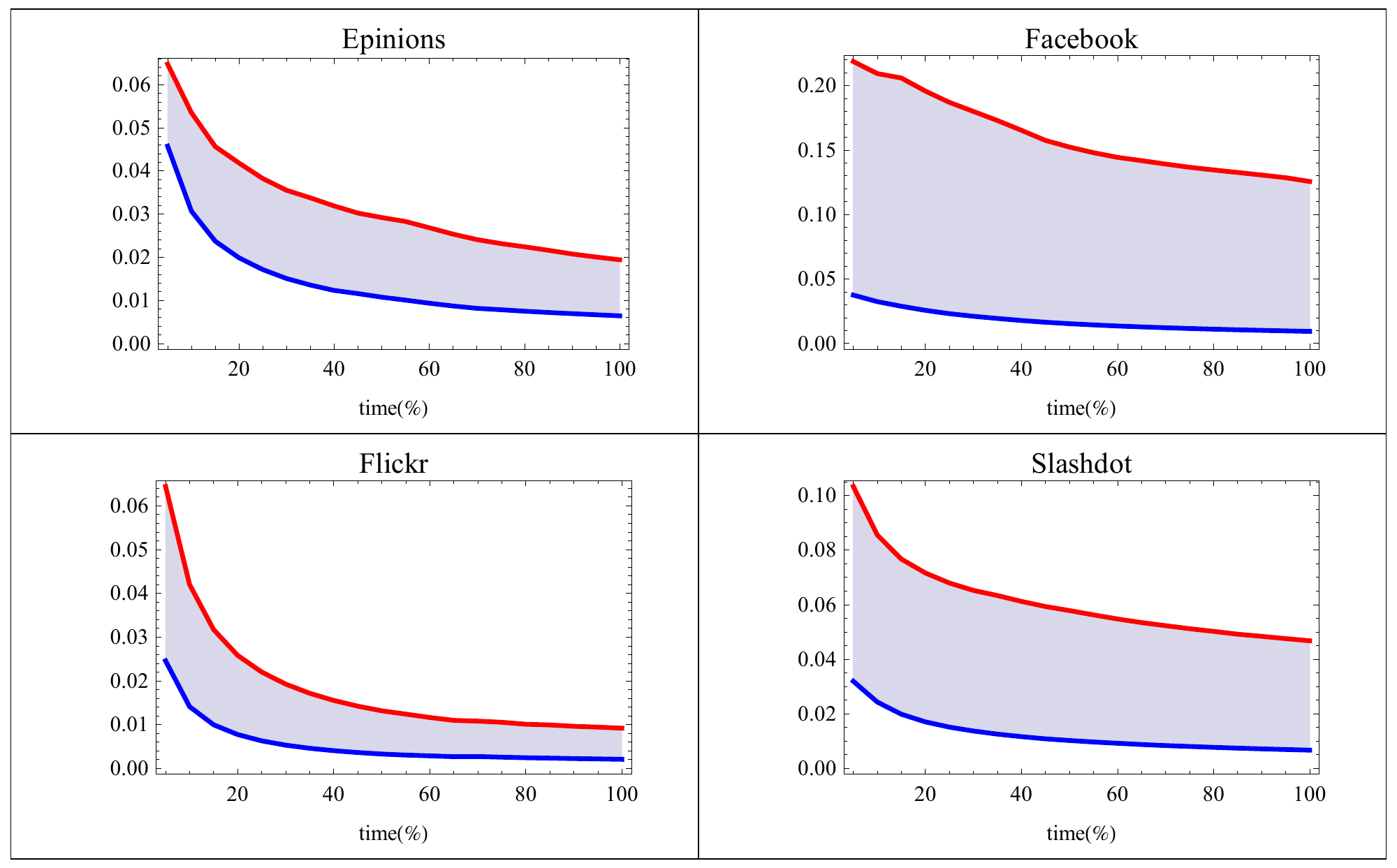}\\
(a)\\
\begin{tabular}{cc}	
{\small 
\begin{tabular}{|l|r|}
\hline 
\textbf{Network} & \multicolumn{1}{m{2cm}|}{\textbf{Duration (months)}} \\ \hline
%\multicolumn{1}{m{1.3cm|}{\begin{tabular}[c]{@{}c@{}}\footnotesize{Duration}\\(\footnotesize{months})\end{tabular}} \\ \hline 
Epinions & 31   \\ \hline
Facebook & 52 \\ \hline
Flickr & 7 \\ \hline
Slashdot & 33 \\ \hline
\multicolumn{2}{c}{ }
\end{tabular}}
 & 
\raisebox{-.5\height}{\includegraphics[width=0.5\textwidth]{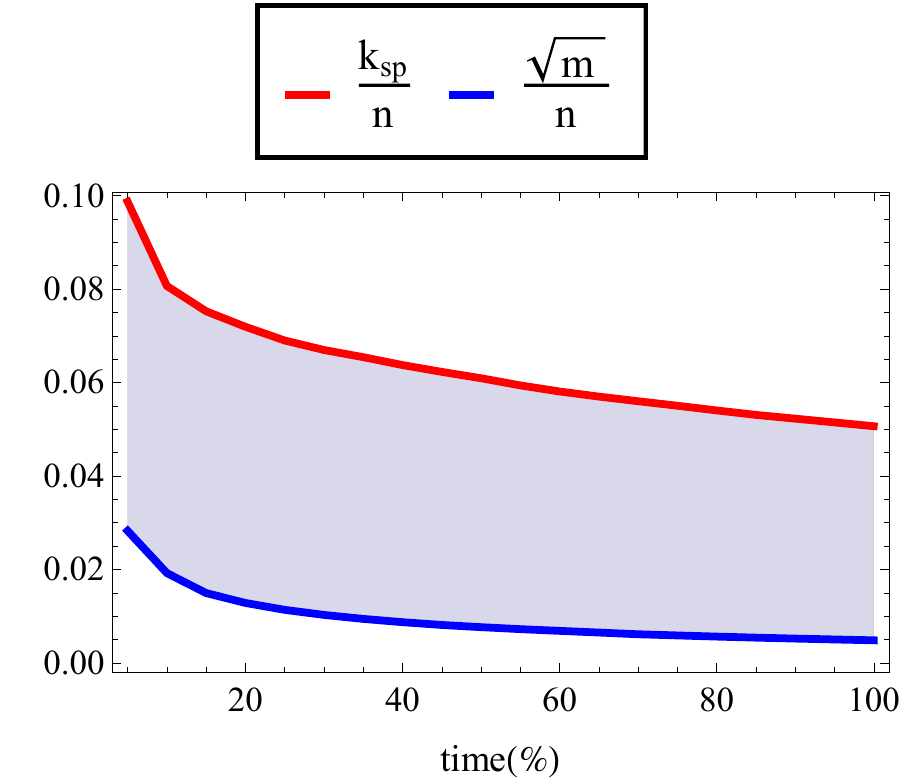}
} \\ %figures/symmetry/Sqrt_symmetry_point_Legned.pdf} \\
			(b) & (c)
\end{tabular}
\end{center}

\bigskip
\bigskip
\caption{(a) The fraction of the elite at the symmetry point, from the whole population, as the networks evolve over time. 4 examples.
(b) Duration of the example networks
(c) The median out of $\NND$ networks. %of the fraction of the elite at the symmetry point, from the whole population, as the networks grows in size over time}
}
\label{fig:symmetry_point_four}
\end{figure*}

Observed robustness ratios exhibit a similar behavior to observed dominance ratios. For small elites of size $k=\sqrt{m}$, all the example networks have observed robustness ratios well bounded away from zero, and so is the median value for all $\NN$ networks. 
The lowest value of any of the example networks, observed for 'Flixter',
is $\mathrm{rob}(\cE_{\sqrt{m}})=0.02$.
The highest value, observed for 'Digg', is $0.13$. The median observed robustness ratio at that point is about $0.1$. At the symmetry point $k_{sp}$ the observed ratios are higher. For example, 'Pokec' has an observed robustness ratio of $\mathrm{rob}(\cE_{sp})=0.65$, and the smallest value, observed for 'Flixter', is $0.18$. The median value at that point is about $0.45$. Again, it follows that elites of size greater than $\sqrt{m}$ and smaller than the symmetry point in social networks have a high robustness. Similar results, in fact somewhat stronger, were obtained for the $c$-core elites.
% (Appendix~\ref{ss:kcore_results}). 
In general, elites produced by $c$-core exhibit lower dominance but higher robustness than those based on $k$-rich clubs. 

Another interesting observation that can be deduced from Figures~\ref{fig:domrob_krichclub_nine} is that in almost all the social networks that we tested, viewing elites of increasing size, we note that the elites attain dominance {\em before} attaining robustness.

Turning to density, the figures clearly reveal that the density of the elite, 
$\delta(\cE_k)$, is significantly higher than the density of the whole graph. 
The median results for $\NN$ networks exhibit that the elite at the symmetry point is about $20\%$ denser than the whole graph. The density at $k=\sqrt{m}$ is even higher. Interestingly, $\cE_k$ seems to reach its maximum density around this point, when $k=\sqrt{m}$.

To conclude the discussion of Figure \ref{fig:domrob_krichclub_nine}, we state that our empirical results provide strong evidence that it is a universal property that the $k$-rich-club elite, at its symmetry point, satisfies Axioms (A1), (A2) and (A4). From Theorem \ref{thm:density} we therefor conclude that $k_{sp}$, the elite size at the symmetry point, is a sublinear function of $n$.

%%%%%%%%%%%%%%%%%%%%%%%
%\subsection*{The Scaling Law: Elite Size at the Symmetry Point}
%\subsection{The Scaling Law: Elite Size}
%\label{ss:sym_point_dynamics}
%\paragraph{The Scaling Law: Elite Size in growing networks}
\subsection{The Scaling Law: Elite Size in growing networks}
%%%%%%%%%%%%%%%%%%%%%%%
%
%(* (strong) Evidence that elites at the symmetry point are small (TBD), \\
%conclude that elites in social networks are small (TBD) *)
%
%In the previous section we presented evidence that elites 
%at the symmetry point exhibit both dominance and robustness. 
In the networks we examined so far, the sizes of the elites at the symmetry point (in both the $k$-rich-club and the $c$-core methods) appear to be sublinear, and specifically, between $n^{0.6}$ and $n^{0.85}$. 
Although this result was obtained for networks of different sizes, all of them were static, and therefore the data collection does not allow us to ascertain whether the size of the elite in the network's symmetry point is indeed asymptotically sublinear. 

%
% Dynamic figure was here; moved to allow page flexibility
%

%Theses sizes are sub-linear, which mean that their porion of the network grows smaller as the network size increases. However, it is hard to demostrate this asymptotic claim, due to limitations of the available data. Altough \hilight{all} networks' symmetry points lie between the mentioned sizes, it is hard to claim for certainty that thses are indeed non linear size (e.g. \hilight{$5\%$  remark: calculate percentage for each symmetry point in table} of the network). 

To study this crucial question more carefully, we turned to data collected on dynamic networks, namely, networks for which information is available on their evolution over time. If the elite size at the symmetry point is indeed sublinear, then one should observe a decrease in the relative size of the elite (its fraction of the network size) as the network grows.

Figure~\ref{fig:symmetry_point_four} presents the data collected for dynamic networks. 
%we try to provide some more evidence of this claim (though, again, due to data limitations, it is not fully decisive). 
We evaluated $\NND$ networks for which information was available about the creation time of each edge. Using this information, we simulated the evolution of the network. As data on the appearance time of each network vertex was not available, we made the assumption that each vertex has joined the network at the same time when the first edge incident to it has appeared. We then divided the evolution time of the network into 20 time frames, each corresponding to a time period during which the network size increased by $5\%$ of the total (final) network size. 
For each time frame $t$, we calculated the elite $\cE_k(t)$ and the symmetry point $k=k_{sp}(t)$ in the snapshot of the network at time $t$. Figure~\ref{fig:symmetry_point_four} shows the ratio 
$r = \card{\cE_{k_{sp}}(t)} / n(t) = k_{sp}(t) / n(t)$ 
% $$r~=~\frac{\card{\cE_{k_{sp}}(t)}}{n(t)}= \frac{k_{sp}(t)}{n(t)}$$
of the number of vertices in the elite at the symmetry point, where $k_{sp}(t)$ is the elite size at the symmetry point at time $t$ and $n(t)$ is the number of vertices in the entire network at time $t$.  In each figure we added the trend of 
$\sqrt{m(t)} / n(t)$,
%$\frac{\sqrt{m(t)}}{n(t)}$ 
where $m(t)$ is the number of edges in the network at time $t$. 
Clearly, unless the network is \emph{dense} (i.e., with an order of $n(t)^2$ edges at time $t$), the ratio $\frac{\sqrt{m(t)}}{n(t)}$ converges to zero.
%Our data with time information available includes $\NND$ networks (See appendix for full information). 
Figure \ref{fig:symmetry_point_four} (a) shows the results for four example networks out of the 8,
% (a different set from the nine networks in previous sections) 
and Figure~ \ref{fig:symmetry_point_four} (b) shows the median result of all $\NND$ networks. These figures demonstrate that the elite size at the symmetry point is a relatively small fraction of the entire network (starting at a median of $10\%$ and ending at a median of $5\%$). Furthermore, it can be seen that as the network evolves and grows, the ratio $r$ decreases, following a pattern similar to the function $\sqrt{m(t)}$, implying that asymptotically indeed the elite size in the symmetry point has a sublinear size.

\section{Related Work}
\label{sec:relwork}

%CosnHidden

As identifying the most influential vertices in a network is crucial 
to understanding its members' behaviour, many studies considered 
a variety of notions related to the elite and core-periphery decompositions (see ~\cite{csermely2013structure} for a recent survey).
Borgatti and Everett~\cite{borgatti2000models} measured the similarity between the adjaceny matrix 
of a graph and the  block matrix ${1~1 \choose 1~0}$.
%$\{\{1,1\},\{1,0\}\}$. 
This captures the intuition that social networks have a dense, 
cohesive core and a sparse, disconnected periphery. 
Core/periphery networks revolve around a set of central vertices 
that are well-connected with each other as well as with the periphery. 
%Peripheral vertices, in contrast, are (sparsely) connected to the core, 
%but not to each other. 
In addition to formalizing these intuitions, Borgatti and Everett devised 
algorithms for detecting core/periphery structures, along with 
statistical tests for verifying a-priori hypotheses~\cite{borgatti2002ucinet}.
Other efforts at identifying such structures and decomposing networks include a coefficient to measure if a network exhibits a clear-cut core-periphery dichotomy~\cite{holme2005core}, a method to extract cores based on a modularity parameter~\cite{da2008centrality} a centrality measure computed as a continuous value along a core-periphery spectrum~\cite{rombach2014core}, a coreness value attributed to each node, qualifying its position and role based on random walks~\cite{della2013profiling}, a detection method using spectral analysis and geodesic paths~\cite{cucuringu2014detection}, and a decomposition method using statistical inference~\cite{Zhang2014Identification}.
Mislove et al.~\cite{mislove-2007-socialnetworks} defined the \emph{core} 
of a network to be any (minimal) set of vertices that satisfies two properties.
First, the core must be essential for ensuring the connectivity of the network 
(i.e., removing it breaks the remainder of the vertices into many small,
disconnected clusters). Second, the core must be strongly connected 
with a relatively small diameter.
%Consequently, a core is a small and well-connected group of vertices 
%that is essential for keeping the rest of the network connected.
%Mislove et al. used an approximation technique previously used in Web graph 
%analysis, based on removing increasing numbers of the highest degree vertices
%and analyzing the connectivity of the remaining graph. 
%The core generated by this technique is thus the largest remaining 
%strongly connected component. 
They observed that for such cores, 
the path lengths increase with the core size when progressively including 
vertices ordered inversely by their degree. 
The graphs studied in \cite{mislove-2007-socialnetworks} have a densely 
connected core comprising of between 1\% and 10\% of the highest degree 
vertices, such that removing this core completely disconnects the graph.
A recent article~\cite{yang2014overlapping} argues that the core/periphery structure is simply the result of several overlapping communities and proposes a community detection method coping with overlap.

One of the first papers to focus on the fact that the highest degree vertices 
are well-connected ~\cite{zhou2004rich} %examined the Autonomous Systems network~\cite{zhou2004rich} 
coined the term {\em rich-club coefficient} for the density of the vertices
of degree $k$ or more. % \footnote{The density of a vertex set $S$ is the ratio between the number of edges connecting vertices in $S$ and the maximum possible number of edges between these vertices, $|S|(|S|-1)$.}.
Colizza et al.~\cite{colizza2006detecting} refined this notion to account 
for the fact that higher degree vertices have a higher probability of sharing 
an edge  than lower degree vertices, and suggested to use baseline networks 
to avoid a false identification of a rich-club. % More precisely, they proposed to use the rich-club coefficient of random uncorrelated networks and/or networks derived by random rewiring of edges, while maintaining the degree distribution of the network. 
Xu et al.~\cite{xu2010rich} shows that the rich-club connectivity has a strong influence on the assortativity and transitivity of a network.
Weighted and hierarchical versions of the rich-club coefficient have been studied 
in~\cite{mcauley2007rich,opsahl2008prominence,serrano2008rich,zlatic2009rich}.
%The question of how the rich-club phenomenon is manifested across hierarchies is studied in~\cite{mcauley2007rich}.

The {\em nestedness} of a network represents the likelihood of a vertex 
to be connected to the neighbors of higher degree vertices. 
%When examining this property, block modeling of adjacency matrices 
%arranged by the degree of the vertices has also been used. 
%For instance, 
Lee et al~\cite{lee2012scaling} %studied such block diagrams 
%for complex network models, and 
defined a %simple 
nestedness measure 
%for unipartite and bipartite networks, in order to capture
capturing the degree 
to which different groups in networks interact.
Yet another perspective is offered in~\cite{hojman2008core} where a network formation game and its equilibria are studied (benefits from connections exhibit decreasing returns and decay with network distance).

%Apart from analyzing the most influential vertices, a wide range of other 
%properties of real-life social networks were studied. 
%For example, the networks created by YouTube, Flickr, Facebook, Wikipedia 
%and LiveJournal have been analyzed in depth in 
%\cite{mislove-2007-socialnetworks, mislove-2008-flickr,viswanath-2009-activity}.
%%Twitter has been studied for its applicability to spot trends, homophily and rumour spreading \cite{kwak2010twitter,jansen2009Twitter}.
%In addition, there is a large body of work studying the evolution of 
%social networks (cf. \cite{albert2002statistical,leskovec2008microscopic,kumar2010structure,viswanath-2009-activity},
%information dissemination and path lengths \cite{albert2002statistical,kempe2003maximizing, leskovec2008planetary,goel2009social}, and community structure \cite{leskovec2008statistical}
%to name but a few examples).

\section{Discussion}\label{sec:discussion}
In this article we address the forces responsible for the creation of elites 
in social networks. We provide axioms modeling the influence relationships 
between the elite and the periphery. We prove that at the power symmetry point,
the size of the elite is sublinear in the size of the network. 
In particular that means that an elite is much smaller than a constant fraction
of the network, evidence of which is often observed in the widening gap 
between the very rich and the rest of societies. 
To understand better what these axioms mean in practice, we studied 
a multitude of large real-world social networks. 
We approximated the elites by rich-clubs 
%(sets of vertices above a certain degree) 
of various sizes. Our findings indicate that in these networks, 
rich-clubs near the symmetry point exhibit elite 
properties such as disproportionate dominance, robustness and density as stated by the axioms. 

Our results do not only advance the theoretical understanding of the elite of social structures, but may also help to improve infrastructure and algorithms targeted at online social networks (e.g., \cite{Avin2014Distributed}), organize institutions better or identify sources of power in social networks in general.

\bibliographystyle{acm}
\bibliography{social}

\begin{thebibliography}{10}

\bibitem{alberich2002marvel}
{\sc Alberich, R., Miro-Julia, J., and Rossell{\'o}, F.}
\newblock Marvel universe looks almost like a real social network.
\newblock {\em arXiv preprint cond-mat/0202174\/} (2002).

\bibitem{albert2002statistical}
{\sc Albert, R., and Barab{\'a}si, A.}
\newblock {Statistical mechanics of complex networks}.
\newblock {\em Reviews of modern physics 74}, 1 (2002), 47--97.

\bibitem{Avin2014Distributed}
{\sc Avin, C., Borokhovich, M., Lotker, Z., and Peleg, D.}
\newblock Distributed computing on core-periphery networks: Axiom-based design.
\newblock In {\em Automata, Languages, and Programming - 41st International
  Colloquium, {ICALP} 2014, Proceedings, Part {II}\/} (2014), pp.~399--410.

\bibitem{borgatti2000models}
{\sc Borgatti, S., and Everett, M.}
\newblock Models of core/periphery structures.
\newblock {\em Social networks 21}, 4 (2000), 375--395.

\bibitem{borgatti2002ucinet}
{\sc Borgatti, S., Everett, M., and Freeman, L.}
\newblock Ucinet: Software for social network analysis.
\newblock {\em Harvard Analytic Technologies 2006\/} (2002).

\bibitem{Borgatti2013Analyzing}
{\sc Borgatti~S.P., E. M. . J.~J.}
\newblock {\em Analyzing Social Networks}.
\newblock London: Sage Publications, 2013.

\bibitem{colizza2006detecting}
{\sc Colizza, V., Flammini, A., Serrano, M., and Vespignani, A.}
\newblock Detecting rich-club ordering in complex networks.
\newblock {\em Nature Physics 2}, 2 (2006), 110--115.

\bibitem{csermely2013structure}
{\sc Csermely, P., London, A., Wu, L.-Y., and Uzzi, B.}
\newblock Structure and dynamics of core/periphery networks.
\newblock {\em Journal of Complex Networks 1}, 2 (2013), 93--123.

\bibitem{cucuringu2014detection}
{\sc Cucuringu, M., Rombach, M.~P., Lee, S.~H., and Porter, M.~A.}
\newblock Detection of core-periphery structure in networks using spectral
  methods and geodesic paths.
\newblock {\em arXiv preprint arXiv:1410.6572\/} (2014).

\bibitem{da2008centrality}
{\sc Da~Silva, M.~R., Ma, H., and Zeng, A.-P.}
\newblock Centrality, network capacity, and modularity as parameters to analyze
  the core-periphery structure in metabolic networks.
\newblock {\em Proceedings of the IEEE 96}, 8 (2008), 1411--1420.

\bibitem{della2013profiling}
{\sc Della~Rossa, F., Dercole, F., and Piccardi, C.}
\newblock Profiling core-periphery network structure by random walkers.
\newblock {\em Scientific reports 3\/} (2013).

\bibitem{dorogovtsev2006k}
{\sc Dorogovtsev, S.~N., Goltsev, A.~V., and Mendes, J. F.~F.}
\newblock K-core organization of complex networks.
\newblock {\em Physical review letters 96}, 4 (2006), 040601.

\bibitem{Facundo2013The-World}
{\sc Facundo, A., Atkinson, A.~B., Piketty, T., and Saez, E.}
\newblock The world top incomes database, 2013.

\bibitem{FW-92}
{\sc Faust, K., and Wasserman, S.}
\newblock Blockmodels: Interpretation and evaluation.
\newblock {\em Social Networks 14\/} (1992), 5--61.

\bibitem{hojman2008core}
{\sc Hojman, D.~A., and Szeidl, A.}
\newblock Core and periphery in networks.
\newblock {\em Journal of Economic Theory 139}, 1 (2008), 295--309.

\bibitem{holme2005core}
{\sc Holme, P.}
\newblock Core-periphery organization of complex networks.
\newblock {\em Physical Review E 72}, 4 (2005), 046111.

\bibitem{lee2012scaling}
{\sc Lee, D., Maeng, S., and Lee, J.}
\newblock Scaling of nestedness in complex networks.
\newblock {\em Journal of the Korean Physical Society 60}, 4 (2012), 648--656.

\bibitem{leskovec2007graph}
{\sc Leskovec, J., Kleinberg, J., and Faloutsos, C.}
\newblock {Graph evolution: Densification and shrinking diameters}.
\newblock {\em Transactions on Knowledge Discovery from Data (TKDD) 1}, 1
  (2007), 2.

\bibitem{mcauley2007rich}
{\sc McAuley, J., da~Fontoura~Costa, L., and Caetano, T.}
\newblock Rich-club phenomenon across complex network hierarchies.
\newblock {\em Applied Physics Letters 91\/} (2007), 084103.

\bibitem{mislove-2007-socialnetworks}
{\sc Mislove, A., Marcon, M., Gummadi, K.~P., Druschel, P., and Bhattacharjee,
  B.}
\newblock {Measurement and Analysis of Online Social Networks}.
\newblock In {\em Internet Measurement Conference (IMC'07)\/} (2007).

\bibitem{newman2010networks}
{\sc Newman, M.}
\newblock {\em Networks: an introduction}.
\newblock Oxford University Press, 2010.

\bibitem{opsahl2008prominence}
{\sc Opsahl, T., Colizza, V., Panzarasa, P., and Ramasco, J.}
\newblock Prominence and control: The weighted rich-club effect.
\newblock {\em Physical review letters 101}, 16 (2008), 168702.

\bibitem{Oxfam-International2014}
{\sc {Oxfam International}}.
\newblock Working for the few. political capture and economic inequality,
  January 2014.

\bibitem{pareto1935mind}
{\sc Pareto, V.}
\newblock {\em The mind and society: Trattato di sociologia generale}.
\newblock AMS Press, 1935.

\bibitem{piketty2014capital}
{\sc Piketty, T.}
\newblock {\em Capital in the Twenty-first Century}.
\newblock Harvard University Press, 2014.

\bibitem{rombach2014core}
{\sc Rombach, M.~P., Porter, M.~A., Fowler, J.~H., and Mucha, P.~J.}
\newblock Core-periphery structure in networks.
\newblock {\em SIAM Journal on Applied mathematics 74}, 1 (2014), 167--190.

\bibitem{serrano2008rich}
{\sc Serrano, M.}
\newblock Rich-club vs rich-multipolarization phenomena in weighted networks.
\newblock {\em Physical Review E 78}, 2 (2008), 026101.

\bibitem{whitehouse2013}
{\sc {The White House, Office of the Press Secretary}}.
\newblock Remarks by the president on economic mobility, December 2013.

\bibitem{watts1998collective}
{\sc Watts, D., and Strogatz, S.}
\newblock Collective dynamics of `small-world' networks.
\newblock {\em nature 393}, 6684 (1998), 440--442.

\bibitem{xu2010rich}
{\sc Xu, X., Zhang, J., and Small, M.}
\newblock Rich-club connectivity dominates assortativity and transitivity of
  complex networks.
\newblock {\em Phys. Review E 82}, 4 (2010), 046117.

\bibitem{yang2014overlapping}
{\sc Yang, J., and Leskovec, J.}
\newblock Overlapping communities explain core-periphery organization of
  networks.

\bibitem{Zhang2014Identification}
{\sc Zhang, X., Martin, T., and Newman, M. E.~J.}
\newblock Identification of core-periphery structure in networks.
\newblock {\em CoRR abs/1409.4813\/} (2014).

\bibitem{zhou2004rich}
{\sc Zhou, S., and Mondrag{\'o}n, R.}
\newblock The rich-club phenomenon in the internet topology.
\newblock {\em Communications Letters, IEEE 8}, 3 (2004), 180--182.

\bibitem{zlatic2009rich}
{\sc Zlatic, V., Bianconi, G., Diaz-Guilera, A., Garlaschelli, D., Rao, F., and
  Caldarelli, G.}
\newblock On the rich-club effect in dense and weighted networks.
\newblock {\em European Physical Journal B-Condensed Matter and Complex Systems
  67}, 3 (2009), 271--275.

\end{thebibliography}


\begin{thebibliography}{100}

\bibitem{BGU-Academia}
Academia. \url{http://proj.ise.bgu.ac.il/sns/academia.html}.

\bibitem{SNAP-Amazon}
Amazon. \url{http://snap.stanford.edu/data/com-Amazon.html}.

\bibitem{SNAP-amazon0302}
amazon0302. \url{http://snap.stanford.edu/data/amazon0302.html}.

\bibitem{SNAP-amazon0312}
amazon0312. \url{http://snap.stanford.edu/data/amazon0312.html}.

\bibitem{BGU-AnyBeat}
Anybeat. \url{http://proj.ise.bgu.ac.il/sns/anybeat.html}.

\bibitem{Konect-arXivHep-thKDDCup-Reference}
arxivhep-thkddcup-reference.
  \url{http://konect.uni-koblenz.de/networks/hep-th-citations}.

\bibitem{SNAP-as-Caida}
as-caida. \url{http://snap.stanford.edu/data/as-caida.html}.

\bibitem{SNAP-as-Skitter}
as-skitter. \url{http://snap.stanford.edu/data/as-skitter.html}.

\bibitem{Konect-BaiduInternal-Reference}
Baiduinternal-reference.
  \url{http://konect.uni-koblenz.de/networks/zhishi-baidu-internallink}.

\bibitem{Konect-BaiduRelated-Reference}
Baidurelated-reference.
  \url{http://konect.uni-koblenz.de/networks/zhishi-baidu-relatedpages}.

\bibitem{ASU-Blog}
Blog. \url{http://socialcomputing.asu.edu/datasets/BlogCatalog}.

\bibitem{ASU-Blog2}
Blog2. \url{http://socialcomputing.asu.edu/datasets/BlogCatalog2}.

\bibitem{ASU-Blog3}
Blog3. \url{http://socialcomputing.asu.edu/datasets/BlogCatalog3}.

\bibitem{ASU-Buzznet}
Buzznet. \url{http://socialcomputing.asu.edu/datasets/Buzznet}.

\bibitem{SNAP-ca-AstroPh}
ca-astroph. \url{http://snap.stanford.edu/data/ca-AstroPh.html}.

\bibitem{SNAP-ca-CondMat}
ca-condmat. \url{http://snap.stanford.edu/data/ca-CondMat.html}.

\bibitem{SNAP-ca-HepPh}
ca-hepph. \url{http://snap.stanford.edu/data/ca-HepPh.html}.

\bibitem{Konect-Catster-Social}
Catster-social.
  \url{http://konect.uni-koblenz.de/networks/petster-friendships-cat}.

\bibitem{Konect-CatsterDogster-Social}
Catsterdogster-social.
  \url{http://konect.uni-koblenz.de/networks/petster-carnivore}.

\bibitem{SNAP-cit-HepPh}
cit-hepph. \url{http://snap.stanford.edu/data/cit-HepPh.html}.

\bibitem{SNAP-cit-HepTh}
cit-hepth. \url{http://snap.stanford.edu/data/cit-HepTh.html}.

\bibitem{SNAP-cit-Patents}
cit-patents. \url{http://snap.stanford.edu/data/cit-Patents.html}.

\bibitem{Konect-CiteSeer-Reference}
Citeseer-reference. \url{http://konect.uni-koblenz.de/networks/citeseer}.

\bibitem{Konect-CoraCitation-Reference}
Coracitation-reference.
  \url{http://konect.uni-koblenz.de/networks/subelj-cora}.

\bibitem{Konect-DBLP-Contact}
Dblp-contact. \url{http://konect.uni-koblenz.de/networks/dblp-coauthor}.

\bibitem{SNAP-DBLP}
Dblp. \url{http://snap.stanford.edu/data/com-DBLP.html}.

\bibitem{ASU-Delicious}
Delicious. \url{http://socialcomputing.asu.edu/datasets/Delicious}.

\bibitem{Konect-Digg-Communication}
Digg-communication.
  \url{http://konect.uni-koblenz.de/networks/munmun-digg-reply}.

\bibitem{ASU-Digg}
Digg. \url{http://socialcomputing.asu.edu/datasets/Digg}.

\bibitem{Konect-Dogster-Social}
Dogster-social.
  \url{http://konect.uni-koblenz.de/networks/petster-friendships-dog}.

\bibitem{ASU-Douban}
Douban. \url{http://socialcomputing.asu.edu/datasets/Douban}.

\bibitem{SNAP-ego-Gplus}
ego-gplus. \url{http://snap.stanford.edu/data/egonets-Gplus.html}.

\bibitem{SNAP-ego-Twitter}
ego-twitter. \url{http://snap.stanford.edu/data/egonets-Twitter.html}.

\bibitem{SNAP-email-Enron}
email-enron. \url{http://snap.stanford.edu/data/email-Enron.html}.

\bibitem{SNAP-email-EuAll}
email-euall. \url{http://snap.stanford.edu/data/email-EuAll.html}.

\bibitem{Konect-Epinions-Social}
Epinions-social. \url{http://konect.uni-koblenz.de/networks/epinions}.

\bibitem{SNAP-epinions}
epinions. \url{http://snap.stanford.edu/data/soc-Epinions1.html}.

\bibitem{Konect-Facebook-Communication}
Facebook-communication.
  \url{http://konect.uni-koblenz.de/networks/facebook-wosn-wall}.

\bibitem{Konect-Facebook-WOSN--Social}
Facebook-wosn--social.
  \url{http://konect.uni-koblenz.de/networks/facebook-wosn-links}.

\bibitem{Konect-Flickr-Social}
Flickr-social. \url{http://konect.uni-koblenz.de/networks/flickr-growth}.

\bibitem{SNAP-flickr}
flickr. \url{http://snap.stanford.edu/data/web-flickr.html}.

\bibitem{ASU-Flickr}
Flickr. \url{http://socialcomputing.asu.edu/datasets/Flickr}.

\bibitem{Konect-FlickrLinks-Social}
Flickrlinks-social. \url{http://konect.uni-koblenz.de/networks/flickr-links}.

\bibitem{ASU-Flixster}
Flixster. \url{http://socialcomputing.asu.edu/datasets/Flixster}.

\bibitem{ASU-Foursquare}
Foursquare. \url{http://socialcomputing.asu.edu/datasets/Foursquare}.

\bibitem{ASU-Friendster}
Friendster. \url{http://socialcomputing.asu.edu/datasets/Friendster}.

\bibitem{BGU-GooglePlus}
Googleplus. \url{http://proj.ise.bgu.ac.il/sns/googlep.html}.

\bibitem{SNAP-higgs-twitter-friendship}
higgs-twitter-friendship.
  \url{http://snap.stanford.edu/data/higgs-twitter.html}.

\bibitem{SNAP-higgs-twitter-mention}
higgs-twitter-mention. \url{http://snap.stanford.edu/data/higgs-twitter.html}.

\bibitem{SNAP-higgs-twitter-retweet}
higgs-twitter-retweet. \url{http://snap.stanford.edu/data/higgs-twitter.html}.

\bibitem{Konect-HudongInternal-Reference}
Hudonginternal-reference.
  \url{http://konect.uni-koblenz.de/networks/zhishi-hudong-internallink}.

\bibitem{Konect-HudongRelated-Reference}
Hudongrelated-reference.
  \url{http://konect.uni-koblenz.de/networks/zhishi-hudong-relatedpages}.

\bibitem{ASU-Hyves}
Hyves. \url{http://socialcomputing.asu.edu/datasets/Hyves}.

\bibitem{ASU-LastFm}
Lastfm. \url{http://socialcomputing.asu.edu/datasets/Last.fm}.

\bibitem{Konect-LibimsetiCZ-Social}
Libimseticz-social. \url{http://konect.uni-koblenz.de/networks/libimseti}.

\bibitem{SNAP-LiveJournal}
Livejournal. \url{http://snap.stanford.edu/data/soc-LiveJournal1.html}.

\bibitem{ASU-LiveJournal}
Livejournal. \url{http://socialcomputing.asu.edu/datasets/LiveJournal}.

\bibitem{SNAP-LiveJournalCom}
Livejournalcom. \url{http://snap.stanford.edu/data/com-LiveJournal.html}.

\bibitem{ASU-Livemocha}
Livemocha. \url{http://socialcomputing.asu.edu/datasets/Livemocha}.

\bibitem{SNAP-loc-brightkite}
loc-brightkite. \url{http://snap.stanford.edu/data/loc-brightkite.html}.

\bibitem{SNAP-loc-gowalla}
loc-gowalla. \url{http://snap.stanford.edu/data/loc-gowalla.html}.

\bibitem{SNAP-Oregon-1-1}
Oregon-1-1. \url{http://snap.stanford.edu/data/oregon1.html}.

\bibitem{SNAP-Oregon-2-1}
Oregon-2-1. \url{http://snap.stanford.edu/data/oregon2.html}.

\bibitem{SNAP-p2p-Gnutella31}
p2p-gnutella31. \url{http://snap.stanford.edu/data/p2p-Gnutella31.html}.

\bibitem{SNAP-Pokec}
Pokec. \url{http://snap.stanford.edu/data/soc-pokec.html}.

\bibitem{Konect-PrettyGoodPrivacy-Contact}
Prettygoodprivacy-contact.
  \url{http://konect.uni-koblenz.de/networks/arenas-pgp}.

\bibitem{SNAP-roadNet-CA}
roadnet-ca. \url{http://snap.stanford.edu/data/roadNet-CA.html}.

\bibitem{SNAP-roadNet-PA}
roadnet-pa. \url{http://snap.stanford.edu/data/roadNet-PA.html}.

\bibitem{SNAP-roadNet-TX}
roadnet-tx. \url{http://snap.stanford.edu/data/roadNet-TX.html}.

\bibitem{Konect-Slashdot-Communication}
Slashdot-communication.
  \url{http://konect.uni-koblenz.de/networks/slashdot-threads}.

\bibitem{SNAP-slashdot1}
slashdot1. \url{http://snap.stanford.edu/data/soc-Slashdot0811.html}.

\bibitem{Konect-SlashdotZoo-Social}
Slashdotzoo-social. \url{http://konect.uni-koblenz.de/networks/slashdot-zoo}.

\bibitem{BGU-TheMarkerCafe}
Themarkercafe. \url{http://proj.ise.bgu.ac.il/sns/themarker.html}.

\bibitem{Konect-TRECWT10g-Reference}
Trecwt10g-reference. \url{http://konect.uni-koblenz.de/networks/trec-wt10g}.

\bibitem{ASU-Twitter}
Twitter. \url{http://socialcomputing.asu.edu/datasets/Twitter}.

\bibitem{Konect-TwitterICWSM-Social}
Twittericwsm-social.
  \url{http://konect.uni-koblenz.de/networks/munmun-twitter-social}.

\bibitem{Konect-USpatents-Reference}
Uspatents-reference. \url{http://konect.uni-koblenz.de/networks/patentcite}.

\bibitem{SNAP-web-BerStan}
web-berstan. \url{http://snap.stanford.edu/data/web-BerkStan.html}.

\bibitem{SNAP-web-Google}
web-google. \url{http://snap.stanford.edu/data/web-Google.html}.

\bibitem{SNAP-web-NotreDame}
web-notredame. \url{http://snap.stanford.edu/data/web-NotreDame.html}.

\bibitem{SNAP-web-Stanford}
web-stanford. \url{http://snap.stanford.edu/data/web-Stanford.html}.

\bibitem{SNAP-wiki-talk}
wiki-talk. \url{http://snap.stanford.edu/data/wiki-Talk.html}.

\bibitem{Konect-WikipediaChinese-Reference}
Wikipediachinese-reference.
  \url{http://konect.uni-koblenz.de/networks/zhishi-zhwiki-internallink}.

\bibitem{Konect-WikipediaEnglish-Reference}
Wikipediaenglish-reference.
  \url{http://konect.uni-koblenz.de/networks/wikipedia-growth}.

\bibitem{Konect-WikipediaLinksDE-Reference}
Wikipedialinksde-reference.
  \url{http://konect.uni-koblenz.de/networks/wikipedia-link-de}.

\bibitem{Konect-WikipediaLinksFR-Reference}
Wikipedialinksfr-reference.
  \url{http://konect.uni-koblenz.de/networks/wikipedia-link-fr}.

\bibitem{Konect-WikipediaLinksIT-Reference}
Wikipedialinksit-reference.
  \url{http://konect.uni-koblenz.de/networks/wikipedia-link-it}.

\bibitem{Konect-WikipediaLinksJA-Reference}
Wikipedialinksja-reference.
  \url{http://konect.uni-koblenz.de/networks/wikipedia-link-ja}.

\bibitem{Konect-WikipediaLinksPL-Reference}
Wikipedialinkspl-reference.
  \url{http://konect.uni-koblenz.de/networks/wikipedia-link-pl}.

\bibitem{Konect-WikipediaLinksPT-Reference}
Wikipedialinkspt-reference.
  \url{http://konect.uni-koblenz.de/networks/wikipedia-link-pt}.

\bibitem{Konect-WikipediaLinksRU-Reference}
Wikipedialinksru-reference.
  \url{http://konect.uni-koblenz.de/networks/wikipedia-link-ru}.

\bibitem{ASU-YouTube-2}
Youtube-2. \url{http://socialcomputing.asu.edu/datasets/YouTube}.

\bibitem{ASU-YouTube-3}
Youtube-3. \url{http://socialcomputing.asu.edu/datasets/YouTube}.

\bibitem{ASU-YouTube-4}
Youtube-4. \url{http://socialcomputing.asu.edu/datasets/YouTube}.

\bibitem{ASU-YouTube-5}
Youtube-5. \url{http://socialcomputing.asu.edu/datasets/YouTube}.

\bibitem{Konect-YouTube-Social}
Youtube-social. \url{http://konect.uni-koblenz.de/networks/youtube-u-growth}.

\bibitem{SNAP-YouTube}
Youtube. \url{http://snap.stanford.edu/data/com-Youtube.html}.

\bibitem{ASU-YouTube}
Youtube. \url{http://socialcomputing.asu.edu/datasets/YouTube}.

\bibitem{albert1999internet}
{\sc Albert, R., Jeong, H., and Barab{\'a}si, A.-L.}
\newblock Internet: Diameter of the world-wide web.
\newblock {\em Nature 401}, 6749 (1999), 130--131.

\bibitem{backstrom-group-2006}
{\sc Backstrom, L., Huttenlocher, D., Kleinberg, J., and Lan, X.}
\newblock Group formation in large social networks: membership, growth, and
  evolution.
\newblock In {\em Proc. 12th {ACM} {SIGKDD} Int. Conf. on Knowledge discovery
  and data mining\/} (2006), pp.~44--54.

\bibitem{b397}
{\sc Bailey, P., Craswell, N., and Hawking, D.}
\newblock Engineering a multi-purpose test collection for {W}eb retrieval
  experiments.
\newblock {\em Information Processing and Management 39}, 6 (2003), 853--871.

\bibitem{konect:boguna}
{\sc Bogu{\~A}$\pm${\~A}{!'}, M., Pastor-Satorras, R., az~Guilera, A.~D., and
  Arenas, A.}
\newblock Models of social networks based on social distance attachment.
\newblock {\em Phys. Rev. E 70}, 5 (2004), 056122.

\bibitem{b524}
{\sc Bollacker, K., Lawrence, S., and Giles, C.~L.}
\newblock {CiteSeer}: An autonomous {Web} agent for automatic retrieval and
  identification of interesting publications.
\newblock In {\em Proc. Int. Conf. on Autonomous Agents\/} (1998),
  pp.~116--123.

\bibitem{konect:brozovsky07}
{\sc Bro{\v z}ovsk{\'y}, L., and Pet{\v r}{\'\i}{\v c}ek, V.}
\newblock Recommender system for online dating service.
\newblock In {\em Proc. Znalosti\/} (2007), pp.~29--40.

\bibitem{cho2011friendship}
{\sc Cho, E., Myers, S.~A., and Leskovec, J.}
\newblock Friendship and mobility: user movement in location-based social
  networks.
\newblock In {\em Proc. 17th ACM SIGKDD Int. Conf. on Knowledge discovery and
  data mining\/} (2011), ACM, pp.~1082--1090.

\bibitem{konect:choudhury10}
{\sc Choudhury, M.~D., Lin, Y.-R., Sundaram, H., Candan, K.~S., Xie, L., and
  Kelliher, A.}
\newblock How does the data sampling strategy impact the discovery of
  information diffusion in social media?
\newblock In {\em ICWSM\/} (2010), pp.~34--41.

\bibitem{konect:choudhury09}
{\sc Choudhury, M.~D., Sundaram, H., John, A., and Seligmann, D.~D.}
\newblock Social synchrony: Predicting mimicry of user actions in online social
  media.
\newblock In {\em Proc. Int. Conf. on Computational Science and Engineering\/}
  (2009), pp.~151--158.

\bibitem{DMM2013}
{\sc De~Domenico, M., Lima, A., Mougel, P., and Musolesi, M.}
\newblock The anatomy of a scientific rumor.
\newblock {\em SCIENTIFIC REPORTS 3\/} (2013).

\bibitem{bguAcademia1}
{\sc Fire, M., Tenenboim, L., Lesser, O., Puzis, R., Rokach, L., and Elovici,
  Y.}
\newblock Link prediction in social networks using computationally efficient
  topological features.
\newblock In {\em Privacy, Security, Risk and Trust (PASSAT), 2011 IEEE Third
  International Conference on and 2011 IEEE Third International Confernece on
  Social Computing (SocialCom)\/} (2011), IEEE, pp.~73--80.

\bibitem{bguAcademia2}
{\sc Fire, M., Tenenboim, L., Puzis, R., Lesser, O., Rokach, L., and Elovici,
  Y.}
\newblock Computationally efficient link prediction in variety of social
  networks.

\bibitem{konect:slashdot-threads}
{\sc G{\~A}³mez, V., Kaltenbrunner, A., and L{\~A}³pez, V.}
\newblock Statistical analysis of the social network and discussion threads in
  {Slashdot}.
\newblock In {\em Proc. Int. World Wide Web Conf.\/} (2008), pp.~645--654.

\bibitem{b619}
{\sc Gehrke, J., Ginsparg, P., and Kleinberg, J.}
\newblock Overview of the 2003 {KDD} {Cup}.
\newblock {\em SIGKDD Explorations 5}, 2 (2003), 149--151.

\bibitem{gehrke03kddcup}
{\sc Gehrke, J., Ginsparg, P., and Kleinberg, J.~M.}
\newblock Overview of the 2003 kdd cup.
\newblock {\em SIGKDD Explorations 5}, 2 (2003), 149--151.

\bibitem{b376}
{\sc Hall, B.~H., Jaffe, A.~B., and Trajtenberg, M.}
\newblock The {NBER} patent citations data file: Lessons, insights and
  methodological tools.
\newblock In {\em {NBER} Working Papers 8498, National Bureau of Economic
  Research, Inc\/} (2001).

\bibitem{klimt2004introducing}
{\sc Klimt, B., and Yang, Y.}
\newblock Introducing the enron corpus.
\newblock In {\em CEAS\/} (2004).

\bibitem{Konect}
{\sc Kunegis, J.}
\newblock Konect - the koblenz network collection.
\newblock pp.~1343--1350.

\bibitem{konect:kunegis2012}
{\sc Kunegis, J., Gr\"{o}ner, G., and Gottron, T.}
\newblock Online dating recommender systems: The split-complex number approach.
\newblock In {\em Proc. RecSys Workshop on Recommender Systems and the Social
  Web\/} (2012), pp.~37--44.

\bibitem{kunegis:slashdot-zoo}
{\sc Kunegis, J., Lommatzsch, A., and Bauckhage, C.}
\newblock The {Slashdot Zoo}: Mining a social network with negative edges.
\newblock In {\em Proc. Int. World Wide Web Conf.\/} (2009), pp.~741--750.

\bibitem{leskovec2010predicting}
{\sc Leskovec, J., Huttenlocher, D., and Kleinberg, J.}
\newblock Predicting positive and negative links in online social networks.
\newblock In {\em Proc. 19th Int. Conf. on World wide web\/} (2010), ACM,
  pp.~641--650.

\bibitem{leskovec2010signed}
{\sc Leskovec, J., Huttenlocher, D., and Kleinberg, J.}
\newblock Signed networks in social media.
\newblock In {\em Proc. SIGCHI Conf. on Human Factors in Computing Systems\/}
  (2010), ACM, pp.~1361--1370.

\bibitem{leskovec2005graphs}
{\sc Leskovec, J., Kleinberg, J., and Faloutsos, C.}
\newblock Graphs over time: densification laws, shrinking diameters and
  possible explanations.
\newblock In {\em Proc. 11th ACM SIGKDD Int. Conf. on Knowledge discovery in
  data mining\/} (2005), ACM, pp.~177--187.

\bibitem{Leskovec:2005:GOT:1081870.1081893}
{\sc Leskovec, J., Kleinberg, J., and Faloutsos, C.}
\newblock Graphs over time: Densification laws, shrinking diameters and
  possible explanations.
\newblock In {\em Proc. 11th ACM SIGKDD Int. Conf. on Knowledge Discovery in
  Data Mining\/} (New York, NY, USA, 2005), KDD '05, ACM, pp.~177--187.

\bibitem{leskovec2007graph}
{\sc Leskovec, J., Kleinberg, J., and Faloutsos, C.}
\newblock {Graph evolution: Densification and shrinking diameters}.
\newblock {\em Trans. Knowledge Discovery from Data (TKDD) 1}, 1 (2007), 2.

\bibitem{Leskovec:2007:GED:1217299.1217301}
{\sc Leskovec, J., Kleinberg, J., and Faloutsos, C.}
\newblock Graph evolution: Densification and shrinking diameters.
\newblock {\em ACM Trans. Knowl. Discov. Data 1}, 1 (Mar. 2007).

\bibitem{DBLP:journals/corr/abs-0810-1355}
{\sc Leskovec, J., Lang, K.~J., Dasgupta, A., and Mahoney, M.~W.}
\newblock Community structure in large networks: Natural cluster sizes and the
  absence of large well-defined clusters.
\newblock {\em CoRR abs/0810.1355\/} (2008).

\bibitem{leskovec2009community}
{\sc Leskovec, J., Lang, K.~J., Dasgupta, A., and Mahoney, M.~W.}
\newblock Community structure in large networks: Natural cluster sizes and the
  absence of large well-defined clusters.
\newblock {\em Internet Mathematics 6}, 1 (2009), 29--123.

\bibitem{konect:DBLP}
{\sc Ley, M.}
\newblock The {DBLP} computer science bibliography: Evolution, research issues,
  perspectives.
\newblock In {\em Proc. Int. Symposium on String Processing and Information
  Retrieval\/} (2002), pp.~1--10.

\bibitem{bguAnyBeat}
{\sc M., F., R., P., and Y., E.}
\newblock Link prediction in highly fractional data sets.
\newblock {\em Handbook of Computational Approaches to Counterterrorism\/}
  (2012).

\bibitem{konect:massa05}
{\sc Massa, P., and Avesani, P.}
\newblock Controversial users demand local trust metrics: an experimental study
  on {epinions.com} community.
\newblock In {\em Proc. American Association for Artificial Intelligence
  Conf.\/} (2005), pp.~121--126.

\bibitem{mcauley2012image}
{\sc McAuley, J., and Leskovec, J.}
\newblock Image labeling on a network: using social-network metadata for image
  classification.
\newblock In {\em Computer Vision--ECCV 2012}. Springer, 2012, pp.~828--841.

\bibitem{mcauley2012learning}
{\sc McAuley, J.~J., and Leskovec, J.}
\newblock Learning to discover social circles in ego networks.
\newblock In {\em NIPS\/} (2012), vol.~272, pp.~548--556.

\bibitem{konect:mislove2}
{\sc Mislove, A.}
\newblock {\em Online Social Networks: Measurement, Analysis, and Applications
  to Distributed Information Systems}.
\newblock PhD thesis, Rice University, 2009.

\bibitem{b494}
{\sc Mislove, A., Koppula, H.~S., Gummadi, K.~P., Druschel, P., and
  Bhattacharjee, B.}
\newblock Growth of the {Flickr} social network.
\newblock In {\em Proc. Workshop on Online Social Networks\/} (2008),
  pp.~25--30.

\bibitem{konect:mislove}
{\sc Mislove, A., Marcon, M., Gummadi, K.~P., Druschel, P., and Bhattacharjee,
  B.}
\newblock Measurement and analysis of online social networks.
\newblock In {\em Proc. Internet Measurement Conf.\/} (2007).

\bibitem{konect:zhishi}
{\sc Niu, X., Sun, X., Wang, H., Rong, S., Qi, G., and Yu, Y.}
\newblock {Zhishi.me} -- weaving {Chinese} linking open data.
\newblock In {\em Proc. Int. Semantic Web Conf.\/} (2011), pp.~205--220.

\bibitem{richardson2003trust}
{\sc Richardson, M., Agrawal, R., and Domingos, P.}
\newblock Trust management for the semantic web.
\newblock In {\em Proc. 2nd Int. Semantic Web Conf. (ISWC), Sanibel Island, FL,
  USA, October 20-23, 2003\/} (2003), vol.~2, Springer, p.~351.

\bibitem{ripeanu2002mapping}
{\sc Ripeanu, M., Foster, I., and Iamnitchi, A.}
\newblock Mapping the gnutella network: Properties of large-scale peer-to-peer
  systems and implications for system design.
\newblock {\em arXiv preprint cs/0209028\/} (2002).

\bibitem{konect:dependency4}
{\sc {\v S}ubelj, L., and Bajec, M.}
\newblock Model of complex networks based on citation dynamics.
\newblock In {\em Proc. {WWW} Workshop on Large Scale Network Analysis\/}
  (2013), pp.~527--530.

\bibitem{takac2012data}
{\sc Takac, L., and Zabovsky, M.}
\newblock Data analysis in public social networks.
\newblock In {\em International Scientific Conference and International
  Workshop Present Day Trends of Innovations\/} (2012).

\bibitem{viswanath09}
{\sc Viswanath, B., Mislove, A., Cha, M., and Gummadi, K.~P.}
\newblock On the evolution of user interaction in {Facebook}.
\newblock In {\em Proc. Workshop on Online Social Networks\/} (2009),
  pp.~37--42.

\bibitem{DBLP:journals/corr/abs-1205-6233}
{\sc Yang, J., and Leskovec, J.}
\newblock Defining and evaluating network communities based on ground-truth.
\newblock {\em CoRR abs/1205.6233\/} (2012).

\bibitem{Zafarani+Liu:2009}
{\sc Zafarani, R., and Liu, H.}
\newblock Social computing data repository at {ASU}, 2009.

\end{thebibliography}

\newpage
\appendix 
\section*{Appendix}

%%%%%%%%%%%%%%%%%%%%%
\section{Extreme Elite Size}
\label{sec:extreme}
%%%%%%%%%%%%%%%%%%%%%

We have seen that a compact elite (satisfying axioms A1, A2, A3) has
$\cI(\cE,\cE)=\Omega(m)$ and $\cI(\cE,\cP)=\Omega(m)$.
A natural question is whether it is also guaranteed that
$\cI(\cP,\cP)=\Omega(m)$ (which may imply a stronger version of a symmetry 
point than the one defined earlier). We now demonstrate that unfortunately
this is not the case. This is done by showing an example of a family of graphs 
with elites satisfying Axioms (A1), (A2), (A3), 
where $\cI(\cE,\cE)=\Omega(m)$ and $\cI(\cE,\cP)=\Omega(m)$ but
$\cI(\cP,\cP)=O(n)$.

We assume that the axioms are defined with constants $c_r,c_d \le 1$
and for simplicity we assume also that $c_r = 1/ b$ for constant integer $b$
(can easily be modified to a rational $c_r$, i.e., $c_r = a/ b$ 
for constant integers $a,b$).

The construction of the graph uses a parameter $Z$ 
(to be thought of as roughly $n^{1/4}$), namely, the graph family contains 
one graph $G_Z$ for any integer $Z>0$.

The elite $\cE$ consists of a complete network of $\EE=|\cE|=4Z^3-1$ vertices.
the periphery $\cP$ consists of $|\cP| = 2bZ\EE = 2bZ(4Z^3-1)$ vertices
with no connections between them (except for a self-loop for each vertex).
Hence altogether the graph $G_Z$ contains $n=8bZ^4+4Z^3-2bZ-1$ vertices.

Every vertex in $\cE$ is connected to $2bZ^3$ vertices in $\cP$, and
every vertex in $\cP$ is connected to $Z^2$ vertices in $\cE$.
Hence we have 
\begin{eqnarray}
\cI(\cE,\cE) &=& \EE(\EE+1)/2 ~=~ 2Z^3(4Z^3-1) \nonumber \\
\cI(\cE,\cP) &=& Z^2 |\cP| ~=~ 2bZ^3\EE ~=~ 2bZ^3(4Z^3-1) \nonumber \\
\cI(\cP,\cP) &=& |\cP| ~=~ 2bZ(4Z^3-1) ~. 
\label{eq:I-sizes}
\end{eqnarray}
Note that $n=\Theta(Z^4)$ and $m=\Theta(Z^6)=\Theta(n^{3/2})$, and indeed
$\cI(\cE,\cE), \cI(\cE,\cP)=\Omega(m)$ and $\cI(\cP,\cP)=O(n)$.

We need to verify that this construction satisfies the three axioms.
Axiom (A1) says that $\cI(\cE,\cP) \ge c_d \cdot \cI(\cP,\cP)$.
This follows readily from Eq. (\ref{eq:I-sizes}) and the assumption 
that $c_d\le 1$.

Axiom (A2) says that $\cI(\cE,\cE) \ge \cI(\cE,\cP) / b$.
In fact, Eq. (\ref{eq:I-sizes}) establishes equality here.

To prove Axiom (A3), we need to show that every subset 
$\emptyset\subset\cE'\subset\cE$ violates either axiom (A1) or (A2).
Concretely, we show that it violates Axiom (A2), i.e., that
$$\cI(\cE',\cE') ~<~ \cI(\cE',\cP') / b.$$

Consider some $\emptyset\subset\cE'\subset\cE$ and let $\cX=\cE\setminus\cE'$.
Note that
\begin{eqnarray*}
\cI(\cE',\cE') &=& \cI(\cE,\cE) - \cI(\cX,\cE') - \cI(\cX,\cX) \\
\cI(\cE',\cP') &=& \cI(\cE,\cP) + \cI(\cX,\cE') - \cI(\cX,\cP)
\end{eqnarray*}
hence we need to prove that
$$b\cdot(\cI(\cE,\cE) - \cI(\cX,\cE') - \cI(\cX,\cX)) ~<~ 
\cI(\cE,\cP) - \cI(\cX,\cE') - \cI(\cX,\cP)~.$$
Let $\XX=|\cX|$ and $\EE'=|cE'| = \EE-\XX$. Then plugging the quantities 
of Eq. (\ref{eq:I-sizes}), it remains to prove that
$$b(2Z^3(4Z^3-1) - \XX(4Z^3-1-\XX) - \XX(\XX+1)/2) ~<~ 
2bZ^3(4Z^3-1)  + \XX(4Z^3-1-\XX) - 2bZ^3\XX~.$$
Simplifying, we get that the above indeed holds for every $\XX<\EE$.

\begin{figure}[h!]
\begin{center}
	\begin{tabular}{cc}
		\includegraphics[width=3in]{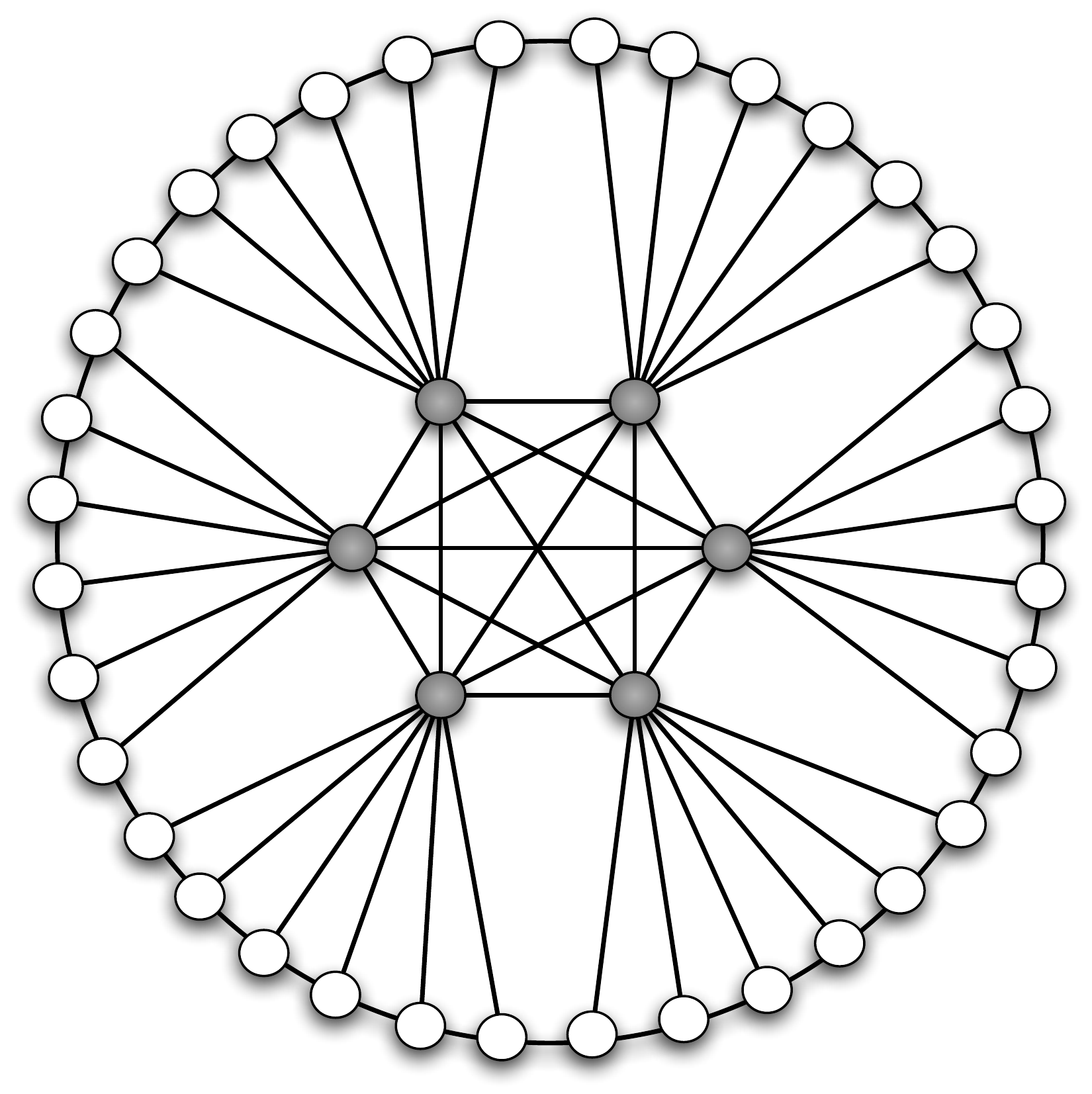} &
		\includegraphics[width=3in]{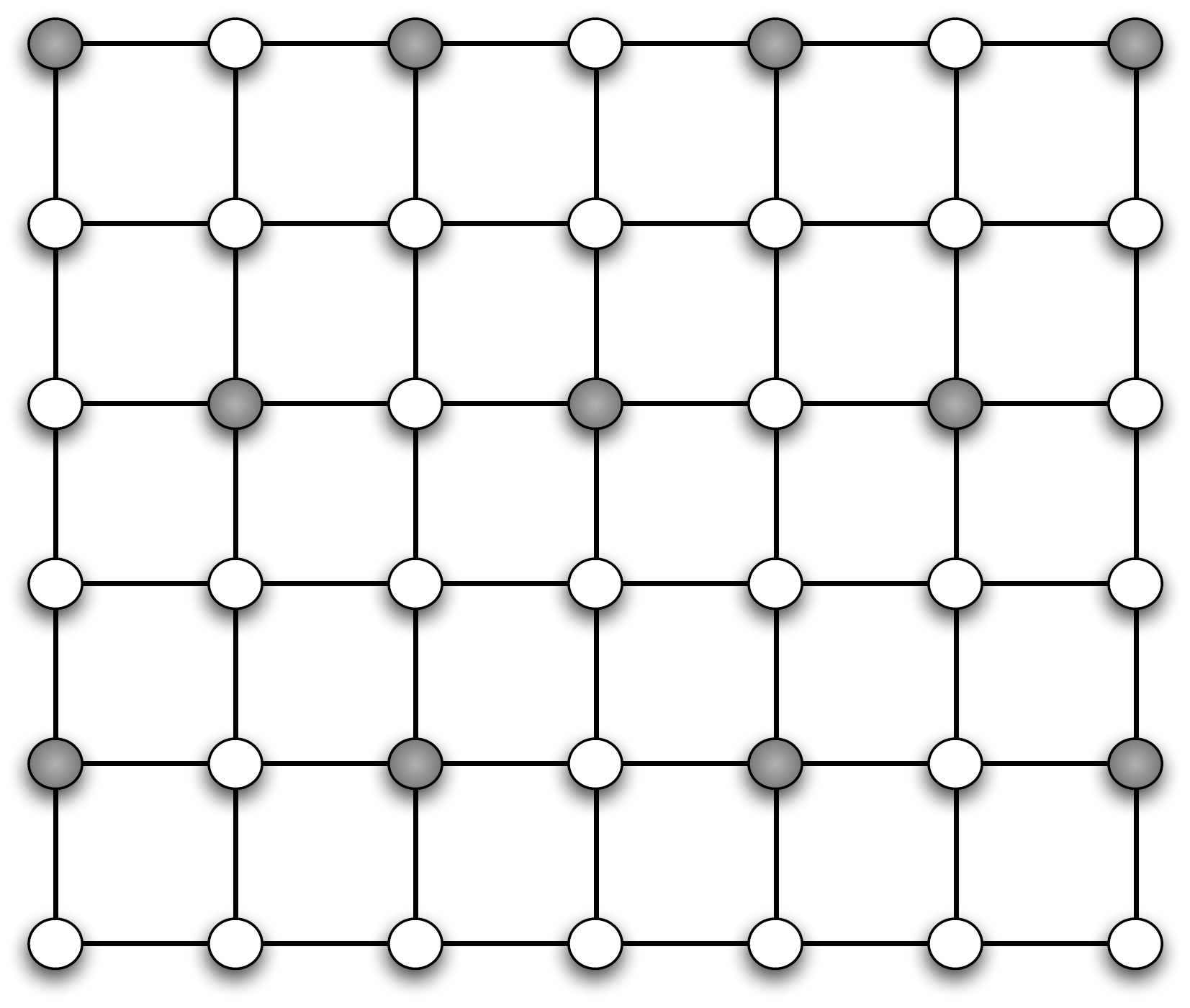} \\
		(a) & (b) \\
		\end{tabular}
\caption{Extreme examples of elite-periphery partitions that satisfy 
our axioms and are close to the symmetry point. 
In both networks we assume each vertex has a self-loop edge that represents 
self-influence and is not drown for simplicity. 
Elite vertices are gray.
%highlighted in gray.
(a) A ``purely elitistic society'' example with 41 vertices of which 6 
are the elite. One can extend this example to a network of $n$ vertices
where the number of edges is $m< 2n$ edges. The elite then is of \emph{sub-linear} size and consists of 
about $\sqrt{m} \approx \sqrt{n}$ vertices.  But it is dense, namely, the elite contains about $m$ edges. The periphery also contains about $m$ edges 
so the partition is at a symmetry point. (b) A ``purely egalitarian society''
example of a grid network with 49 vertices from which 11 are the elite. 
This example can be extended to a network with $n$ vertices and $m<4n$ edges.
It can be shown then that every dominant elite must be of size 
\emph{linear} in $n$ (and $m$). It follows that both the elite and the 
periphery will have a linear number of edges and therefore the partition 
is close to the symmetry point.
Most social networks observed in reality exhibit neither of these 
extreme behaviors.
}
\label{fig:sizeexample}
\end{center}
\end{figure}
\clearpage
%%%%%%%%%%%%%%%%%%%%%%%
\section{Symmetry in Random Networks}
\label{sec:symm-rg}
%%%%%%%%%%%%%%%%%%%%%%%
%%%%%%%%%%%%%%%%%%%%%%%
%\subsection{Random Configuration Graphs}
%\label{sec:rgc}
%\paragraph{Random Configuration Graphs}
%%%%%%%%%%%%%%%%%%%%%%%

Let $\mathbf{d}= d_1, d_2, \dots ,d_n$, where $1 \le d_i \le n-1$ be a positive degree sequence.
A \emph{random configuration graph} $\cG(n,\mathbf{d})$ is constructed in the following way over the set of vertices $[n]$. Let $W=[2m]$ be the set of \emph{configuration points} 
where $m=\sum_1^n d_i/2$ is the number of edges. Define the ranges 
$W_j = [1+ \sum_1^{j-1} d_i ~,~  \sum_1^{j} d_i]$ for $j \in [n]$. 
Given a pairing $F$ (i.e., a partition of $W$ into $m$ pairs) we obtain a (multi) graph $G_F$ with vertex set $[n]$ and for each pair $(u,v) \in F$ we add an edge $(i,j)$ to $G_F$ where $u \in W_i$ and $v \in W_j$. Choosing $F$ uniformly from all possible pairings of $W$ we obtain a random (multi-)graph $\cG(n,\mathbf{d})$.

%\subsection{The Symmetry Point}
%\paragraph{The Symmetry Point}

For a given degree sequence $\mathbf{d}$, define its \emph{symmetry point} 
$\kappa(\mathbf{d})$ as follows:
$$ \kappa(\mathbf{d}) = \arg\min_j \left \lvert \sum_{i=1}^j d_i - \sum_{i=j+1}^n d_i \right \rvert.$$
Given $\mathbf{d}$ and an index $1 \le i \le n$, let $\cE_i$ denote the core as the vertex set $[i] = \{1, \dots, i\}$ and the periphery $\cP_i$ as the vertex set $\{i+1, \dots, n\}$.

\begin{theorem}\label{thm:maxflowsym}
Let $\cG(n,\mathbf{d})$ be a random configuration graph of given positive degree sequence $\mathbf{d}$ and let $\kappa$ be the symmetry point of $\mathbf{d}$ i.e., $ \kappa(\mathbf{d}) = \arg\min_j \left \lvert \sum_{i=1}^j d_i - \sum_{i=j+1}^n d_i \right \rvert$. Then,
\begin{enumerate}
	\item $\bE[\cI(\cE_i,\cP_i)] \le \bE[\cI(\cE_\kappa,\cP_\kappa)]$ 
               for every $i$
	\item $\bE[\cI(\cE_\kappa,\cP_\kappa)] \ge 4m/9$
	\item $\bE[\cI(\cE_\kappa,\cE_\kappa)] \ge m/9$
	\item $\bE[\cI(\cP_\kappa,\cP_\kappa)] \ge m/9$
\end{enumerate}
\end{theorem}

\begin{proofsketch}
Let $D(k,l) = \sum_{i=k}^l d_i/2m$. 
Given a partition index $i$, the expected number of edges of each component is the following: 
\begin{eqnarray*}
\bE[\cI(\cE_i,\cP_i)] &=& 2D(1,i)D(i+1,n)m, 
\\
\bE[\cI(\cE_i,\cE_i)] &=& D(1,i)^2m,
\\
\bE[\cI(\cE_i,\cE_i)] &=& D(i+1,n)^2m.
\end{eqnarray*}
Now, since $D(1,i) +  D(i+1,n) = 2m$, the maximum value of $D(1,i)D(i+1,n)$ is when $i=\arg \min_j\card{D(1,i)-D(i+1,n)}$ which is exactly $\kappa(\mathbf{d})$.
What are lower bounds for the three influences? Note that  $\card{D(1,\kappa)-D(\kappa+1,n)}$ will be maximized when $d_\kappa = n-1$ and 
$D(1,\kappa-1)=D(\kappa+1,n)$ so the worst case is when $D(1,\kappa-1)=D(\kappa+1,n)=n-1$ and $d_\kappa = n-1$, yielding the claimed result.
\end{proofsketch}

%%%%%%%%%%%%%%%%%%%%%%%
\section{$C$-Core Results}
\label{sec:kcore_results}
%%%%%%%%%%%%%%%%%%%%%%%

%%%%%%%%%%%%%%%%%%%%%%%%%%%%%%%%%
\begin{figure}[!ht]
\centering
\includegraphics[width=0.84\textwidth]{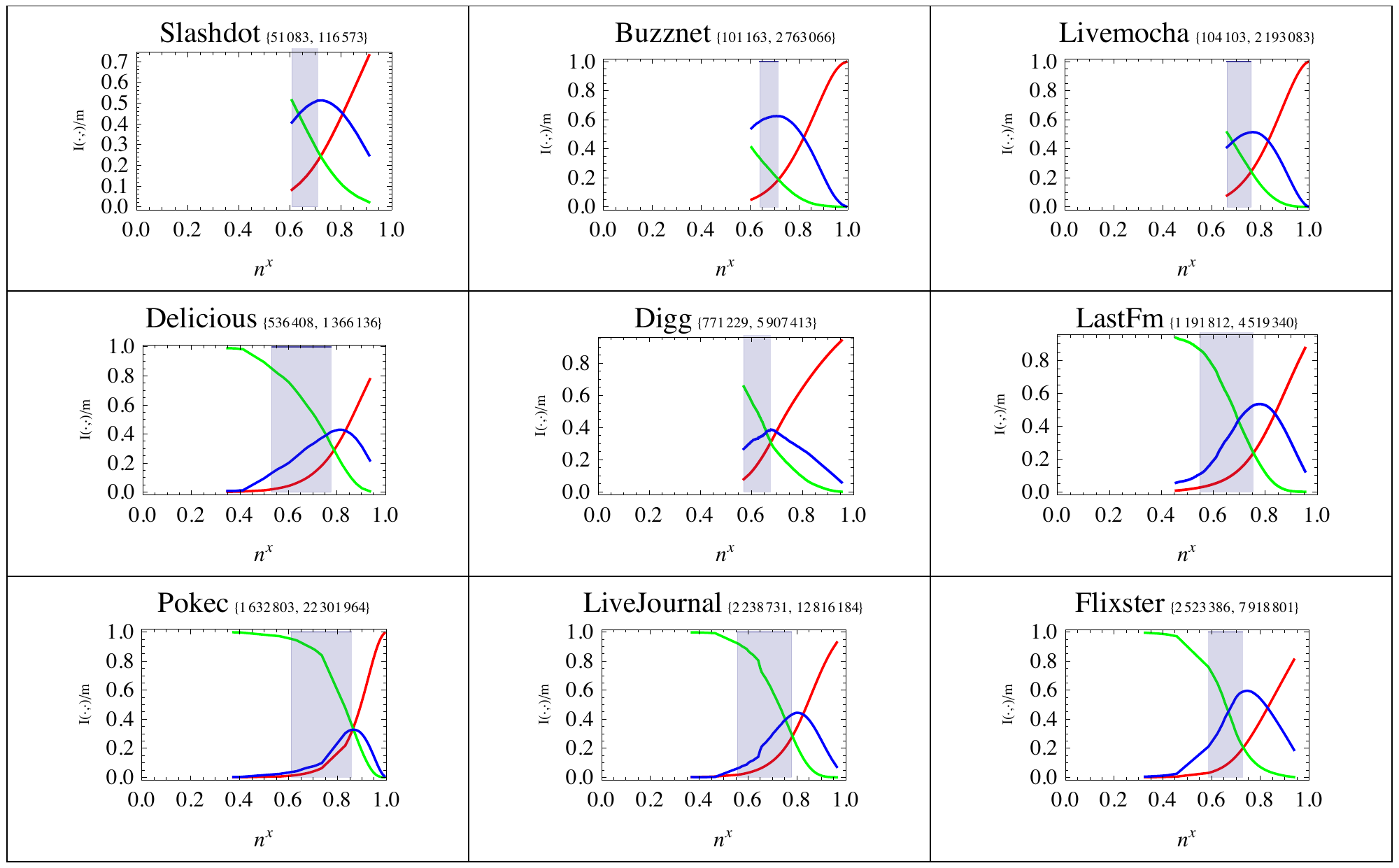}	
\includegraphics[width=0.8\columnwidth]{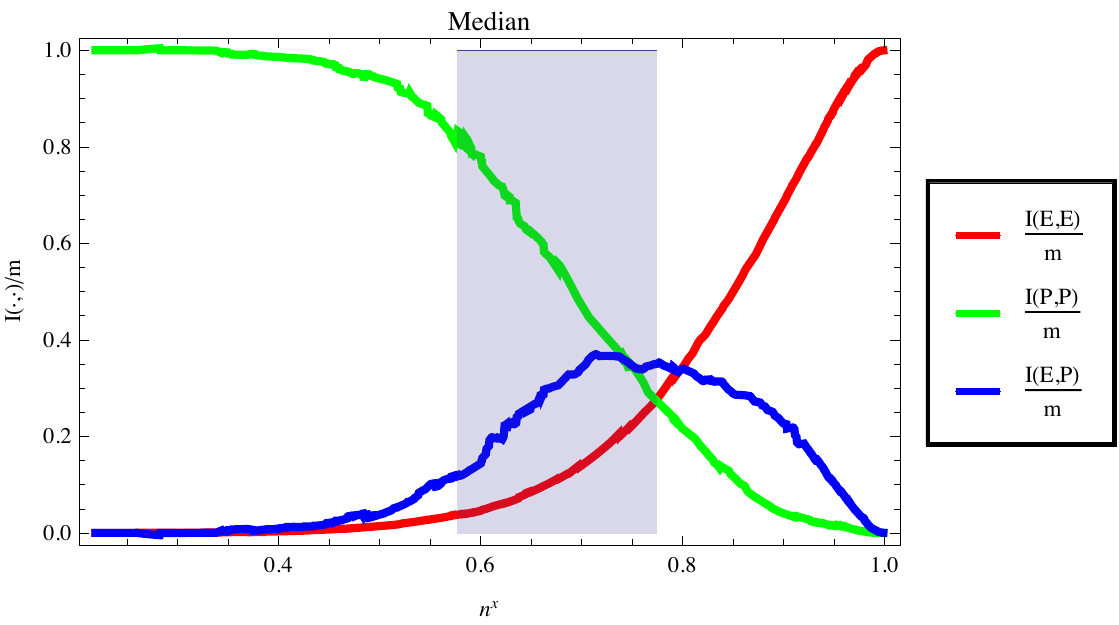}
	
\caption{$c$-core elites: $\cI(\cE,\cE)/m$, $\cI(\cP,\cP)/m$ and 
$\cI(\cE,\cP)/m$ for 9 networks for increasing $k_c$ and their medians for the $\NN$ networks in our measurements.
A point $x$ on the $X$ axis shows a $c$-core of size $k_c = n^x$, for $x$ 
in the range [0,1].
The $Y$ axis gives the percentage of the measured value from the total number 
of edges. The range of $x$ values between $n^x=\sqrt{m}$ and the symmetry point
is highlighted.}
\label{fig:symmetry_kcore_nine}
\end{figure}
%%%%%%%%%%%%%%%%%%%%

Section \ref{sec:empirical} described the results obtained when we examined 
the changes in $\cI(\cE,\cE), \cI(\cP,\cP)$ and $\cI(\cE,\cP)$ for various 
elite sizes when the elites were chosen using the $k$-rich-club method. 
In this section we consider the setting when the elites are chosen using
the $c$-core mothod, and show the results obtained in this setting.

As in Figures~\ref{fig:symmetry_krichclub_nine}, we show the changes in 
$\cI(\cE,\cE)$, $\cI(\cP,\cP)$ and $\cI(\cE,\cP)$ as the elite grows 
from its minimum to its maximum possible size. 
In the $c$-core case, these sizes are determined by the network structure, 
and cannot be set by us. Specifically, the minimum size is the smallest 
possible $k_c>0$ and the maximum size is $k_1$.
 
Figure \ref{fig:symmetry_kcore_nine} shows our results for nine example networks
and the median results 
for the $c$-cores of all $\NN$ networks included in our experiments.
Each graph contains three plots. The first is for ${\cI(\cE,\cE)}/{m}$, the 
second is for ${\cI(\cP,\cP)}/{m}$ and the third is for ${\cI(\cE,\cP)}/{m}$.
The X-axis for each graph is on a logarithmic scale, 
where an $x$ value represents a $c$-core of size $k_c = n^x$. 
%(note that $x=0.5$ is a $c$-core of size $k_c = \sqrt{n}$). 
Although the results in the setting of $c$-core elites are not as consistent 
and smooth for all networks as in the setting of $k$-rich-club elites, 
one can observe that the results here are similar to those obtained 
in the $k$-rich-club setting. Here, too, most of the networks exhibit 
similar patterns: as the elite size grows, 
%the number of internal edges, 
$\cI(\cE,\cE)$ grows as well and 
%the number of edges in the periphery, 
$\cI(\cP,\cP)$ decreases, while
%Both trends are obvious. 
the number of crossing edges 
%between the elite and the periphery, 
$\cI(\cE,\cP)$ grows with the elite size up to some maximum and then 
starts decreasing. 
Here too, $\cI(\cE,\cP)$ is larger than $\cI(\cE,\cE)$ (almost) right from 
the beginning and remains larger until the maximum point,
and this relation
% between these numbers 
changes only after $\cI(\cE,\cP)$ begins to decrease 
(while $\cI(\cE,\cE)$ continues to grow).

Figures \ref{fig:symmetry_kcore_nine}
also show that in the setting of $c$-core elites, just as with $k$-rich-club 
elites (as mentioned in Section~\ref{sec:symmetry}), $\cI(\cE,\cP)$ attains
its maximum at (or very close to) the symmetry point.

%%%%%%%%%%%%%%%%%%%%%%%%%%%%%%%%%%%%%%
\begin{figure}[!ht]
\centering
\subfigure{	
\includegraphics[width=0.84\textwidth]{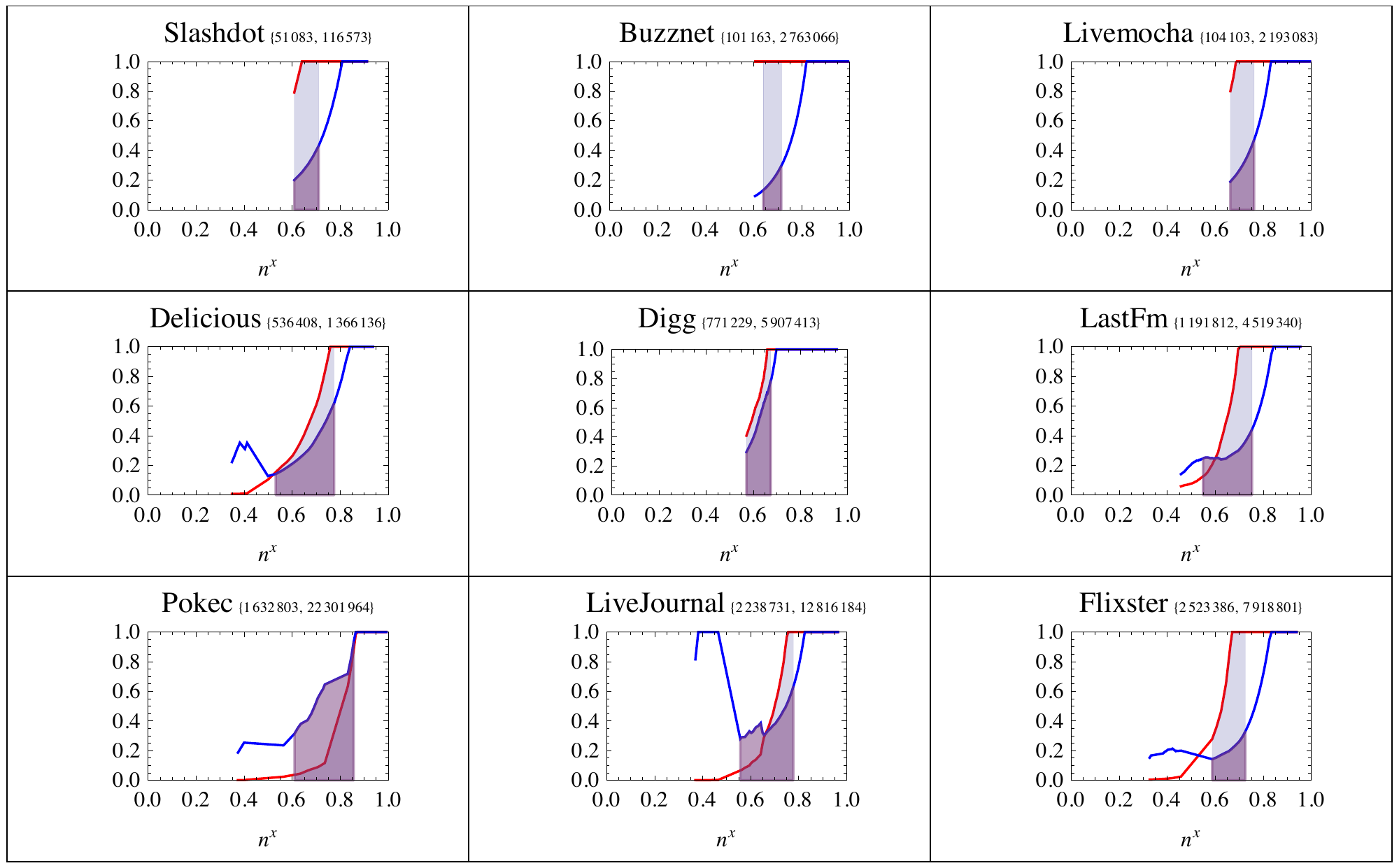}
}	\includegraphics[width=0.7\columnwidth]{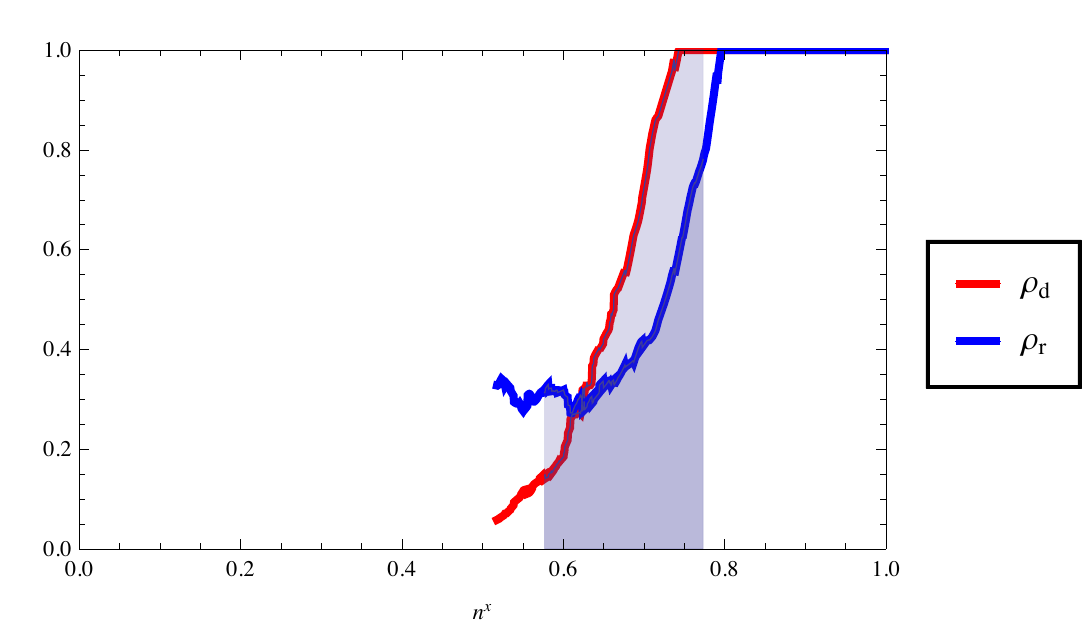}
\caption{Dominance and robustness constants for the nine example networks and their median for the $\NN$ networks. 
The x axis is logarithmic and represents the sizes of $c$-cores with $k=n^x$. 
The higlighted area of $x$ values corresponds to sizes between 
$k_c=\sqrt{m}$ and the symmetry point.}
\label{fig:domrob_kcore_nine}
\end{figure}
%%%%%%%%%%%%%%%%%%%%%%%%%%%%%%%%%%%%%%%

Section \ref{sec:empirical} presented also the values of the dominance and 
robustness constants,  $\rho_d$ and $\rho_r$ respectively, for various elite sizes 
in the setting of $k$-rich-club elites. Here, we present the corresponding
results for $c$-core elites.

As in Figure~\ref{fig:domrob_krichclub_nine} we show the changes in $\rho_d$ and $\rho_r$ 
as the elite grows from its minimum to its maximum possible size. 
%As mentioned earlier, in the $c$-core setting these sizes are determined 
%by the network structure and cannot be set by us. The minimum size 
%is the smallest possible $k_c>0$ and the maximum size is $k_1$.
%
Figure \ref{fig:domrob_kcore_nine} shows results for our nine example networks 
and %Figure \ref{fig:domrob_kcore_median} presents 
the median results for 
the $c$-core elites of all the $\NN$ networks included in our experiments.
The X-axis for each graph is, again, on a logarithmic scale, 
where an $x$ value represents a $c$-core of size $k_c = n^x$. 
There are two plots in each graph, one for the dominance constant $\rho_d$ 
and the other for the robustness constant $\rho_r$. 
We focus on values of $\rho_d$ and $\rho_r$ up to 1, and ignore higher values. 
As before, we highlight in each figure the $c$-cores of sizes in the range
$\sqrt{m} \leq k_c \leq SP$. 

Clearly, elites chosen by the $c$-core method are different from same size 
elites chosen by the $k$-rich-club method. Indeed, the results presented
for the $c$-core setting are not as consistent as in the $k$-rich-club setting. 
Nevertheless, our results show that the same basic characteristics, namely, 
high dominance and robustness in elites of sizes in the range of
$\sqrt{m} \leq k_c \leq SP$, hold also for $c$-core elites. 
It is worth mentioning that in the setting of $c$-core elites, 
it does not hold for all networks that dominance is achieved before robustness.
 
%Similar results, but somewhat weaker were obtained for the $c$-core elites (Appendix~\ref{ss:kcore_results}).   

%Similar results, but somewhat stronger were obtained for the $c$-core elites (Appendix~\ref{ss:kcore_results}). In general elites produced by $c$-core have less domination but are more robust. 

%%%%%%%%%%%%%%%%%
%, and we show them in comparison to the $k$-rich-club method results.

%Figures~\ref{fig:symmetry_kcore_nine} and~\ref{fig:symmetry_kcore_mean} are actually figures~\ref{fig:symmetry_krichclub_nine} and \ref{fig:symmetry_krichclub_mean} respectively, with the addition of the $c$-core results.
%%%%%%%%%%%%%%%%%%%%%%%%%%%

%%%%%%%%%%%%%%%%%%%%
\begin{figure}[!h]
	\centering
			\includegraphics[width=0.7\textwidth]{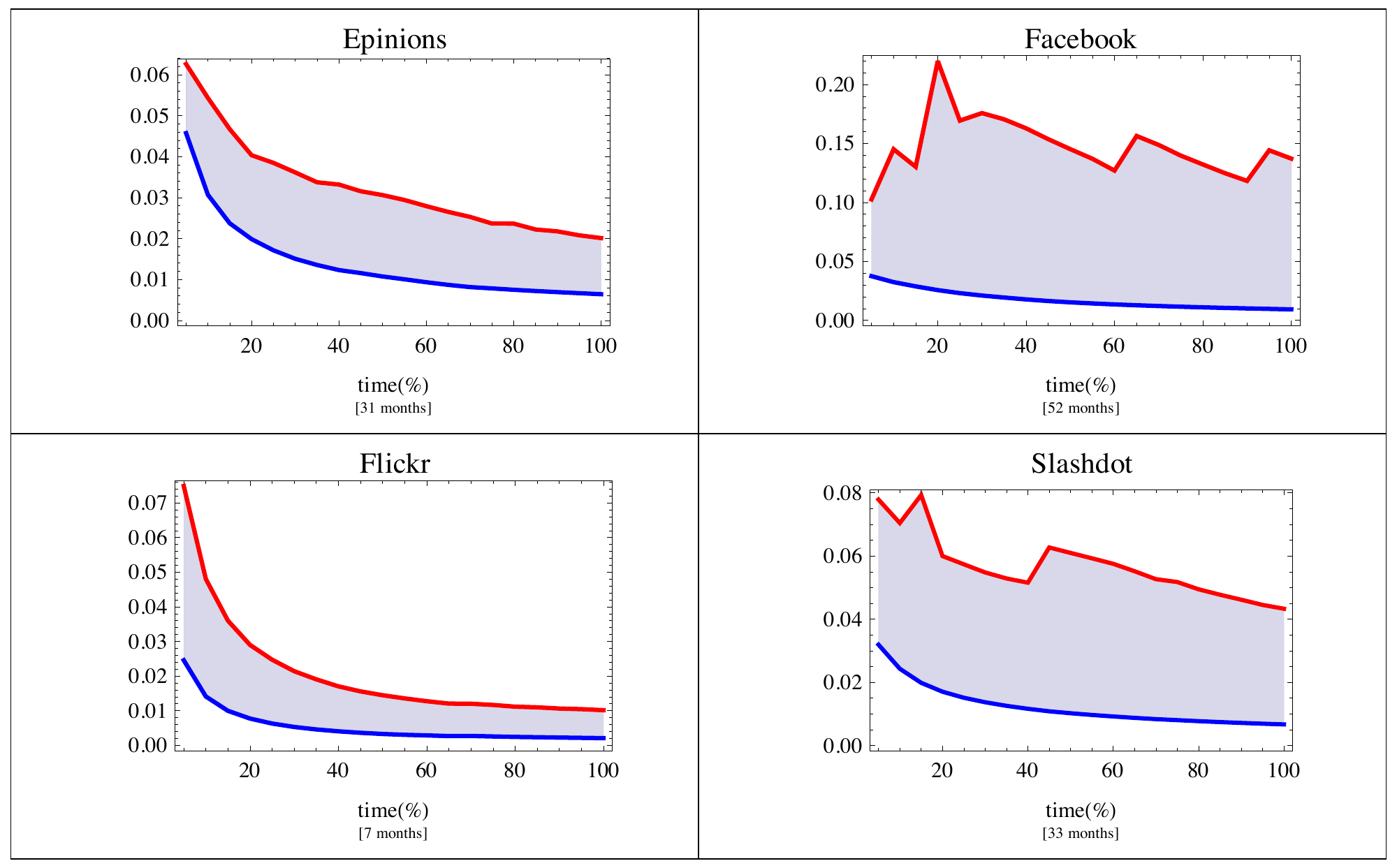}		
\includegraphics[width=0.7\textwidth]{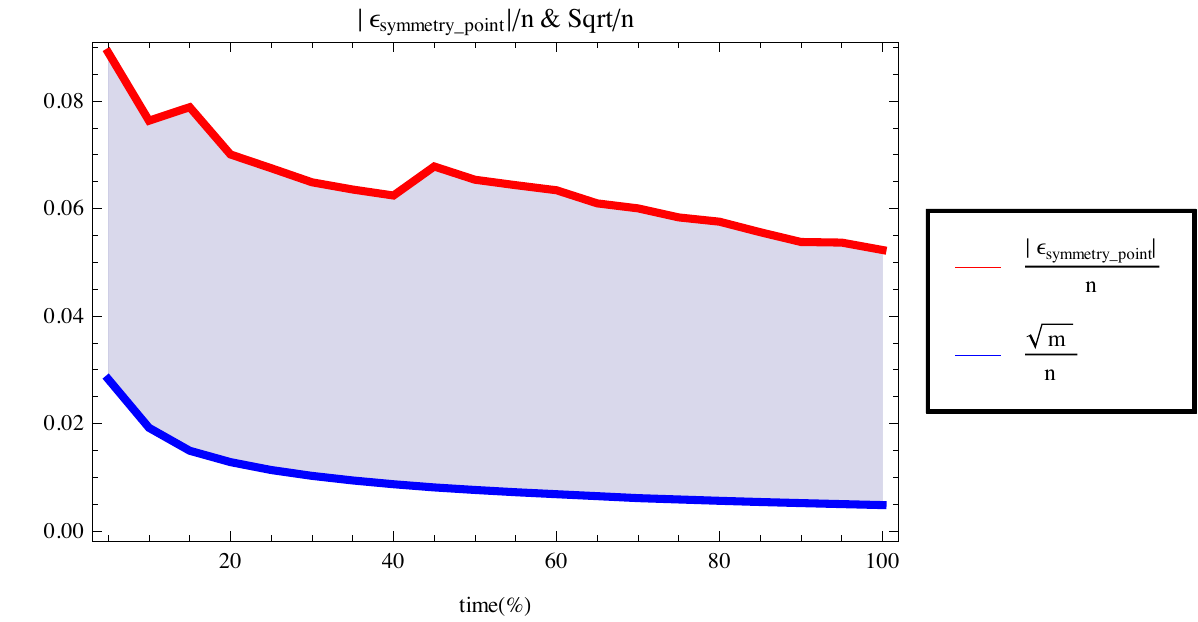}	

\caption{The fraction of the elite at the symmetry point, from the whole population, as the networks evolve over time. 4 examples and the  median out of $\NND$ networks. Elites are chosen by $c$-core method.}
\label{fig:symmetry_point_four_kcore}
\end{figure}
%%%%%%%%%%%%%%%%%%%%

Finally, turning to dynamic data, in Fig.~\ref{fig:symmetry_point_four_kcore} 
we repeated in the $c$-core setting the analysis of
Fig.~\ref{fig:symmetry_point_four}
from Section \ref{sec:empirical} for the $k$-rich-club setting.
Specifically, we looked at the percentage of the number of vertices in the elite, 
$r~=~\frac{\card{\cE_{sp}}}{n}$, 
for elite sizes $\card{\cE_{sp}}$ at the symmetry point.
We used the same assumptions made in Section \ref{sec:empirical} about 
the time information and followed the same procedure as described therein.
Fig.~\ref{fig:symmetry_point_four_kcore}
depicts the ratio $r$ as it evolves in time. 
Figure \ref{fig:symmetry_point_four_kcore} shows the results for four 
example networks (this is a different set from the nine networks studied above,
and the median result 
of all $\NND$ networks for which evolutionary data was available to us. 
Once again, the results in the setting of $c$-core elites are not as smooth 
as in the setting of $k$-rich-club elites. Nevertheless, here too 
the figures demonstrate that the elite size at the symmetry point 
is a relatively small portion of the entire network 
(starting at median of $9\%$ and ending at median of $6\%$). 
Furthermore, elites chosen by the $c$-core method exhibit similar behavior 
of evolution and growth as the $k$-rich-club setting. Here too the ratio $r$ decrease, implying that 
asymptotically, the elite size at the symmetry point is indeed sub-linear.

\begin{landscape}

%%%%%%%%%%%%%%%%%%%%%%%
\section{Datasets}
%%%%%%%%%%%%%%%%%%%%%%%

\begin{center}
\begin{longtable}{|l|c|r|r|r|r|c|c|p{5cm}|}

%\caption[d]{d.} \label{dd} \\

\endfirsthead

\multicolumn{9}{c}%
{{\bfseries \tablename\ \thetable{} -- continued from previous page}} \\

%\hline \multicolumn{1}{|c|}{\textbf{Time (s)}} &
%\multicolumn{1}{c|}{\textbf{Triple chosen}} &
%\multicolumn{1}{c|}{\textbf{Other feasible triples}} \\ \hline 

\hline
 
%Data & \multicolumn{1}{c|}{$n$} 	&	\multicolumn{1}{c|}{$m$}	&	\multicolumn{1}{c|}{\pbox{30cm}{Duration \\ (months)} }  \\ \hline
Data &  \multicolumn{1}{c|}{\footnotesize{Repository}} 	&  \multicolumn{1}{c|}{$n$} 	&	\multicolumn{1}{c|}{$m$} & \begin{tabular}[c]{@{}c@{}}\footnotesize{Average}\\\footnotesize{Degree}\end{tabular} & \begin{tabular}[c]{@{}c@{}}\footnotesize{Max}\\\footnotesize{Degree}\end{tabular} 	& \begin{tabular}[c]{@{}c@{}}\footnotesize{Undirected}\\\footnotesize{Directed}\end{tabular}    &	\multicolumn{1}{c|}{   \begin{tabular}[c]{@{}c@{}}\footnotesize{Duration}\\(\footnotesize{months})\end{tabular} } & \multicolumn{1}{p{5cm}|}{Description} \\ \hline

\endhead

\hline \multicolumn{9}{|r|}{{Continued on next page}} \\ \hline
\endfoot

\hline \hline

\caption[List of All Tested Networks]{List of All Tested Networks} 
\label{table:allNetworks} %\\

\endlastfoot

\hline
%Data & \multicolumn{1}{c|}{$n$} 	&	\multicolumn{1}{c|}{$m$}	&	\multicolumn{1}{c|}{\pbox{30cm}{Duration \\ (months)} }  \\ \hline
Data &  \multicolumn{1}{c|}{\footnotesize{Repository}} 	&  \multicolumn{1}{c|}{$n$} 	&	\multicolumn{1}{c|}{$m$} & \begin{tabular}[c]{@{}c@{}}\footnotesize{Average}\\\footnotesize{Degree}\end{tabular} & \begin{tabular}[c]{@{}c@{}}\footnotesize{Max}\\\footnotesize{Degree}\end{tabular} 	& \begin{tabular}[c]{@{}c@{}}\footnotesize{Undirected}\\\footnotesize{Directed}\end{tabular}    &	\multicolumn{1}{c|}{   \begin{tabular}[c]{@{}c@{}}\footnotesize{Duration}\\(\footnotesize{months})\end{tabular} } & \multicolumn{1}{p{5cm}|}{Description} \\ \hline

\begin{tabular}[l]{@{}l@{}}\normalsize{Academia}\\\normalsize{~\citesup{BGU-Academia}\citesup{ bguAcademia1}\citesup{bguAcademia2}}\end{tabular}&BGU&200169&1022441&5.1&10693&D&& Platform for academics to share research papers. \\
\hline
\begin{tabular}[l]{@{}l@{}}\normalsize{Blog}\\\normalsize{~\citesup{ASU-Blog}\citesup{Zafarani+Liu:2009}}\end{tabular}&ASU&88784&2093195&23.6&9444&U&&Social blog directory which manages bloggers their blog \\
\hline
\begin{tabular}[l]{@{}l@{}}\normalsize{Blog2}\\\normalsize{~\citesup{ASU-Blog2}\citesup{Zafarani+Liu:2009}}\end{tabular}&ASU&97884&1668647&17.0&27849&U&&Social blog directory which manages bloggers their blog \\
\hline
\begin{tabular}[l]{@{}l@{}}\normalsize{Blog3}\\\normalsize{~\citesup{ASU-Blog3}\citesup{Zafarani+Liu:2009}}\end{tabular}&ASU&10312&333983&32.4&3992&U&&Social blog directory which manages bloggers their blog \\
\hline
\begin{tabular}[l]{@{}l@{}}\normalsize{Buzznet}\\\normalsize{~\citesup{ASU-Buzznet}\citesup{Zafarani+Liu:2009}}\end{tabular}&ASU&101163&2763066&27.3&64289&U&&Photo, journal, and video-sharing social media network \\
\hline
\begin{tabular}[l]{@{}l@{}}\normalsize{Delicious}\\\normalsize{~\citesup{ASU-Delicious}\citesup{Zafarani+Liu:2009}}\end{tabular}&ASU&536408&1366136&2.5&3216&U&&Social bookmarking web service for storing  sharing  and discovering web bookmarks \\
\hline
\begin{tabular}[l]{@{}l@{}}\normalsize{Digg}\\\normalsize{~\citesup{ASU-Digg}\citesup{Zafarani+Liu:2009}}\end{tabular}&ASU&771229&5907413&7.7&17643&U&&Social news website \\
\hline
\begin{tabular}[l]{@{}l@{}}\normalsize{Douban}\\\normalsize{~\citesup{ASU-Douban}\citesup{Zafarani+Liu:2009}}\end{tabular}&ASU&154908&327162&2.1&287&U&&Chinese website providing user review and recommendation services for movies  books  and music \\
\hline
\begin{tabular}[l]{@{}l@{}}\normalsize{Flickr}\\\normalsize{~\citesup{ASU-Flickr}\citesup{Zafarani+Liu:2009}}\end{tabular}&ASU&80513&5899882&73.3&5706&U&&An image hosting and video hosting website  web services suite  and online community \\
\hline
\begin{tabular}[l]{@{}l@{}}\normalsize{Flixster}\\\normalsize{~\citesup{ASU-Flixster}\citesup{Zafarani+Liu:2009}}\end{tabular}&ASU&2523386&7918801&3.1&1474&U&&Social movie site allowing users to share movie ratings  discover new movies and meet others with similar movie taste \\
\hline
\begin{tabular}[l]{@{}l@{}}\normalsize{Foursquare}\\\normalsize{~\citesup{ASU-Foursquare}\citesup{Zafarani+Liu:2009}}\end{tabular}&ASU&639014&3214986&5.0&106218&U&&Location-based social networking website  software for mobile devices. This service is available to users with GPS enabled mobile devices  such as iPhones and Blackberries \\
\hline
\begin{tabular}[l]{@{}l@{}}\normalsize{Friendster}\\\normalsize{~\citesup{ASU-Friendster}\citesup{Zafarani+Liu:2009}}\end{tabular}&ASU&5689498&14067887&2.5&4423&U&&Social networking website.The service allows users to contact other members  maintain those contacts  and share online content and media with those contacts.  \\
\hline
\begin{tabular}[l]{@{}l@{}}\normalsize{Hyves}\\\normalsize{~\citesup{ASU-Hyves}\citesup{Zafarani+Liu:2009}}\end{tabular}&ASU&1402673&2777419&2.0&31883&U&&The most popular social networking site in the Netherlands with mainly Dutch visitors and members and competes in this country with sites such as Facebook and MySpace. \\
\hline
\begin{tabular}[l]{@{}l@{}}\normalsize{LastFm}\\\normalsize{~\citesup{ASU-LastFm}\citesup{Zafarani+Liu:2009}}\end{tabular}&ASU&1191812&4519340&3.8&5150&U&&Music website  founded in the United Kingdom in 2002. It has claimed over 40 million active users based in more than 190 countries. \\
\hline
\begin{tabular}[l]{@{}l@{}}\normalsize{LiveJournal}\\\normalsize{~\citesup{ASU-LiveJournal}\citesup{Zafarani+Liu:2009}}\end{tabular}&ASU&2238731&12816184&5.7&5873&U&&Virtual community where Internet users can keep a blog  journal or diary \\
\hline
\begin{tabular}[l]{@{}l@{}}\normalsize{Livemocha}\\\normalsize{~\citesup{ASU-Livemocha}\citesup{Zafarani+Liu:2009}}\end{tabular}&ASU&104103&2193083&21.1&2980&U&&The world's largest online language learning community  offering free and paid online language courses in 35 languages to more than 6 million members from over 200 countries around the world.  \\
\hline
\begin{tabular}[l]{@{}l@{}}\normalsize{Twitter}\\\normalsize{~\citesup{ASU-Twitter}\citesup{Zafarani+Liu:2009}}\end{tabular}&ASU&11316811&63555749&5.6&564795&D&&Social news website. It can be viewed as a hybrid of email  instant messaging and sms messaging all rolled into one neat and simple package. It's a new and easy way to discover the latest news related to subjects you care about. \\
\hline
\begin{tabular}[l]{@{}l@{}}\normalsize{YouTube}\\\normalsize{~\citesup{ASU-YouTube}\citesup{Zafarani+Liu:2009}}\end{tabular}&ASU&13723&76765&5.6&534&U&&Video-sharing website on which users can upload  share  and view videos \\
\hline
\begin{tabular}[l]{@{}l@{}}\normalsize{YouTube-2}\\\normalsize{~\citesup{ASU-YouTube-2}\citesup{Zafarani+Liu:2009}}\end{tabular}&ASU&13242&1940806&146.6&3068&U&&Video-sharing website on which users can upload  share  and view videos \\
\hline
\begin{tabular}[l]{@{}l@{}}\normalsize{YouTube-3}\\\normalsize{~\citesup{ASU-YouTube-3}\citesup{Zafarani+Liu:2009}}\end{tabular}&ASU&11765&5574249&473.8&6745&U&&Video-sharing website on which users can upload  share  and view videos \\
\hline
\begin{tabular}[l]{@{}l@{}}\normalsize{YouTube-4}\\\normalsize{~\citesup{ASU-YouTube-4}\citesup{Zafarani+Liu:2009}}\end{tabular}&ASU&10455&2239440&214.2&5958&U&&Video-sharing website on which users can upload  share  and view videos \\
\hline
\begin{tabular}[l]{@{}l@{}}\normalsize{YouTube-5}\\\normalsize{~\citesup{ASU-YouTube-5}\citesup{Zafarani+Liu:2009}}\end{tabular}&ASU&13160&3797635&288.6&5759&U&&Video-sharing website on which users can upload  share  and view videos \\
\hline
%\begin{tabular}[l]{@{}l@{}}\normalsize{YouTube2}\\\normalsize{~\citesup{ASU-YouTube2}\citesup{Zafarani+Liu:2009}}\end{tabular}&ASU&1138499&2990443&2.6&28754&U&&Video-sharing website on which users can upload  share  and view videos \\
%\hline
\begin{tabular}[l]{@{}l@{}}\normalsize{AnyBeat}\\\normalsize{~\citesup{BGU-AnyBeat}\citesup{bguAnyBeat}}\end{tabular}&BGU&12645&49132&3.9&4800&D&&Online community.  A public gathering place where you can interact with people from around your neighborhood or across the world \\
\hline
\begin{tabular}[l]{@{}l@{}}\normalsize{GooglePlus}\\\normalsize{~\citesup{BGU-GooglePlus}\citesup{bguAcademia2}}\end{tabular}&BGU&211186&1141650&5.4&1790&D&&Google+ is a social networking service and website offered by Google \\
\hline
\begin{tabular}[l]{@{}l@{}}\normalsize{TheMarkerCafe}\\\normalsize{~\citesup{BGU-TheMarkerCafe}\citesup{bguAcademia1}\citesup{bguAcademia2}}\end{tabular}&BGU&69413&1644843&23.7&8930&U&& Israeli online social network site that allows its member to connect and interact \\
\hline
%\begin{tabular}[l]{@{}l@{}}\normalsize{Advogato}\\\normalsize{~\citesup{Konect-Advogato-Social}\citesup{Konect}\citesup{konect:massa09}}\end{tabular}&Konect&5167&39432&7.6&807&D&&Trust network of Advogato, an online community platform for developers of free software. \\
%\hline
\begin{tabular}[l]{@{}l@{}}\normalsize{arXivHep-thKDDCup}\\\normalsize{~\citesup{Konect-arXivHep-thKDDCup-Reference}\citesup{Konect}\citesup{b619}}\end{tabular}&Konect&27769&352285&12.7&2468&D&&Citation network from arXiv's section on high energy physics theory (hep-th)  as used in the KDD cup 2003.  \\
\hline
\begin{tabular}[l]{@{}l@{}}\normalsize{BaiduInternal}\\\normalsize{~\citesup{Konect-BaiduInternal-Reference}\citesup{Konect}\citesup{konect:zhishi}}\end{tabular}&Konect&2140198&17014946&8.0&97848&D&&Hyperlinks between the articles of the Chinese online encyclopedia Baidu. \\
\hline
\begin{tabular}[l]{@{}l@{}}\normalsize{BaiduRelated}\\\normalsize{~\citesup{Konect-BaiduRelated-Reference}\citesup{Konect}\citesup{konect:zhishi}}\end{tabular}&Konect&415624&2374044&5.7&127066&D&&"Related to" links between articles of the Chinese online encyclopedia Baidu. \\
\hline
%\begin{tabular}[l]{@{}l@{}}\normalsize{CaenorhabditisElegans}\\\normalsize{~\citesup{Konect-CaenorhabditisElegans-Contact}\citesup{Konect}\citesup{konect:duch05}}\end{tabular}&Konect&453&2025&4.5&237&U&&Metabolic network of the roundworm Caenorhabditis elegans. \\
%\hline
\begin{tabular}[l]{@{}l@{}}\normalsize{Catster}\\\normalsize{~\citesup{Konect-Catster-Social}\citesup{Konect}}\end{tabular}&Konect&149684&5448196&36.4&80634&U&&This Network contains friendships between users of the website catster.com. \\
\hline
\begin{tabular}[l]{@{}l@{}}\normalsize{CatsterDogster}\\\normalsize{~\citesup{Konect-CatsterDogster-Social}\citesup{Konect}}\end{tabular}&Konect&623748&15695166&25.2&80636&U&&Familylinks between cats and cats  cats and dogs  as well as dogs and dogs from the social websites catster.com and dogster.com. Also included are cat-cat and dog-dog friendships. \\
\hline
\begin{tabular}[l]{@{}l@{}}\normalsize{CiteSeer}\\\normalsize{~\citesup{Konect-CiteSeer-Reference}\citesup{Konect}\citesup{b524}}\end{tabular}&Konect&384054&1736172&4.5&1739&D&&Citation network extracted from the CiteSeer digital library. \\
\hline
\begin{tabular}[l]{@{}l@{}}\normalsize{CoraCitation}\\\normalsize{~\citesup{Konect-CoraCitation-Reference}\citesup{Konect}\citesup{konect:dependency4}}\end{tabular}&Konect&23166&89157&3.8&377&D&&Cora citation network. \\
\hline
\begin{tabular}[l]{@{}l@{}}\normalsize{DBLP}\\\normalsize{~\citesup{Konect-DBLP-Contact}\citesup{Konect}\citesup{konect:DBLP}}\end{tabular}&Konect&1103412&4225686&3.8&1189&U&913&Collaboration graph of authors of scientific papers from DBLP computer science bibliography \\
\hline
%\begin{tabular}[l]{@{}l@{}}\normalsize{DBLP}\\\normalsize{~\citesup{Konect-DBLP-Reference}\citesup{Konect}\citesup{konect:DBLP}}\end{tabular}&Konect&12591&49636&3.9&713&D&&Citation network of DBLP. \\
%\hline
\begin{tabular}[l]{@{}l@{}}\normalsize{Digg}\\\normalsize{~\citesup{Konect-Digg-Communication}\citesup{Konect}\citesup{konect:choudhury09}}\end{tabular}&Konect&30360&85155&2.8&283&D&&Reply network of the social news website Digg \\
\hline
\begin{tabular}[l]{@{}l@{}}\normalsize{Dogster}\\\normalsize{~\citesup{Konect-Dogster-Social}\citesup{Konect}}\end{tabular}&Konect&426816&8543548&20.0&46503&U&&This Network contains friendships between users of the website dogster.com. \\
\hline
\begin{tabular}[l]{@{}l@{}}\normalsize{Epinions}\\\normalsize{~\citesup{Konect-Epinions-Social}\citesup{Konect}\citesup{konect:massa05}}\end{tabular}&Konect&131580&711210&5.4&3558&D&31&Trust and distrust network of Epinions,  an online product rating site \\
\hline
%\begin{tabular}[l]{@{}l@{}}\normalsize{Euroroad}\\\normalsize{~\citesup{Konect-Euroroad-Physical}\citesup{Konect}\citesup{konect:eroads}}\end{tabular}&Konect&1174&1417&1.2&10&U&&The international E-road network,  a road network located mostly in Europe.  \\
%\hline
\begin{tabular}[l]{@{}l@{}}\normalsize{Facebook}\\\normalsize{~\citesup{Konect-Facebook-Communication}\citesup{Konect}\citesup{viswanath09}}\end{tabular}&Konect&45813&183412&4.0&223&D&52&Directed network of a small subset of posts to other user's wall on Facebook \\
\hline
\begin{tabular}[l]{@{}l@{}}\normalsize{Facebook-WOSN}\\\normalsize{~\citesup{Konect-Facebook-WOSN--Social}\citesup{Konect}\citesup{viswanath09}}\end{tabular}&Konect&63731&817090&12.8&1098&U&29& Friendship data of facebook users \\
\hline
\begin{tabular}[l]{@{}l@{}}\normalsize{Flickr}\\\normalsize{~\citesup{Konect-Flickr-Social}\citesup{Konect}\citesup{b494}}\end{tabular}&Konect&2302925&22838276&9.9&27937&D&7&Social network of users and their friendship connections \\
\hline
\begin{tabular}[l]{@{}l@{}}\normalsize{FlickrLinks}\\\normalsize{~\citesup{Konect-FlickrLinks-Social}\citesup{Konect}\citesup{konect:mislove}}\end{tabular}&Konect&1715255&15555041&9.1&27236&U&&Social network of Flickr users and their connections. \\
\hline
\begin{tabular}[l]{@{}l@{}}\normalsize{HudongInternal}\\\normalsize{~\citesup{Konect-HudongInternal-Reference}\citesup{Konect}\citesup{konect:zhishi}}\end{tabular}&Konect&1974655&14428382&7.3&61440&D&&Hyperlinks between articles of the Chinese online encyclopedia Hudong. \\
\hline
\begin{tabular}[l]{@{}l@{}}\normalsize{HudongRelated}\\\normalsize{~\citesup{Konect-HudongRelated-Reference}\citesup{Konect}\citesup{konect:zhishi}}\end{tabular}&Konect&2452673&18691099&7.6&204282&D&&"Related to" links between articles of the Chinese online encyclopedia Hudong. \\
\hline
%\begin{tabular}[l]{@{}l@{}}\normalsize{Hypertext2009}\\\normalsize{~\citesup{Konect-Hypertext2009-Contact}\citesup{Konect}\citesup{konect:sociopatterns}}\end{tabular}&Konect&113&2196&19.4&98&U&1&Network of face-to-face contacts of the attendees of the ACM Hypertext 2009 conference.  \\
%\hline
%\begin{tabular}[l]{@{}l@{}}\normalsize{Infectious}\\\normalsize{~\citesup{Konect-Infectious-Contact}\citesup{Konect}\citesup{konect:sociopatterns}}\end{tabular}&Konect&410&2765&6.7&50&U&1&Network of face-to-face contacts between people during the exhibition INFECTIOUS: STAY AWAY in 2009 at the Science Gallery in Dublin.  \\
%\hline
%\begin{tabular}[l]{@{}l@{}}\normalsize{JDKDependency}\\\normalsize{~\citesup{Konect-JDKDependency-Reference}\citesup{Konect}}\end{tabular}&Konect&6434&53658&8.3&5923&D&&Software class dependency network of the JDK 1.6.0.7 framework. \\
%\hline
%\begin{tabular}[l]{@{}l@{}}\normalsize{JUNGDependency}\\\normalsize{~\citesup{Konect-JUNGDependency-Reference}\citesup{Konect}\citesup{konect:dependency1}}\end{tabular}&Konect&6120&50290&8.2&5655&D&&Software class dependency network of the JUNG 2.0.1 and javax 1.6.0.7 libraries  namespaces edu.uci.ics.jung and java/javax. \\
%\hline
\begin{tabular}[l]{@{}l@{}}\normalsize{LibimsetiCZ}\\\normalsize{~\citesup{Konect-LibimsetiCZ-Social}\citesup{Konect}\citesup{konect:kunegis2012}\citesup{konect:brozovsky07}}\end{tabular}&Konect&220970&17233144&78.0&33389&D&&Czech dating site. This is the network of ratings given by users to other users. \\
\hline
%\begin{tabular}[l]{@{}l@{}}\normalsize{ManufacturingEmails}\\\normalsize{~\citesup{Konect-ManufacturingEmails-Communication}\citesup{Konect}\citesup{konect:2014:radoslaw-email}}\end{tabular}&Konect&167&3250&19.5&139&D&1&Internal email communication network between employees of a mid-sized manufacturing company. \\
%\hline
%\begin{tabular}[l]{@{}l@{}}\normalsize{OpenFlights}\\\normalsize{~\citesup{Konect-OpenFlights-Physical}\citesup{Konect}\citesup{konect:opsahl2010b}}\end{tabular}&Konect&2939&15677&5.3&242&D&&Flights between airports of the world.  \\
%\hline
\begin{tabular}[l]{@{}l@{}}\normalsize{PrettyGoodPrivacy}\\\normalsize{~\citesup{Konect-PrettyGoodPrivacy-Contact}\citesup{Konect}\citesup{konect:boguna}}\end{tabular}&Konect&10680&24316&2.3&205&U&&Interaction network of users of the Pretty Good Privacy (PGP) algorithm. \\
\hline
\begin{tabular}[l]{@{}l@{}}\normalsize{Slashdot}\\\normalsize{~\citesup{Konect-Slashdot-Communication}\citesup{Konect}\citesup{konect:slashdot-threads}}\end{tabular}&Konect&51083&116573&2.3&2915&D&33&Reply network of technology website  \\
\hline
\begin{tabular}[l]{@{}l@{}}\normalsize{SlashdotZoo}\\\normalsize{~\citesup{Konect-SlashdotZoo-Social}\citesup{Konect}\citesup{kunegis:slashdot-zoo}}\end{tabular}&Konect&79120&467869&5.9&2537&D&&Signed social network of users of the technology news site Slashdot connected by directed "friend" and "foe" relations. \\
\hline
\begin{tabular}[l]{@{}l@{}}\normalsize{TRECWT10g}\\\normalsize{~\citesup{Konect-TRECWT10g-Reference}\citesup{Konect}\citesup{b397}}\end{tabular}&Konect&1601787&6679248&4.2&25609&D&&Web Research Collections (TREC Web, Terabyte and Blog Tracks) \\
\hline
\begin{tabular}[l]{@{}l@{}}\normalsize{TwitterICWSM}\\\normalsize{~\citesup{Konect-TwitterICWSM-Social}\citesup{Konect}\citesup{konect:choudhury10}}\end{tabular}&Konect&465017&833541&1.8&677&D&&Who follows whom on Twitter. \\
\hline
%\begin{tabular}[l]{@{}l@{}}\normalsize{UCIrvineMessages}\\\normalsize{~\citesup{Konect-UCIrvineMessages-Communication}\citesup{Konect}\citesup{konect:opsahl09}}\end{tabular}&Konect&899&7019&7.8&126&D&5&Messages between the users of an online community of students from the University of California  Irvine. \\
%\hline
%\begin{tabular}[l]{@{}l@{}}\normalsize{URoviraIVirgili}\\\normalsize{~\citesup{Konect-URoviraIVirgili-Communication}\citesup{Konect}\citesup{konect:guimera03}}\end{tabular}&Konect&1133&5451&4.8&71&D&&Email communication network at the University Rovira i Virgili in Tarragona in the south of Catalonia in Spain. \\
%\hline
%\begin{tabular}[l]{@{}l@{}}\normalsize{USAirports}\\\normalsize{~\citesup{Konect-USAirports-Physical}\citesup{Konect}}\end{tabular}&Konect&1574&17215&10.9&314&D&&Flights between US airports in 2010.  \\
%\hline
\begin{tabular}[l]{@{}l@{}}\normalsize{USpatents}\\\normalsize{~\citesup{Konect-USpatents-Reference}\citesup{Konect}\citesup{b376}}\end{tabular}&Konect&3774768&16518947&4.4&793&D&&Citation network of patents registered with the United States Patent and Trademark Office. \\
\hline
%\begin{tabular}[l]{@{}l@{}}\normalsize{USPowerGrid}\\\normalsize{~\citesup{Konect-USPowerGrid-Physical}\citesup{Konect}\citesup{konect:duncan98}}\end{tabular}&Konect&4941&6594&1.3&19&U&&The power grid of the Western States of the United States of America.  \\
%\hline
\begin{tabular}[l]{@{}l@{}}\normalsize{WikipediaChinese}\\\normalsize{~\citesup{Konect-WikipediaChinese-Reference}\citesup{Konect}\citesup{konect:zhishi}}\end{tabular}&Konect&1930270&8956902&4.6&29005&D&&Network of hyperlinks between the articles of Wikipedia in Chinese. \\
\hline
\begin{tabular}[l]{@{}l@{}}\normalsize{WikipediaEnglish}\\\normalsize{~\citesup{Konect-WikipediaEnglish-Reference}\citesup{Konect}\citesup{konect:mislove2}}\end{tabular}&Konect&1870709&36532531&19.5&226073&D&75&Hyperlink network of the English Wikipedia with edge arrival times. \\
\hline
\begin{tabular}[l]{@{}l@{}}\normalsize{WikipediaLinksDE}\\\normalsize{~\citesup{Konect-WikipediaLinksDE-Reference}\citesup{Konect}}\end{tabular}&Konect&3783012&68714064&18.2&437732&D&&Wikilinks inside the German Wikipedia. \\
\hline
\begin{tabular}[l]{@{}l@{}}\normalsize{WikipediaLinksFR}\\\normalsize{~\citesup{Konect-WikipediaLinksFR-Reference}\citesup{Konect}}\end{tabular}&Konect&4905934&104591689&21.3&1274642&D&&Wikilinks inside the French Wikipedia.  \\
\hline
\begin{tabular}[l]{@{}l@{}}\normalsize{WikipediaLinksIT}\\\normalsize{~\citesup{Konect-WikipediaLinksIT-Reference}\citesup{Konect}}\end{tabular}&Konect&2790019&86754664&31.1&825147&D&&Wikilinks inside the Italian Wikipedia. \\
\hline
\begin{tabular}[l]{@{}l@{}}\normalsize{WikipediaLinksJA}\\\normalsize{~\citesup{Konect-WikipediaLinksJA-Reference}\citesup{Konect}}\end{tabular}&Konect&2140579&58200970&27.2&390239&D&&Wikilinks inside the Japanese Wikipedia. \\
\hline
\begin{tabular}[l]{@{}l@{}}\normalsize{WikipediaLinksPL}\\\normalsize{~\citesup{Konect-WikipediaLinksPL-Reference}\citesup{Konect}}\end{tabular}&Konect&1646203&41216900&25.0&215361&D&&Wikilinks inside the Polish Wikipedia. \\
\hline
\begin{tabular}[l]{@{}l@{}}\normalsize{WikipediaLinksPT}\\\normalsize{~\citesup{Konect-WikipediaLinksPT-Reference}\citesup{Konect}}\end{tabular}&Konect&2804569&51539953&18.4&628617&D&&Wikilinks inside the Portuguese Wikipedia. \\
\hline
\begin{tabular}[l]{@{}l@{}}\normalsize{WikipediaLinksRU}\\\normalsize{~\citesup{Konect-WikipediaLinksRU-Reference}\citesup{Konect}}\end{tabular}&Konect&2819989&64066427&22.7&587438&D&&Wikilinks inside the Russian Wikipedia. \\
\hline
\begin{tabular}[l]{@{}l@{}}\normalsize{YouTube}\\\normalsize{~\citesup{Konect-YouTube-Social}\citesup{Konect}\citesup{konect:mislove2}}\end{tabular}&Konect&3223589&9376594&2.9&91751&U&7& Video-sharing website on which users can upload share and view videos. Social network of users and their friendship connections \\
\hline
%\begin{tabular}[l]{@{}l@{}}\normalsize{ZacharyKarateClub}\\\normalsize{~\citesup{Konect-ZacharyKarateClub-Social}\citesup{Konect}\citesup{konect:ucidata-zachary}}\end{tabular}&Konect&34&78&2.3&17&U&&The well-known Zachary karate club network.  \\
%\hline
\begin{tabular}[l]{@{}l@{}}\normalsize{Amazon}\\\normalsize{~\citesup{SNAP-Amazon}\citesup{DBLP:journals/corr/abs-1205-6233}}\end{tabular}&SNAP&334863&925872&2.8&549&U&&Based on 'Customers Who Bought This Item Also Bought' feature of the Amazon website. If a product i is frequently co-purchased with product j  the graph contains an undirected edge from i to j. \\
\hline
\begin{tabular}[l]{@{}l@{}}\normalsize{amazon0302}\\\normalsize{~\citesup{SNAP-amazon0302}\citesup{leskovec2007graph}}\end{tabular}&SNAP&262111&899792&3.4&420&D&&Amazon product co-purchasing network from March 2 2003 \\
\hline
\begin{tabular}[l]{@{}l@{}}\normalsize{amazon0312}\\\normalsize{~\citesup{SNAP-amazon0312}\citesup{leskovec2007graph}}\end{tabular}&SNAP&400727&2349869&5.9&2747&D&&Amazon product co-purchasing network from March 12 2003 \\
\hline
%\begin{tabular}[l]{@{}l@{}}\normalsize{amazon0505}\\\normalsize{~\citesup{SNAP-amazon0505}\citesup{leskovec2007graph}}\end{tabular}&SNAP&410236&2439437&5.9&2760&D&&Amazon product co-purchasing network from May 5 2003 \\
%\hline
%\begin{tabular}[l]{@{}l@{}}\normalsize{amazon0601}\\\normalsize{~\citesup{SNAP-amazon0601}\citesup{leskovec2007graph}}\end{tabular}&SNAP&403394&2443408&6.1&2752&D&&Amazon product co-purchasing network from June 1 2003 \\
%\hline
%\begin{tabular}[l]{@{}l@{}}\normalsize{as-733}\\\normalsize{~\citesup{SNAP-as-733}\citesup{leskovec2005graphs}}\end{tabular}&SNAP&6474&12572&1.9&1458&U&&733 daily instances(graphs) from November 8 1997 to January 2 2000 \\
%\hline
\begin{tabular}[l]{@{}l@{}}\normalsize{as-Caida}\\\normalsize{~\citesup{SNAP-as-Caida}\citesup{leskovec2005graphs}}\end{tabular}&SNAP&26475&53381&2.0&2628&D&&The CAIDA AS Relationships Datasets, from January 2004 to November 2007 \\
\hline
\begin{tabular}[l]{@{}l@{}}\normalsize{as-Skitter}\\\normalsize{~\citesup{SNAP-as-Skitter}\citesup{leskovec2005graphs}}\end{tabular}&SNAP&1696415&11095298&6.5&35455&U&&Internet topology graph, from traceroutes run daily in 2005 \\
\hline
\begin{tabular}[l]{@{}l@{}}\normalsize{ca-AstroPh}\\\normalsize{~\citesup{SNAP-ca-AstroPh}\citesup{Leskovec:2007:GED:1217299.1217301}}\end{tabular}&SNAP&18771&198050&10.6&504&U&&Arxiv ASTRO-PH (Astro Physics) collaboration network \\
\hline
\begin{tabular}[l]{@{}l@{}}\normalsize{ca-CondMat}\\\normalsize{~\citesup{SNAP-ca-CondMat}\citesup{Leskovec:2007:GED:1217299.1217301}}\end{tabular}&SNAP&23133&93439&4.0&279&U&&Arxiv COND-MAT (Condense Matter Physics) collaboration network \\
\hline
%\begin{tabular}[l]{@{}l@{}}\normalsize{ca-GrQc}\\\normalsize{~\citesup{SNAP-ca-GrQc}\citesup{Leskovec:2007:GED:1217299.1217301}}\end{tabular}&SNAP&5241&14484&2.8&81&U&&Arxiv GR-QC (General Relativity and Quantum Cosmology) collaboration network \\
%\hline
\begin{tabular}[l]{@{}l@{}}\normalsize{ca-HepPh}\\\normalsize{~\citesup{SNAP-ca-HepPh}\citesup{Leskovec:2007:GED:1217299.1217301}}\end{tabular}&SNAP&12006&118489&9.9&491&U&&Arxiv HEP-PH (High Energy Physics - Phenomenology) collaboration network \\
\hline
%\begin{tabular}[l]{@{}l@{}}\normalsize{ca-HepTh}\\\normalsize{~\citesup{SNAP-ca-HepTh}\citesup{Leskovec:2007:GED:1217299.1217301}}\end{tabular}&SNAP&9875&25973&2.6&65&U&&Arxiv HEP-TH (High Energy Physics - Theory) collaboration network \\
%\hline
\begin{tabular}[l]{@{}l@{}}\normalsize{cit-Patents}\\\normalsize{~\citesup{SNAP-cit-Patents}\citesup{Leskovec:2005:GOT:1081870.1081893}}\end{tabular}&SNAP&3774768&16518947&4.4&793&D&& Citations made by patents granted between 1975 and 1999 \\
\hline
\begin{tabular}[l]{@{}l@{}}\normalsize{cit-HepPh}\\\normalsize{~\citesup{SNAP-cit-HepPh}\citesup{Leskovec:2005:GOT:1081870.1081893}\citesup{gehrke03kddcup}}\end{tabular}&SNAP&34546&420877&12.2&846&D&& Arxiv  HEP-PH (high energy physics phenomenology) citation graph. If a paper i cites paper j  the graph contains a directed edge from i to j \\
\hline
\begin{tabular}[l]{@{}l@{}}\normalsize{cit-HepTh}\\\normalsize{~\citesup{SNAP-cit-HepTh}\citesup{Leskovec:2005:GOT:1081870.1081893}\citesup{gehrke03kddcup}}\end{tabular}&SNAP&27769&352285&12.7&2468&D&& Arxiv  HEP-TH (high energy physics theory) citation graph. If a paper i cites paper j  the graph contains a directed edge from i to j \\
\hline
\begin{tabular}[l]{@{}l@{}}\normalsize{DBLP}\\\normalsize{~\citesup{SNAP-DBLP}\citesup{DBLP:journals/corr/abs-1205-6233}}\end{tabular}&SNAP&317080&1049866&3.3&343&U&&Co-authorship network from DBLP computer science bibliography, where two authors are connected if they publish at least one paper together.  \\
\hline
%\begin{tabular}[l]{@{}l@{}}\normalsize{ego-Facebook}\\\normalsize{~\citesup{SNAP-ego-Facebook}\citesup{mcauley2012learning}}\end{tabular}&SNAP&4039&88234&21.8&1045&U&&Social circles from Facebook (anonymized) \\
%\hline
\begin{tabular}[l]{@{}l@{}}\normalsize{ego-Gplus}\\\normalsize{~\citesup{SNAP-ego-Gplus}\citesup{mcauley2012learning}}\end{tabular}&SNAP&107614&12238285&113.7&20127&D&&Social circles from Google+ \\
\hline
\begin{tabular}[l]{@{}l@{}}\normalsize{ego-Twitter}\\\normalsize{~\citesup{SNAP-ego-Twitter}\citesup{mcauley2012learning}}\end{tabular}&SNAP&81306&1342296&16.5&3383&D&&Social circles from Twitter \\
\hline
\begin{tabular}[l]{@{}l@{}}\normalsize{email-Enron}\\\normalsize{~\citesup{SNAP-email-Enron}\citesup{leskovec2009community}\citesup{klimt2004introducing}}\end{tabular}&SNAP&36692&183831&5.0&1383&U&&Email communication network from Enron \\
\hline
\begin{tabular}[l]{@{}l@{}}\normalsize{email-EuAll}\\\normalsize{~\citesup{SNAP-email-EuAll}\citesup{leskovec2007graph}}\end{tabular}&SNAP&265009&364481&1.4&7636&D&&Email network from a EU research institution \\
\hline
\begin{tabular}[l]{@{}l@{}}\normalsize{epinions}\\\normalsize{~\citesup{SNAP-epinions}\citesup{richardson2003trust}}\end{tabular}&SNAP&75879&405740&5.3&3044&D&&General consumer review site. An online social network of Who-trust-whom. \\
\hline
\begin{tabular}[l]{@{}l@{}}\normalsize{flickr}\\\normalsize{~\citesup{SNAP-flickr}\citesup{mcauley2012image}}\end{tabular}&SNAP&105943&2316952&21.9&5425&U&&Images sharing common metadata on Flickr \\
\hline
\begin{tabular}[l]{@{}l@{}}\normalsize{higgs-twitter-friendship}\\\normalsize{~\citesup{SNAP-higgs-twitter-friendship}\citesup{DMM2013}}\end{tabular}&SNAP&456631&12508442&27.4&51386&D&&Spreading processes of the announcement of the discovery of a new particle with the features of the Higgs boson on 4th July 2012. \\
\hline
\begin{tabular}[l]{@{}l@{}}\normalsize{higgs-twitter-mention}\\\normalsize{~\citesup{SNAP-higgs-twitter-mention}\citesup{DMM2013}}\end{tabular}&SNAP&302523&436816&1.4&22790&D&&Spreading processes of the announcement of the discovery of a new particle with the features of the Higgs boson on 4th July 2012. \\
\hline
%\begin{tabular}[l]{@{}l@{}}\normalsize{higgs-twitter-reply}\\\normalsize{~\citesup{SNAP-higgs-twitter-reply}\citesup{DMM2013}}\end{tabular}&SNAP&37145&28105&0.8&1177&D&&Spreading processes of the announcement of the discovery of a new particle with the features of the Higgs boson on 4th July 2012. \\
%\hline
\begin{tabular}[l]{@{}l@{}}\normalsize{higgs-twitter-retweet}\\\normalsize{~\citesup{SNAP-higgs-twitter-retweet}\citesup{DMM2013}}\end{tabular}&SNAP&425008&732790&1.7&31556&D&&Spreading processes of the announcement of the discovery of a new particle with the features of the Higgs boson on 4th July 2012. \\
\hline
\begin{tabular}[l]{@{}l@{}}\normalsize{LiveJournal}\\\normalsize{~\citesup{SNAP-LiveJournal}\citesup{backstrom-group-2006}\citesup{DBLP:journals/corr/abs-0810-1355}}\end{tabular}&SNAP&4846609&42851237&8.8&20333&D&&Virtual community where Internet users can keep a blog  journal or diary \\
\hline
\begin{tabular}[l]{@{}l@{}}\normalsize{LiveJournalCom}\\\normalsize{~\citesup{SNAP-LiveJournalCom}\citesup{DBLP:journals/corr/abs-1205-6233}}\end{tabular}&SNAP&3997962&34681189&8.7&14815&U&&Virtual community where Internet users can keep a blog  journal or diary \\
\hline
\begin{tabular}[l]{@{}l@{}}\normalsize{loc-brightkite}\\\normalsize{~\citesup{SNAP-loc-brightkite}\citesup{cho2011friendship}}\end{tabular}&SNAP&58228&214078&3.7&1134&U&&Location-based social networking service provider where users shared their locations by checking-in.  \\
\hline
\begin{tabular}[l]{@{}l@{}}\normalsize{loc-gowalla}\\\normalsize{~\citesup{SNAP-loc-gowalla}\citesup{cho2011friendship}}\end{tabular}&SNAP&196591&950327&4.8&14730&U&&Location-based social networking website where users share their locations by checking-in.  \\
\hline
\begin{tabular}[l]{@{}l@{}}\normalsize{Oregon-1-1}\\\normalsize{~\citesup{SNAP-Oregon-1-1}\citesup{leskovec2005graphs}}\end{tabular}&SNAP&10670&22002&2.1&2312&U&&AS peering information inferred from Oregon route-views between March 31 and May 26 2001 \\
\hline
\begin{tabular}[l]{@{}l@{}}\normalsize{Oregon-2-1}\\\normalsize{~\citesup{SNAP-Oregon-2-1}\citesup{leskovec2005graphs}}\end{tabular}&SNAP&10900&31180&2.9&2343&U&&AS peering information inferred from Oregon route-views between March 31 and May 26 2001 \\
\hline
\begin{tabular}[l]{@{}l@{}}\normalsize{p2p-Gnutella31}\\\normalsize{~\citesup{SNAP-p2p-Gnutella31}\citesup{leskovec2007graph}\citesup{ripeanu2002mapping}}\end{tabular}&SNAP&62586&147892&2.4&95&D&&Gnutella peer to peer network from August 31 2002 \\
\hline
\begin{tabular}[l]{@{}l@{}}\normalsize{Pokec}\\\normalsize{~\citesup{SNAP-Pokec}\citesup{takac2012data}}\end{tabular}&SNAP&1632803&22301964&13.7&14854&D&&Pokec is the most popular on-line social network in Slovakia.  \\
\hline
\begin{tabular}[l]{@{}l@{}}\normalsize{roadNet-CA}\\\normalsize{~\citesup{SNAP-roadNet-CA}\citesup{leskovec2009community}}\end{tabular}&SNAP&1965206&2766607&1.4&12&U&&Road network of California \\
\hline
\begin{tabular}[l]{@{}l@{}}\normalsize{roadNet-PA}\\\normalsize{~\citesup{SNAP-roadNet-PA}\citesup{leskovec2009community}}\end{tabular}&SNAP&1088092&1541898&1.4&9&U&&Road network of Pennsylvania \\
\hline
\begin{tabular}[l]{@{}l@{}}\normalsize{roadNet-TX}\\\normalsize{~\citesup{SNAP-roadNet-TX}\citesup{leskovec2009community}}\end{tabular}&SNAP&1379917&1921660&1.4&12&U&&Road network of Texas \\
\hline
\begin{tabular}[l]{@{}l@{}}\normalsize{slashdot1}\\\normalsize{~\citesup{SNAP-slashdot1}\citesup{DBLP:journals/corr/abs-0810-1355}}\end{tabular}&SNAP&77360&469180&6.1&2539&D&&Technology-related news website. Friend/foe network. Obtained in November 2008. \\
\hline
%\begin{tabular}[l]{@{}l@{}}\normalsize{slashdot2}\\\normalsize{~\citesup{SNAP-slashdot2}\citesup{DBLP:journals/corr/abs-0810-1355}}\end{tabular}&SNAP&82168&504230&6.1&2552&D&&Technology-related news website. Friend/foe network. Obtained in February 2009. \\
%\hline
%\begin{tabular}[l]{@{}l@{}}\normalsize{soc-sign-epinions}\\\normalsize{~\citesup{SNAP-soc-sign-epinions}\citesup{leskovec2010signed}}\end{tabular}&SNAP&131580&711210&5.4&3558&D&&Epinions signed social network \\
%\hline
%\begin{tabular}[l]{@{}l@{}}\normalsize{soc-sign-Slashdot081106}\\\normalsize{~\citesup{SNAP-soc-sign-Slashdot081106}\citesup{leskovec2010signed}}\end{tabular}&SNAP&77350&468554&6.1&2537&D&&Slashdot Zoo signed social network from November 6 2008 \\
%\hline
%\begin{tabular}[l]{@{}l@{}}\normalsize{soc-sign-Slashdot090216}\\\normalsize{~\citesup{SNAP-soc-sign-Slashdot090216}\citesup{leskovec2010signed}}\end{tabular}&SNAP&81867&497672&6.1&2546&D&&Slashdot Zoo signed social network from February 16 2009 \\
%\hline
%\begin{tabular}[l]{@{}l@{}}\normalsize{soc-sign-Slashdot090221}\\\normalsize{~\citesup{SNAP-soc-sign-Slashdot090221}\citesup{leskovec2010signed}}\end{tabular}&SNAP&82140&500481&6.1&2548&D&&Slashdot Zoo signed social network from February 21 2009 \\
%\hline
\begin{tabular}[l]{@{}l@{}}\normalsize{web-BerStan}\\\normalsize{~\citesup{SNAP-web-BerStan}\citesup{DBLP:journals/corr/abs-0810-1355}}\end{tabular}&SNAP&685230&6649470&9.7&84230&D&&Nodes represent pages from berkely.edu and stanford.edu domains and directed edges represent hyperlinks between them. The data was collected in 2002. \\
\hline
\begin{tabular}[l]{@{}l@{}}\normalsize{web-Google}\\\normalsize{~\citesup{SNAP-web-Google}\citesup{DBLP:journals/corr/abs-0810-1355}}\end{tabular}&SNAP&875713&4322051&4.9&6332&D&&Nodes represent web pages and directed edges represent hyperlinks between them. The data was released in 2002 by Google as a part of Google Programming Contest \\
\hline
\begin{tabular}[l]{@{}l@{}}\normalsize{web-NotreDame}\\\normalsize{~\citesup{SNAP-web-NotreDame}\citesup{albert1999internet}}\end{tabular}&SNAP&325729&1090108&3.3&10721&D&&Nodes represent pages from University of Notre Dame (domain nd.edu) and directed edges represent hyperlinks between them. The data was collected in 1999 by Albert  Jeong and Barabasi. \\
\hline
\begin{tabular}[l]{@{}l@{}}\normalsize{web-Stanford}\\\normalsize{~\citesup{SNAP-web-Stanford}\citesup{DBLP:journals/corr/abs-0810-1355}}\end{tabular}&SNAP&281903&1992636&7.1&38625&D&&Nodes represent pages from Stanford University (stanford.edu) and directed edges represent hyperlinks between them.  \\
\hline
\begin{tabular}[l]{@{}l@{}}\normalsize{wiki-talk}\\\normalsize{~\citesup{SNAP-wiki-talk}\citesup{leskovec2010signed}\citesup{leskovec2010predicting}}\end{tabular}&SNAP&2394385&4659565&1.9&100029&D&&Wikipedia's  registered users talk pages.  A directed edge from node i to node j represents that user i at least once edited a talk page of user j. \\
\hline
%\begin{tabular}[l]{@{}l@{}}\normalsize{wiki-vote}\\\normalsize{~\citesup{SNAP-wiki-vote}\citesup{leskovec2010signed}\citesup{leskovec2010predicting}}\end{tabular}&SNAP&7115&100762&14.2&1065&D&&Wikipedia administrators elections voting graph. A directed edge from node i to node j represents that user i voted on user j. \\
%\hline
\begin{tabular}[l]{@{}l@{}}\normalsize{YouTube}\\\normalsize{~\citesup{SNAP-YouTube}\citesup{DBLP:journals/corr/abs-1205-6233}}\end{tabular}&SNAP&1134890&2987624&2.6&28754&U&& Video-sharing website on which users can upload share and view videos. Social network of users and their friendship connections \\
\hline

\hline \hline

%\end{tabular}
%\caption{Basic properties of the examined networks and models 
%(number of vertices, number of edges) and the Axiom constants $c_1, c_2, c_3$ 
%when the elite is the $\sqrt{m}$-rich-club. 
%The average and standard deviation pertain to the real networks only. 
%A gray cell indicates a large deviation of the model from the average 
%of the real data. 
%}
%

%\label{table:prop}
%\end{table*}
\label{table:prop}
\end{longtable}
\end{center}

\end{landscape}

%%%%%%%%%%%%%%%%%%%%%%%
\section{Dominance and robustness}
%%%%%%%%%%%%%%%%%%%%%%%
%\begin{table*}[htb!]
	%\centering \footnotesize
		%\begin{tabular}{|l| r |r||r|r||r|r||c|}
	
\begin{center}
\begin{longtable}{|l||r|r||r|r||}

%\caption[d]{d.} \label{dd} \\

\endfirsthead

\multicolumn{5}{c}%
{{\bfseries \tablename\ \thetable{} -- continued from previous page}} \\

%\hline \multicolumn{1}{|c|}{\textbf{Time (s)}} &
%\multicolumn{1}{c|}{\textbf{Triple chosen}} &
%\multicolumn{1}{c|}{\textbf{Other feasible triples}} \\ \hline 

\hline
\multicolumn{1}{|c||}{$$} &  \multicolumn{2}{c||}{$\rho_d$} &	 \multicolumn{2}{c||}{$\rho_r$}  \\ \hline

\multicolumn{1}{|c||}{Data}  &  \multicolumn{1}{c|}{$\sqrt{m}$} &	 \multicolumn{1}{c||}{SP}&	 \multicolumn{1}{c|}{$\sqrt{m}$} &	 \multicolumn{1}{c||}{SP} \\ \hline

\endhead

\hline \multicolumn{5}{|r|}{{Continued on next page}} \\ \hline
\endfoot

\hline \hline

\caption[Domination and Robustness Values of All Networks]{Domination and robustness values at the symmetry point and $\sqrt{m}$ of all Networks.} 
\label{table:dom_rob_all_networks} \\

\endlastfoot

\hline
\multicolumn{1}{|c||}{$$} &  \multicolumn{2}{c||}{$\rho_d$} &	 \multicolumn{2}{c||}{$\rho_r$}  \\ \hline

\multicolumn{1}{|c||}{Data}  &  \multicolumn{1}{c|}{$\sqrt{m}$} &	 \multicolumn{1}{c||}{SP}&	 \multicolumn{1}{c|}{$\sqrt{m}$} &	 \multicolumn{1}{c||}{SP} \\ \hline

Blog&2.64&3.54&0.19&0.28 \\
Blog2&6.52&6.16&0.17&0.16 \\
Blog3&3.23&3.58&0.25&0.28 \\
Buzznet&2.47&4.30&0.10&0.23 \\
Delicious&0.35&2.62&0.04&0.38 \\
Digg&0.88&1.96&0.13&0.51 \\
Douban&0.18&4.50&0.07&0.22 \\
Flickr&0.72&1.38&0.31&0.73 \\
Flixster&0.40&5.50&0.02&0.18 \\
Foursquare&1.20&2.35&0.04&0.43 \\
Friendster&0.54&13.79&0.01&0.07 \\
Hyves&0.32&6.28&0.02&0.16 \\
LastFm&0.44&4.06&0.03&0.25 \\
LiveJournal&0.33&3.07&0.02&0.33 \\
Livemocha&0.81&2.57&0.10&0.39 \\
Twitter&3.15&8.40&0.04&0.12 \\
YouTube&0.51&1.89&0.07&0.53 \\
%YouTube2&0.56&1.96&0.04&0.51 \\
YouTube 2&1.40&1.72&0.43&0.58 \\
YouTube 3&2.66&2.16&0.60&0.46 \\
YouTube 4&1.49&1.52&0.62&0.66 \\
YouTube 5&3.21&2.39&0.60&0.42 \\
Academia&0.27&1.65&0.06&0.61 \\
AnyBeat&2.17&3.11&0.19&0.32 \\
GooglePlus&0.38&0.91&0.19&1.10 \\
TheMarkerCafe&1.46&2.60&0.18&0.38 \\
%Advogato Social&1.07&2.17&0.18&0.46 \\
arXivHep-thKDDCup Reference&0.45&1.34&0.10&0.75 \\
BaiduInternal Reference&0.54&2.63&0.03&0.38 \\
BaiduRelated Reference&1.02&2.86&0.19&0.35 \\
%CaenorhabditisElegans Contact&2.47&2.84&0.25&0.35 \\
CatsterDogster Social&1.08&2.32&0.17&0.43 \\
Catster Social&3.57&4.12&0.20&0.24 \\
CiteSeer Reference&0.19&2.12&0.01&0.47 \\
CoraCitation Reference&0.30&1.83&0.05&0.55 \\
DBLP Contact&0.09&1.36&0.07&0.73 \\
%DBLP Reference&0.49&2.01&0.09&0.50 \\
Digg Communication&0.34&1.99&0.09&0.50 \\
Dogster Social&1.32&2.93&0.11&0.34 \\
Epinions Social&0.88&1.85&0.18&0.54 \\
%Euroroad Physical&0.17&1.61&0.07&0.62 \\
Facebook Communication&0.18&1.22&0.10&0.82 \\
Facebook WOSN  Social&0.25&1.13&0.16&0.89 \\
FlickrLinks Social&0.62&1.19&0.22&0.84 \\
Flickr Social&0.60&1.22&0.20&0.82 \\
%FloridaEcosystemDry Physical&2.83&2.50&0.45&0.40 \\
%FloridaEcosystemWet Physical&2.83&2.50&0.45&0.41 \\
%GoogleComInternal Reference&1.38&1.97&0.16&0.51 \\
%HamstersterFriendships Social&1.17&2.14&0.23&0.47 \\
%HamstersterFull Social&0.79&1.40&0.29&0.72 \\
%HighlandTribes Social&2.70&1.93&0.78&0.63 \\
HudongInternal Reference&0.81&3.89&0.01&0.26 \\
HudongRelated Reference&0.55&9.61&0.01&0.10 \\
%Hypertext2009 Contact&3.76&2.38&0.66&0.44 \\
%Infectious Contact&0.58&1.21&0.35&0.84 \\
%JDKDependency Reference&3.27&3.25&0.20&0.31 \\
%JUNGDependency Reference&3.03&2.99&0.24&0.34 \\
LibimsetiCZ Social&1.15&3.18&0.08&0.31 \\
%ManufacturingEmails Communication&5.42&2.77&0.69&0.37 \\
%OpenFlights Physical&1.18&1.80&0.35&0.56 \\
PrettyGoodPrivacy Contact&0.26&0.85&0.36&1.18 \\
SlashdotZoo Social&0.63&2.14&0.11&0.47 \\
Slashdot Communication&0.55&2.40&0.09&0.42 \\
TRECWT10g Reference&0.54&3.46&0.01&0.29 \\
TwitterICWSM Social&1.17&57.05&0.01&0.02 \\
%UCIrvineMessages Communication&1.16&2.40&0.22&0.42 \\
%URoviraIVirgili Communication&0.60&1.69&0.17&0.60 \\
%USAirports Physical&2.51&2.10&0.71&0.49 \\
USpatents Reference&0.03&1.40&0.04&0.71 \\
%USPowerGrid Physical&0.13&1.83&0.06&0.55 \\
WikipediaChinese Reference&0.54&2.54&0.06&0.39 \\
WikipediaEnglish Reference&0.73&3.07&0.03&0.33 \\
WikipediaLinksDE Reference&0.58&2.62&0.01&0.38 \\
WikipediaLinksFR Reference&1.00&3.16&0.01&0.32 \\
WikipediaLinksIT Reference&0.99&2.62&0.03&0.38 \\
WikipediaLinksJA Reference&0.72&2.57&0.03&0.39 \\
WikipediaLinksPL Reference&0.63&1.97&0.08&0.51 \\
WikipediaLinksPT Reference&1.17&3.19&0.05&0.31 \\
WikipediaLinksRU Reference&0.76&2.24&0.02&0.45 \\
YouTube Social&0.77&2.16&0.03&0.46 \\
%ZacharyKarateClub Social&3.83&4.25&0.43&0.29 \\
Amazon&0.07&3.05&0.01&0.33 \\
amazon0302&0.07&2.14&0.02&0.47 \\
amazon0312&0.11&2.42&0.01&0.41 \\
%amazon0505&0.11&2.38&0.02&0.42 \\
%amazon0601&0.11&2.42&0.02&0.41 \\
%as-733&2.56&6.11&0.08&0.16 \\
as-Caida&2.18&6.03&0.07&0.17 \\
as-Skitter&0.65&2.82&0.03&0.35 \\
ca-AstroPh&0.35&1.17&0.20&0.85 \\
ca-CondMat&0.23&1.41&0.11&0.71 \\
%ca-GrQc&0.10&0.77&1.95&1.30 \\
ca-HepPh&0.34&0.42&1.49&2.41 \\
%ca-HepTh&0.18&1.35&0.25&0.74 \\
cit-HepPh&0.31&1.62&0.07&0.62 \\
cit-HepTh&0.45&1.34&0.10&0.75 \\
cit Patents&0.03&1.40&0.04&0.71 \\
DBLP&0.08&1.43&0.20&0.70 \\
%ego-Facebook&0.43&0.65&0.72&1.54 \\
ego-Gplus&1.31&1.83&0.30&0.55 \\
ego-Twitter&0.42&1.76&0.11&0.57 \\
email-Enron&0.90&1.76&0.16&0.57 \\
email-EuAll&4.13&10.07&0.05&0.10 \\
epinions&0.76&1.65&0.20&0.61 \\
flickr&0.35&0.93&0.55&1.08 \\
higgs-twitter-friendship&1.06&3.45&0.02&0.29 \\
higgs-twitter-mention&1.36&5.63&0.01&0.18 \\
%higgs-twitter-reply&0.34&2.15&0.01&0.46 \\
higgs-twitter-retweet&1.63&6.83&0.01&0.15 \\
LiveJournal&0.12&1.04&0.10&0.96 \\
LiveJournalCom&0.10&1.08&0.11&0.93 \\
loc-brightkite&0.39&1.39&0.18&0.72 \\
loc-gowalla&0.38&1.23&0.13&0.81 \\
Oregon-1-1&2.81&5.86&0.08&0.17 \\
%Oregon-1-2&2.87&6.08&0.08&0.16 \\
%Oregon-1-3&2.91&5.89&0.08&0.17 \\
%Oregon-1-4&2.93&5.81&0.08&0.17 \\
%Oregon-1-5&2.93&6.00&0.08&0.17 \\
%Oregon-1-6&2.92&6.01&0.08&0.17 \\
%Oregon-1-7&2.93&6.01&0.08&0.17 \\
%Oregon-1-8&2.96&6.05&0.08&0.17 \\
%Oregon-1-9&2.92&5.74&0.08&0.17 \\
Oregon-2-1&1.97&3.10&0.18&0.32 \\
%Oregon-2-2&2.04&3.19&0.18&0.31 \\
%Oregon-2-3&2.03&3.07&0.18&0.33 \\
%Oregon-2-4&2.11&3.18&0.17&0.31 \\
%Oregon-2-5&2.10&3.25&0.17&0.31 \\
%Oregon-2-6&2.13&3.39&0.16&0.30 \\
%Oregon-2-7&2.13&3.37&0.16&0.30 \\
%Oregon-2-8&2.13&3.24&0.17&0.31 \\
%Oregon-2-9&2.20&3.22&0.16&0.31 \\
%%p2p-Gnutella04&0.20&1.90&0.03&0.53 \\
%%p2p-Gnutella05&0.24&1.82&0.03&0.55 \\
%%p2p-Gnutella06&0.22&1.87&0.06&0.54 \\
%%p2p-Gnutella08&0.37&1.83&0.08&0.55 \\
%%p2p-Gnutella09&0.31&1.84&0.06&0.54 \\
%%p2p-Gnutella24&0.11&2.33&0.02&0.43 \\
%%p2p-Gnutella25&0.11&2.44&0.03&0.41 \\
%%p2p-Gnutella30&0.10&2.22&0.02&0.45 \\
p2p-Gnutella31&0.09&2.27&0.02&0.44 \\
Pokec&0.08&1.53&0.04&0.65 \\
roadNet-CA&0.01&0.87&0.01&1.15 \\
roadNet-PA&0.01&0.97&0.01&1.03 \\
roadNet-TX&0.01&0.87&0.01&1.16 \\
slashdot1&0.63&2.14&0.11&0.47 \\
%slashdot2&0.61&2.08&0.11&0.48 \\
%soc-sign-epinions&0.88&1.85&0.18&0.54 \\
%soc-sign-Slashdot081106&0.63&2.14&0.11&0.47 \\
%soc-sign-Slashdot090216&0.61&2.08&0.11&0.48 \\
%soc-sign-Slashdot090221&0.61&2.08&0.11&0.48 \\
web-BerStan&1.12&2.30&0.01&0.43 \\
web-Google&0.26&2.31&0.01&0.43 \\
web-NotreDame&0.42&1.66&0.12&0.60 \\
web-Stanford&1.14&2.56&0.01&0.39 \\
wiki-talk&3.06&9.48&0.05&0.11 \\
%wiki-vote&1.30&2.24&0.23&0.45 \\
YouTube&0.57&1.96&0.04&0.51 \\
\hline \hline
Median&0.59&2.28&0.07&0.44 \\
Average&0.95&3.41&0.13&0.51 \\
STD&1.06&5.90&0.19&0.33 \\
Min&0.01&0.42&0.01&0.02 \\

\hline \hline

%\end{tabular}
%\caption{Basic properties of the examined networks and models 
%(number of vertices, number of edges) and the Axiom constants $c_1, c_2, c_3$ 
%when the elite is the $\sqrt{m}$-rich-club. 
%The average and standard deviation pertain to the real networks only. 
%A gray cell indicates a large deviation of the model from the average 
%of the real data. 
%}
%

%\label{table:prop}
%\end{table*}

\end{longtable}
\end{center}

\bibliographystylesup{acm}
\bibliographysup{social}
\end{document}